\documentclass[english, LaM, oneside]{sapthesis}
\usepackage[utf8]{inputenx}
\usepackage{indentfirst}
\usepackage{microtype}
\usepackage{multicol}
\usepackage{multirow}
\usepackage{lettrine}
\usepackage[nottoc, notlof, notlot]{tocbibind}
\usepackage{hyperref}
\hypersetup{
			hyperfootnotes=true,			
			bookmarks=true,			
			colorlinks=true,
			linkcolor=red,
                        linktoc=page,
			anchorcolor=black,
			citecolor=red,
			urlcolor=blue,
			pdftitle={A sample Bachelor's thesis for Sapienza Università di Roma},
			pdfauthor={FirstName LastName},
			pdfkeywords={thesis, sapienza, roma, university}
 }

\title{Memory Storage and Retrieval in \\Sparsely Connected Balanced Networks}
\author{Enrico Ventura}
\IDnumber{1699775}
\course[]{Fisica}
\courseorganizer{Facolt\`a di Scienze Matematiche Fisiche e Naturali}
\submitdate{2019/2020}
\copyyear{2020}
\advisor{Prof. Giancarlo Ruocco}
\coadvisor{Dr. Gianluigi Mongillo}
\authoremail{ventura.1699775@studenti.uniroma1.it}

\begin{document}

\frontmatter
\maketitle

\tableofcontents
\mainmatter
\chapter*{Introduction}
\addcontentsline{toc}{chapter}{Introduction}
Ever since the last two decades of the past century pioneering studies in the field of statistical physics had focused their efforts on developing models of neural networks that could display memory storage and retrieval \cite{ref:amitbook}. These studies have permitted to better understand how to model memory retrieving processes and similar phenomena in networks, contributing to several disciplines, such as Theoretical Neuroscience, Biology, Social Sciences, and many more \cite{ref:hop} \cite{ref:bio} \cite{ref:boschi}.\\ Though many of these models were easy to handle and still quite effective to explain the basic memory retrieval processes in the brain, they were not satisfactory under the biological point of view. It became clear to scientists that a biologically realistic neural network should have respected typical features that were observed in experiments of neurophysiology. One of these qualities regarded the fact that 
many regions of the brain display a net synaptic input whose mean and fluctuations are both on the
order of the spiking threshold, implying the emergence of a typical asynchronous and irregular firing pattern \cite{ref:cats} of the neuron. This observation has led to the introduction of Balanced Networks, systems where  $\textit{excitatory}$ and $\textit{inhibitory}$ neurons balance their effect on each other as an emergent property of the network dynamics \cite{ref:van}. Another peculiarity of such models is the exhibition of a mean level of neuronal activity (namely the average spiking rate of neurons) that is univocally defined by a linear equation in the external input. This aspect might help to reproduce what is measured in particular areas devoted to memory storage (e.g. the pre-frontal cortex), that is a persistent activity during the memory retrieval performance \cite{ref:curtis} \cite{ref:consta}. Even though progresses in the matter of balanced networks where achieved in the last two decades, there is still no complete theory that conciliates memory retrieval and balance in a network of neurons.\\Therefore, the aim of this work is to develop a biologically plausible model that presents both balance and memory retrieval, building on a framework of mean field equations that can predict the theoretical behaviour of the network under the choice of a set of control parameters. We will thus measure the critical capacity of the system as a function of these parameters, comparing the theoretical results with the numerical simulations. 
\\Our research has followed a series of progressive steps: starting from a first introduction and description of a balanced network model for an inhibitory population of neurons we have then stored patterns, developing a proper mean field theory of memory retrieval for a balanced neural network.\\ \\Firstly, in Chapter \ref{chap:1}, an overview of the theory underlying attractor neural networks and balance networks is presented. The goal of the review is to set the basis for the understanding of both memory retrieval processes and balance, two aspects shared by the original model proposed by this work.\\ \\Furthermore, Chapter \ref{chap:2} introduces the model of a random balanced network describing a population of inhibitory neurons in the Derrida-Gardner-Zippelius regime of extreme dilution. In particular, this regime will allow us to justify the balance condition as well as the mean field analysis applied in the rest of our research work. In addition, the stability of the fixed point of the dynamics is investigated as an interesting insight to understand the attractor dynamics of the model.\\ \\ Eventually, the main results of the thesis are contained in Chapter \ref{chap:3}, where the model of Balanced Network characterized in the previous chapter is generalized to store a set of random generated patterns. A set of equations that define the fixed point state of the system is derived and solved to be later compared with the numerical simulations. The maximum storage capacity of the system is expressed as a function of the control parameters of the model and studied in the limit of sparse coding level. 

\chapter{From Attractor Neural Networks to Balanced Networks}
\label{chap:1} 
Since the early decades of the past century the primary goal of theoretical neuroscience has been to build mathematical models to describe neurons. Experiments had revealed the neuronal cell to be composed of a body ($\textit{soma}$), separated from the exterior by the $\textit{membrane}$; the $\textit{axon}$, that is a "wire" responsible for the information going out of the neuron; the $\textit{dendrites}$, branches connected to the soma that let the inputs from other neurons to come in. The junction role between the axon and the dendrites is performed by the so called $\textit{synapses}$. The neuronal membrane is composed of different kinds of ionic channels, gates that let particular ions to be exchanged with the environment, allowing the formation of a difference in potential. Channels are activated (or inactivated) by particular chemical messangers received by the presynaptic neurons, called $\textit{neurotransmitters}$. When the neuronal potential reaches a certain threshold in voltage an action potential, or $\textit{spike}$, is emitted and propagates to the neighbour cells: in this case we say that the neuron $\textit{fires}$, sending voltage signals to other cells. Hence it looked evident that neurons could be model as simple units interacting with each other, that is a picture that fits well with the framework of Statistical Mechanics.\\In more recent times people have been interested in exploiting both system of interacting agents in physics and concepts of neuroscience to build models that could reproduce the memory retrieval processes in the brain. Attractor neural networks were then introduced and developed, reaching a wide fame all over the scientific scenario. On the other side, an increasing complexity in the description of the neuronal states of those regions of the brain dedicated to memory storage forced neuroscientists to elaborate more realistic models, which took into account the different identity of neurons and their role in the network. One result of these particular studies is the balanced network, interesting manifestation of the collective behaviour of different kinds of neurons that cooperate to stabilize the mean presynaptic input around the spiking threshold, as it has been experimentally measured in many areas of the brain. This Chapter is dedicated to an exploration of the world of biological neural networks, beginning from the simplest examples of attractor neural networks and ending up with an introduction to balanced networks, explaining their functioning and purpose. 

\section{Hebbian Theory of Memory and Attractor Neural Networks.}
\label{sec:hebbs}
A central idea in Neuroscience about the the way memories develop in the brain has been introduced by D.O. Hebb in 1949 in his "The Organization of Behaviour". Hebb's intuition considers the formation of \textit{neural routes} \cite{ref:hebb32} through the dynamical organization of synapses. These routes of neuronal activity in the brain take the name of \textit{patterns} and their role is resumed by the celebrated statement \begin{quote}
"When an axon of cell A is near enough to excite a cell B and repeatedly or persistently takes part in firing it, some growth process or metabolic change takes place in one or both cells such that A's efficiency, as one of the cells firing B, is increased." \cite{ref:hebbook}    
\end{quote}
According to this explanation, synapses that link neurons in a biological neural network are the product of a plastic process that increases the contact area between the afferent axon and the soma of certain cells, enforcing their spatial and temporal correlation (\textit{synaptic plasticity}). The presence of this trace favours the emergence of a \textit{cell-assembly} that is a particular configuration of firing cells in the network associated to one memory or an idea.\\The power of the Hebbian theory thus consists of considering the memories to be stored in the inner synaptic architecture of the brain and not in the neurons themselves. The structure of the network then influences its firing dynamics and so, at last, the behaviour of the individual.\\This conception was suitable enough to be translated into the mathematical framework of the neural networks that had already been introduced by McCulloch and Pitts in 1943 \cite{ref:pitts}. Neurons are treated as simple units that can be active (firing state) or silent. This interpretation is included in the \textit{integrated and fire} picture that will be more extensively described in Subsection (\ref{subsec:LIF}). The McCulloch-Pitts unit $\nu_i$ is linked to other $j$ units through a coupling $w_{ij}$ that we are going to call the $\textit{synaptic efficacy}$ of the $j$ neuron. This quantity recalls the voltage signal emitted by neuron $j$ and directed to $i$ in real neural networks. Each neuron then receives a presynaptic input given by the summation of the contributions by all the other neurons in the network
\begin{equation}
\label{eq:local_field}
h_i = \sum_{j=1}^N w_{ij}\nu_j
\end{equation}
Where $N$ is the number of neurons in the network. This quantity will be called $\textit{local field}$ for the rest of the work. The incoming signal is then interpreted according to a function that compares it to a  threshold $\theta$, such that
\[ 
\nu_i = \phi\left(h_i - \theta\right) = 
\left\{ \begin{array}{l}
0\hspace{1cm}(h_i-\theta)\leq0\\
1\hspace{1cm}(h_i-\theta)>0
\end{array} \right. 
\]
$\nu_i = 0$ represents the silence of the cell, while $\nu_i = 1$ depicts the firing neuron.\\The most popular way to insert memory in the network is to build up the connectivity matrix of synaptic efficacies $w_{ij}$ according to a particular rule that reflects the Hebbian Theory. Let us call pattern a vector $\vec\xi$ such that each entry $ \xi_i = 0,1$ represents the activation state of neurons belonging to a special configuration we have decided \textit{a priori}. One can imagine to train the network with a number $P$ of patterns that are generated independently with each other and which are embedded in the definition of the synaptic efficacies. 
If we build a model where such patterns are fixed points of the network dynamics and these fixed points are stable, a particular initial configuration of the network (\textit{stimulus}) will retrieve an output configuration being close enough to one of the stored patterns, as $P$ is not too large. These type of systems, describing processes related to the \textit{associative memory}, are called Attractor Neural Networks. ANNs basically accomplish a classification task: different classes of similar stimuli correspond to distinct basins of attraction of the patterns.
\begin{figure}[h!!]
    \centering
		\includegraphics[width=12cm]{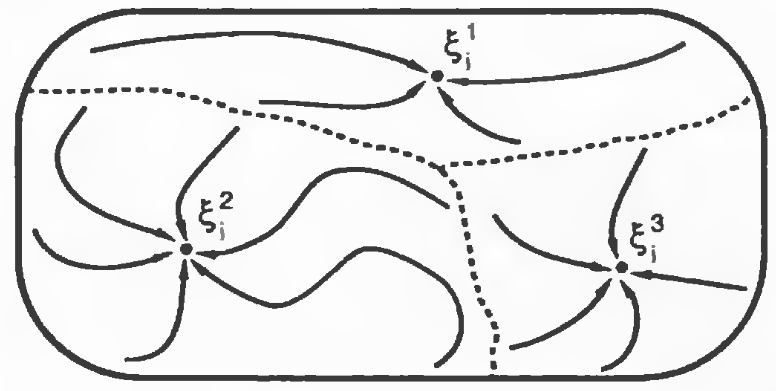}
		\label{attract}
		\caption{\textbf{Intuitive representations of patterns as attractors of the dynamics in the phase space of the network.} Image from \cite{ref:hertz}. Input stimuli that are included in the basins of attractions of a given pattern will retrieve the pattern.}
		\centering
\end{figure}

\subsection{Biological Evidence of the Attractors}
Hebb's hypothesis about the plasticity of synapses and the consequent formation of cell assemblies to explain retrieval of memories by the brain has been repeatedly confirmed by experiments. Works by Miyashita \cite{ref:miya1} \cite{ref:miya2} have proved the emergence of \textit{reverberations}  \cite{ref:amitmiya} of activity locally distributed in the cortex, each one being related to a class of stimuli. During the experiments, sequences of fractal images were displayed to monkeys, training them to recognize identical figures. Afterwards, pictures that were similar, but not coincident, to the initial ones, were exhibited to the animals. At the same time a higher, time-persistent activity in a small region in the cortex was captured by electrodes. The spatial localization of these assemblies is due to the well known tendency of correlated cortical neurons to cluster \cite{ref:perin}. In particular, different assemblies were related to different groups of stimuli, classified depending on their visual correlation, with a few neurons shared among the assemblies. The long-term activity in these clustered groups of neurons is an evidence of the existence of attractors in the network, and a consequence of the formation of patterns of activity \textit{impressed} into synapses in accordance to the Hebbian picture.

\section{The Hopfield Model}
The Hopfield Model \cite{ref:hop} is one of the most celebrated associative memory models. The model consists of a set of variables $\{\sigma_i = 2\nu_i - 1\}$, where $\nu_i$ are the McCulloch-Pitts neurons, such that $\sigma_i = \{-1,+1\}_{i=1}^N$. Let us implement a deterministic dynamics that is given by the following rule
\begin{equation}
\label{eq:dyn}
\sigma_i^{(t+1)} = \text{sgn}\left( \sum_jw_{ij}\sigma_j^{(t)} - \theta\right)
\end{equation}
where $\theta$ is a threshold and 
\[ 
\text{sgn}(x) = 
\left\{ \begin{array}{l}
-1\hspace{1cm}x<0\\
+1\hspace{1cm}x\geq0
\end{array} \right. 
\]
A number $P$ of patterns are stored in the network, as vectors $\vec{\xi^{\mu}}$ having entries $\{\xi_i^{\mu} = \pm 1\}_{\mu = 1}^P$ generated with a probability $p = 1/2$. The memory storage of the Hopfield model is an extensive property of the system, so one can define the $\textit{load parameter}$ $\alpha$ such that
 \begin{equation}
    \label{eq:loadpara}
    \alpha = \frac{P}{N}
 \end{equation}
 representing the memory capacity of the network.
 Synaptic efficacies $w_{ij}$ are chosen to satisfy the so called Hebb's Rule \cite{ref:hertz} reported below 
 \begin{equation}
\label{eq:hebb}
w_{ij} = \frac{1}{N}\sum_{\mu =1}^P \xi_i^{\mu}\xi_j^{\mu}\hspace{2cm}w_{ii} = 0\hspace{0.5cm}\forall i
\end{equation}
The Hopfield model translates the synaptic structure built according to the Hebb's Rule into the quenched disorder we find in frustrated systems in physics. It should be noticed though, that if disordered systems such as Spin Glasses or frustrated Ising Models present a random disorder (couplings are sorted from a random distribution) the Hopfield Model has a structured kind of disorder, that depends on a superposition of the stored patterns.\\Let us analyse the case of one stored pattern in the network, that is $P = 1$. We assume, now on, $\theta = 0$. The synaptic efficacies are now expressed by $$w_{ij}= \frac{1}{N}\xi_i \xi_j$$ and equation (\ref{eq:dyn}) becomes
\begin{equation}
\label{eq:1_dyn}
\sigma_i^{(t+1)} = \text{sgn}\left( \frac{1}{N}\sum_j\xi_i \xi_j \sigma_j^{(t)}\right)
\end{equation}
where the fixed point of the dynamics is given by
\begin{equation}
\label{eq:fixed}
\sigma_i^* = \text{sgn}\left( \frac{1}{N}\sum_j\xi_i\xi_j\sigma_j^*\right)\hspace{0.5cm}\forall i
\end{equation}
that is solved by $$\vec \sigma^* =\pm\vec \xi$$ By applying a perturbation to the fixed point we can prove it is also stable. Hence the system has ended up in the only stored memory or its reverse (due to the inversion symmetry of the dynamics). This result should remind the ferromagnetic behaviour of an Ising model at zero temperature. There are two attractors of the dynamics and the phase space is symmetrically divided in two basins of attraction. \\We now recall the local fields of the neurons as expressed by equation (\ref{eq:local_field}). The $P>1$ case is then evaluated. The stability condition for the generic pattern $\nu$ has become
\begin{equation}
\label{eq:fixed_many}
\text{sgn}\left( h_i^{\nu}\right) = \text{sgn}\left(\sum_j^N w_{ij}\xi_j^{\nu}\right) = \text{sgn}\left(\frac{1}{N}\sum_j^N\sum_{\mu}^P \xi_i^{\mu}\xi_j^{\mu}\xi_j^{\nu}\right) = \xi_i^{\nu}\hspace{0.5cm}\forall i 
\end{equation}
that can be rewritten isolating the so called \textit{noise-to-signal} term
\begin{equation}
\label{eq:nts}
\text{sgn}\left(\xi_i^{\nu} + \frac{1}{N}\sum_j^N\sum_{\mu\neq\nu}^P \xi_i^{\mu}\xi_j^{\mu}\xi_j^{\nu}\right) = \text{sgn}\left(\xi_i^{\nu} + N_i^{\nu}\right) =  \xi_i^{\nu}
\end{equation}
When $|N_i^{\nu}|<1$ $\forall i$ then $\vec \xi^{\nu}$ is a stable fixed point. Since it depends on a sum over the patterns, it can increase when $P$ increases, leading the system to a non-retrieval phase where all the patterns are unstable fixed points of the dynamics. Thus, the noise-to-signal term represents a measure of the interference of the non-retrieved patterns on the retrieved one. In particular, Amit, Gutfreund and Sompolinsky \cite{ref:amitgut} have used a stochastic version of the Hopfield model to trace the full phase diagram of the system as a function of the load parameter and an effective temperature $T$. Since at $T = 0$ one recovers the deterministic model, they have found a first order phase transition at $\alpha_c \simeq 0.138$ from the retrieval to the non-retrieval phase that is valid in our case: for $\alpha > \alpha_c$ all the patterns are unstable and no memory is retrieved.\\Taking inspiration from the noise-to-signal term, it is useful to introduce a quantity representing the correlation of the generic configuration of the network $\vec{\sigma}$ with a pattern $\vec{\xi^{\mu}}$ that is called $\textit{overlap}$
\begin{equation}
\label{eq:overhop}
m_{\mu} = \frac{1}{N}\sum_{i=1}^N\xi_i^{\mu}\sigma_i
\end{equation}
This quantity is $m_1 = 1$ when $\vec{\xi^1}$ is the fixed point, and the network is fully correlated with it, while it is $m_{\mu>1} = O(\frac{1}{\sqrt{N}})$ for the non-retrieved patterns. The overlap is usually adopted as an order parameter to be used to signal a transition from the retrieval phase to the non retrieval one.

\section{Towards a Higher Biological Plausibility}
\label{sec:bio}
Even though the Hopfield Model is a good model of an associative memory, at least in terms of qualitative behaviour of the system, it does not take into account biological details that make the neural network more realistic and predictive.\\We will now describe upgrading concepts that have historically been included in the Hopfield's memory theory to attain a higher biological plausibility. They are: asymmetric and diluted synaptic efficacies, biased patterns, distinction between $\textit{excitatory}$ and $\textit{inhibitory}$ neurons.  
\subsection{Asymmetry and Dilution of the Connectivity Matrix} 
\label{subsec:DGZ}
G\'erard Toulouse considered the choice of symmetric efficacies by Hopfield a \textit{"clever step backwards from biological realism"} (as quoted by \cite{ref:hertz}). This statement refers to the decision of using $w_{ij} = w_{ji}$ that permitted to apply the tools from statistical mechanics at the equilibrium to find the state of the network. In fact, when the connectivity matrix is symmetric and we switch to the stochastic version of the model (that tends to the pure deterministic one when the effective temperature $T\rightarrow0$), detailed balance is satisfied in the evolution of the configurations, and an Hamiltonian can be defined \cite{ref:peretto}. Once one has an Hamiltonian the partition function can be computed and the main thermodynamic quantities, as the free energy of the system, can be also calculated. Local minima of the free energy will correspond to the states at the equilibrium of the network \cite{ref:amitgut}.  
\\However, it's empirically observed that synaptic efficacies are not symmetric in real neural networks. Neuroscientific studies have sure enough demonstrated that neurons are classified depending on their functionality \cite{ref:sporns} and they usually select one kind of neurotransmitter to interact with the other neurons \cite{ref:eccles}. This second property of neurons is referred to as $\textit{Dale's Law}$ and it will be treated further on in this Chapter. Moreover, other studies show that networks of neurons in several parts of the brain yield a certain degree of sparseness. Since synapses evolve in time in order to create activity patterns, it has been observed that potential synapses are not always likely to emerge \cite{ref:stepa} impeding the network to be fully connected. This is probably due to reasons involving volume optimization and again the classification of neurons with different functionalities.
\\Derrida, Gardner and Zippelius have successfully attempted to solve an asymmetric and also diluted version of the Hopfield Model. In this paragraph important hints that will be useful to the rest of our research will be reported neglecting the proper solution to the evolution of the network, which can be read from the original paper \cite{ref:dgz}.\\DGZ have considered a network of binary neurons $\{\sigma_i = \pm 1\}_{i = 1}^N$ where synaptic efficacies are slightly modified with respect to the Hopfield ones. They assume the following expression
\begin{equation}
\label{eq:derrida_synap}
w_{ij} = \frac{c_{ij}}{N}\sum_{\mu = 1}^P\xi_i^{\mu}\xi_j^{\mu}
\end{equation}
where $\xi_i^{\mu} = \pm 1$ for $i = 1,..N$ and $\mu = 1,..,P$. $c_{ij}$ are variables responsible for the asymmetry and dilution of the network and they are indentically and independently generated from the following probability distribution
\begin{equation}
\label{eq:dilu}
P\left( c_{ij}\right) = \frac{C}{N}\delta\left( c_{ij} - 1\right) + \left(1 - \frac{C}{N}\right) \delta\left(c_{ij} \right)
\end{equation}
where $C$ is the $\textit{mean connectivity}$ of the network, that is the average number of neighbours per neuron. The number of patterns in this model will be extensive in the connectivity, that is $$P = \alpha C$$ 
DGZ show that keeping 
\begin{equation}
\label{eq:clog}
    C\ll\ln(N)
\end{equation}
in the thermodynamic limit, so that $C/N\rightarrow0$ when $N\rightarrow\infty$, neurons can be considered fully uncorrelated. This condition is equivalent to performing the thermodynamic limit in two steps: at first, $N\rightarrow\infty$ keeping $C$ fixed, then $C\rightarrow\infty$. 
A naive proof that neurons are uncorrelated is furnished by Amit \cite{ref:amitbook} and is restated in the next paragraph.\\ 
Let us consider a generic choice of the dynamics where time is discrete and at each step neurons $\sigma_j$ fire at neurons $\sigma_i$ $\forall i,j$ such that $c_{ij} = 1$. At time $t$ each cell can count on a tree of $\textit{ancestor neurons}$ that have influenced its final status. In principle these trees can overlap at different time steps yielding mutual correlations among the neurons of the network. Let us assume one tree composed by $M$ neurons. We now suppose to draw other $M$ units at random from the network and compute the probability for the tree of ancestors not to be overlapped with the group of random selected units. This probability must be
\begin{equation}
\label{eq:o_prob}
Prob\left( \text{no overlap}\right) = \left( 1 - \frac{M}{N}\right)^M \simeq \exp\left({-\frac{M^2}{N}}\right)
\end{equation}
where we have already supposed $M \ll N$. Keeping the connectivity finite one can assume that $M$ grows polynomially with $N$ as $M = aN^b$. If this is the case we get
\begin{equation}
\label{eq:o_prob2}
Prob\left( \text{no overlap}\right) = \exp\left( -a^2N^{2b-1}\right)
\end{equation}
Hence, to have $Prob\left( \text{no overlap}\right) \rightarrow 1$ when $N\rightarrow\infty$ we must choose $b < 1/2$. Since after a time $t$ each neuron has an average number of $C^t$ ancestors, one must impose $ C^t < \sqrt{N}$, that leads to $$ C < N^{\frac{1}{2t}} \hspace{1cm} \forall t$$ which is satisfied when relation (\ref{eq:clog}) holds. 
\begin{figure}[h!!]
    \centering
		\includegraphics[width=10cm]{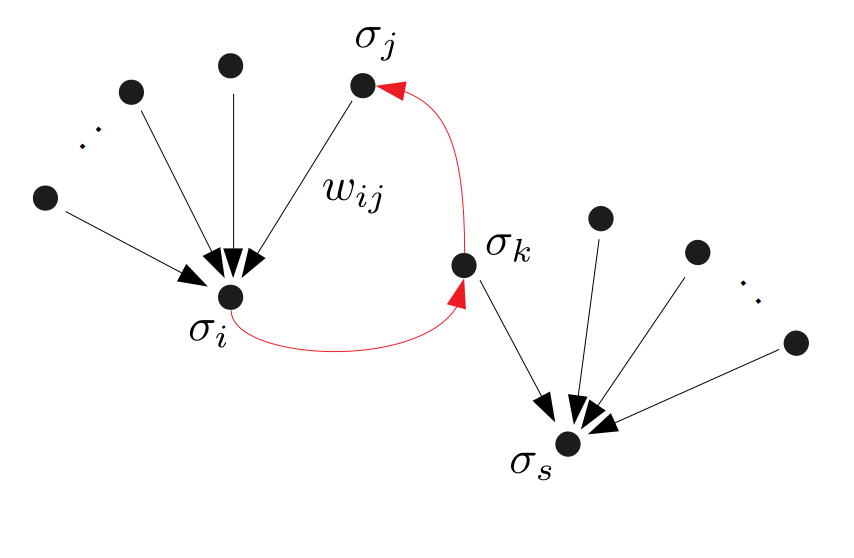}
		\label{fig:lif}
		\caption{ \textbf{Representation of correlated units in a neural network.} $\sigma_j$ affects $\sigma_i$ and $\sigma_k$ fires at $\sigma_s$, nonetheless $\sigma_i$ exerts his action over $\sigma_k$ that is also linked to $\sigma_j$, forming the $\textit{feedback loop}$ drawn in red. As a consequence, units $\sigma_i$ and $\sigma_s$ are mutually correlated.}
		\label{fig:loops}
\end{figure}
\\Notice that this limit is not really biologically plausible, being $C = O(10^3)$ in many areas of the brain and so larger than $\ln(N)$ for any realistic amount of neurons in the network. As a consequence, even in the sparsest networks there will always be a small amount of residual correlation among the neurons. However, if this correlation is due to feedback effects in the network dynamics, through which unit $i$ feels the effect of the unit $j$ that is in turn influenced - even indirectly - by $i$ (see figure (\ref{fig:loops})), in the DGZ regime these $\textit{feedback loops}$ become much larger, dissolving their effect: what in principle is a $\textit{recurrent}$ neural network becomes a $\textit{feed-forward}$ neural network with no loops. A good accordance of simulated finite systems with the theory behind the DGZ limit of extreme dilution has resulted in binary networks \cite{ref:arenzon}. However, the same good consistency will be shown in the biological based model that will be treated in the course of this work.
\subsection{Storing Biased Patterns}
In the Hopfield model patterns were generated at random by means of a Bernoulli process with probability $p = 1/2$. The main advantage of this "symmetric" choice was that the mean of the noise-to-signal term was null. Noise-to-signal terms could then be treated as a Gaussian variable with mean $\mu = 0$ and variance $ \sigma^2 = \alpha$. Since this term is zero on average, the interference of the uncondensed patterns with the retrieved state only depends on the increasing $\alpha$, that is an intensive quantity. If the mean was not null, fixed point would be unstable even at low values of $\alpha$. Moreover, assuming $p = 1/2$ we fix the number of sites which actively partecipate to the pattern (sites $i$ corresponding to $\xi_i^{\mu} = +1$) to be $N/2$ on average. \\The process by which the brain translates stimuli from outside, transforming received signals into configurations of neural activity, is called $\textit{neural coding}$. The main question in the matter of neural coding concerns the finding of a type of coding able to maximize the information learned from the stimulus without implying an astronomical memory storage of the brain or an extreme amount of energy associated to the firing activity of the neurons. A good compromise in this sense is to suppose the patterns to be sparse, meaning that the minority of sites contribute with $\xi^{\mu}_i = +1$. This representation is referred to as $\textit{sparse coding}$ \cite{ref:foldiak}.
Hence, another upgrade for our neural network would be to arbitrarily tune the number of active sites in the pattern without losing the symmetry of the noise. The problem has been solved by Amit, Gutfreund and Sompolinsky in \cite{ref:biased}. \\Remaining in the Hopfield's framework patterns can be sorted from the following probability distribution
\begin{equation}
    \label{eq:bias}
    P(\xi_i^{\mu}) = f\delta\left( \xi_i^{\mu} - 1 \right) + (1-f)\delta\left( \xi_i^{\mu} + 1 \right)
\end{equation}
where $f$ is called $\textit{coding level}$ and it represents the mean number of active sites only if
synaptic efficacies are redefined as it follows
\begin{equation}
\label{eq:bias2}
w_{ij} = \frac{1}{N}\sum_{\mu = 1}^P (\xi_i^{\mu}-f)(\xi_j^{\mu}-f)
\end{equation}
This way the noise-to-signal term is still zero on average and fixed points of the dynamics are not dramatically destabilized at small values of $\alpha$.
\subsection{LIF neurons and the Dale's Law}   
\label{subsec:LIF}
The traditional Hodgkin-Huxley model \cite{ref:hh} \cite{ref:gerstner}, and similar interpretations of the neuron, represent the neuronal membrane as a complex electric circuit, where each different ionic channel contributes through a time-dependent resistance and a tension due to the flow of incoming/outgoing ions. The evolution of the action potential emitted by the neuron is thus recovered from solving a set of at least four non linear coupled differential equations.\\
On the other hand, the \textit{Leaky Integrate and Fire} \cite{ref:gerstner} model is a simplified version of such representations which does not aim to derive the exact firing state of the neuron, but only to determine whether or not the cell fires at a certain given time known the shape of presynaptic input. According to the \textit{LIF} model the neuron membrane is schematized as a RC electrical circuit as in figure (\ref{fig:lif}).
\begin{figure}[h!!]
    \centering
		\includegraphics[width=8cm]{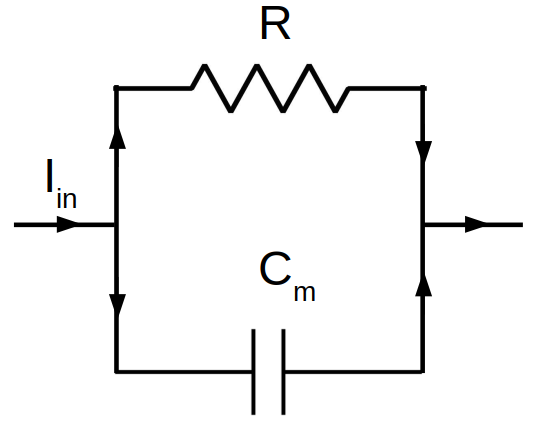}
		\label{fig:lif}
		\caption{ \textbf{RC circuit equivalent to the $LIF$ model for the neuron.} $R$ and $C_m$ are the typical resistance and capacity of the neuronal membrane. The time constant of the circuit is $\tau_m = RC_m$.}
		\label{fig:lif}
\end{figure}
According to this vision of the membrane, the equation of the circuit is 
\begin{equation}
\label{eq:circ}
C_m \frac{dV}{dt} + \frac{V(t)}{R} = I_{in}(t)
\end{equation}
where $V(t)$ represents the voltage difference across the membrane. $C_m$ and $R$ are, respectively, the capacity and the resistance of the neuronal membrane, that we consider to be the same for all the neurons. The time constant of the circuit will be $\tau_m = RC_m$.




In order to determine the best expression to attribute to the input current $I_{in}$ we introduce the Dale's Law of neuroscience. According to the Dale's Law each neuron only emits $\textit{excitatory}$ or $\textit{inhibitory}$ signals to its post-synaptic targets depending on the specific class of neurotransmitters that they are able to diffuse \cite{ref:eccles}. According to this description, excitatory neurons physically transmit positive tension steps to the postsynaptic neuron helping it to make it fire. Inhibitory ones, instead, transmit negative potentials steps, impeding the postsynaptic cell to reach the spiking threshold. Even though recent experimental evidence has shown exceptions that diverged from this description of the neuron \cite{ref:root}, implementing this rule when building up a model for a neural network leads to a more realistic interpretation of the neural system. This biological concept translates into dividing the system in two populations: one made of excitatory neurons, only emitting positive synaptic efficacies, the other made by inhibitory ones, which only emit negative synaptic efficacies. Notice that Hopfield-like models completely neglect Dale's Law, since any unit in the network can emit both positive and negative efficacies. 
The incoming current will be then expressed as 
\begin{equation}
\label{eq:inputcur}
I_{in}(t) = C_m\left[W_E\frac{dN_E}{dt} - W_I\frac{dN_I}{dt}\right]
\end{equation}
with $W_E,W_I > 0$ voltage steps emitted from the neurons belonging to the E, excitatory, and I, inhibitory, populations. $N_E(t),N_I(t)$ are the numbers of, respectively, excitatory and inhibitory inputs that the neuron receives.  
Equation (\ref{eq:circ}) is rewritten as
\begin{equation}
\label{eq:circ2}
\frac{dV}{dt} = -\frac{V(t)}{\tau_m} + W_E\frac{dN_E}{dt} - W_I\frac{dN_I}{dt}    
\end{equation}
Since the neuronal state is usually asynchronous and irregular one assumes the statistics of the incoming excitatory and inhibitory voltage inputs to be Poisson with mean firing rates $\nu_i$, $i = E,I$. We now average equation (\ref{eq:circ2}) according to the Poisson statistics of the incoming inputs, renaming $\langle V \rangle(t) = h(t)$ to recall the local field we introduced in the previous Sections. Notice, however, that quantities in the neural networks of out interest have not physical dimension, for simplicity of the treatment.  
Hence 
\begin{equation}
\label{eq:stein_dyn}
\dot{h} = -h(t) + W_E\nu_E - W_I\nu_I
\end{equation}
where it has been set $\tau_m = 1$. Furthermore, at the fixed point of the dynamics, the average tension across the membrane assumes the following expression
\begin{equation}
\label{eq:eq}
h =  W_E\nu_E  - W_I\nu_I    
\end{equation}

\section{Balanced Networks}
\begin{figure}[h!!]
    \centering
		\includegraphics[width=8cm]{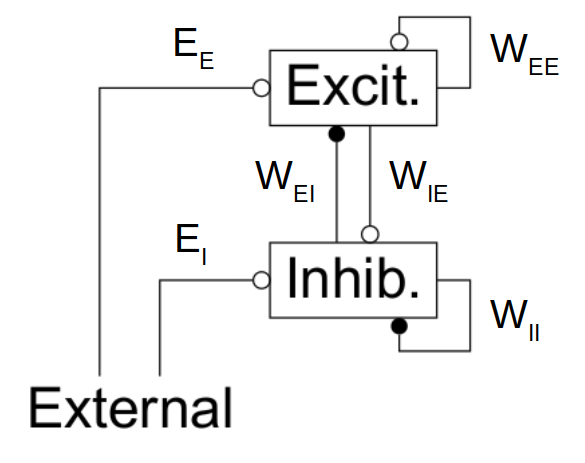}
		\label{fig:architect}
		\caption{\textbf{The architecture of the neural network used by Van Vreeswijk and Sompolinsky.} Image from \cite{ref:van}. Excitatory and Inhibitory populations mutually interact along with an external input coming from a background population. They all contribute to the local field of each neuron in the network. Black circles represent a negative inhibitory efficacy, white circles represent a positive excitatory efficacy. }
		\label{fig:van}
\end{figure}

This section is devoted to explain another important type of neural network used in theoretical neuroscience called $\textit{balanced network}$. Even though this network does not have any memory storage property it is very helpful to reproduce the special condition of particular regions of the brain that are actually devoted to the memory retrieval. In particular, experiments on these areas of the brain have shown two recurrent properties: the mean local field of the neurons is set around the typical spiking threshold of the network, implying the firing state of the neurons to be quite irregular and unpredictable; the mean firing rate of the neurons remains constant while accomplishing memory retrieval tasks. These particular features can be explained by the realisation of a balanced interaction between an excitatory population of neurons and an inhibitory one under particular conditions of the network. \\In order to demonstrate the functioning of balanced networks we present a model that is similar to the one treated by Van Vreeswijk and Sompolinsky in \cite{ref:van}. The architecture of the neural network is depicted by figure (\ref{fig:van}).
\\$N_E$ excitatory neurons and $N_I$ inhibitory ones interact with each other through synaptic efficacies $W_{kl}^{ij}$ where $k,l = E,I$ denote the two populations of neurons, while $i,j = 1,..,N_l$ label the neurons in the whole network. Synaptic efficacies are i.i.d. variables chosen to be 
\[ 
W_{kl}^{ij}  = 
\left\{ \begin{array}{l}
W_{kl}\hspace{1cm}p = \frac{C}{N_l}\\
0\hspace{1cm}\text{otherwise}
\end{array} \right. 
\]
where we put ourselves in the Derrida-Gardner-Zippelius (DGZ) extreme dilution limit (see Subsection \ref{subsec:DGZ}). 
In addition to the $E,I$ populations each neuron receives an input from the external environment $h^{ext}_k = E_k h^{ext}$ only depending on the $k$ label. \\
Each neuron makes experience of a local field $h_k^i(t)$ that evolves according to an arbitrary relaxation dynamics for the network. At the fixed point of the dynamics the field is expressed by 
\begin{equation}
\label{eq:van_lf}
h_k^i = h^{ext}_k + \sum_{l=E,I}\sum_{j=1}^{N_l} W_{kl}^{ij}\nu_l^j
\end{equation}
that is formally analogous to the equation (\ref{eq:eq}) we have found for LIF neurons. 
$\nu_l^j$ physically represents the firing rate of the neuron and it can be chosen to be a continuous function of the local field or a discrete McCulloch-Pitts variable. During the rest of the work this variable will be called $\textit{activity level}$ or $\textit{activity}$ of the neuron.\\
We define $\langle \cdot \rangle$ as the statistical average of a given quantity over the sites of the network when $N_l\rightarrow\infty$. Exploiting the dilution limit to consider synaptic efficacies uncorrelated with respect to the activities, the average field can be factorized. Hence we get
\begin{equation}
\label{eq:van_mean}
\langle h_k \rangle = h^{ext}_k + C\sum_{l=E,I} W_{kl}\langle \nu_l \rangle
\end{equation}
\begin{equation}
\label{eq:van_var}
\sigma_k^2 = C\sum_{l=E,I} W_{kl}^2\langle \nu_l^2 \rangle
\end{equation}
Where it has been used $\sigma_k^2 = \langle h^2_k\rangle - \langle h_k \rangle^2$.
We are now interested in studying a state of the network where the mean local field sets around the threshold level, as any $\theta_k = O(1)$ we might set as a parameter of the dynamics. It is thus required that $$\langle h_k \rangle = O(1)$$
A first way to operate might be to choose $$W_{kl} = \frac{w_{kl}}{C}$$
Nevertheless this takes to $\sigma_k^2 \xrightarrow[]{C\rightarrow \infty} 0$ which is not observed in the experiments.\\Hence, the following new normalization is proposed
\begin{equation}
\label{eq:balance}
W_{kl}= \frac{w_{kl}}{\sqrt{C}} \hspace{2cm}h^{ext}_k = \sqrt{C}E_k h^{ext} 
\end{equation}
such that
\begin{equation}
\label{eq:van_mean2}
\langle h_k \rangle = \sqrt{C}\left(E_k h^{ext} + \sum_{l=E,I} w_{kl}\langle \nu_l \rangle\right) = O(1)
\end{equation}
\begin{equation}
\label{eq:van_var2}
\sigma_k^2 = \sum_{l=E,I} w_{kl}^2\langle \nu_l^2 \rangle = O(1)
\end{equation}
Furthermore, the condition $$\left(E_k h^{ext} + \sum_{l=E,I} w_{kl}\langle \nu_l \rangle\right) = O\left(\frac{1}{\sqrt{C}}\right)\hspace{1cm}k = E,I$$ permits to obtain a linear relation that expresses the activities of the two populations as fixed quantities in the thermodynamic limit, as wanted from the empirical observations.\\It turns out that
\begin{equation}
\label{eq:balance1}
E_E h^{ext} + w_{EE}\langle \nu_E \rangle + w_{EI}\langle \nu_I \rangle = 0
\end{equation}
\begin{equation}
\label{eq:balance2}
E_I h^{ext} + w_{IE}\langle \nu_E \rangle + w_{II}\langle \nu_I \rangle = 0
\end{equation}
By assuming the following values for the synaptic efficacies
$$ E_E = E \hspace{1cm}E_I = I$$
\begin{equation}
\label{eq:coup}
w_{EE} = w_{IE} = 1\hspace{1cm}  w_{EI} = -w_E\hspace{1cm}w_{II} = - w_{I}
\end{equation}
with $E,I,w_E,w_I >0$ one gets
\begin{equation}
\label{eq:balance12}
E h^{ext} + \langle \nu_E \rangle - w_{E}\langle \nu_I \rangle = 0
\end{equation}
\begin{equation}
\label{eq:balance22}
I h^{ext} + \langle \nu_E \rangle - w_I\langle \nu_I \rangle = 0
\end{equation}
that implies the same result obtained by Van Vreeswijk and Sompolinsky
\begin{equation}
\label{eq:balance13}
\langle \nu_E \rangle = \frac{w_IE - w_E I}{w_E - w_I}h^{ext}
\end{equation}
\begin{equation}
\label{eq:balance23}
\langle \nu_I \rangle = \frac{E - I}{w_E - w_I}h^{ext}
\end{equation}
As a result, the mean activities of the two populations are determined by the parameters of the model, and they respond linearly to the external input. Moreover, since the activity level is a non-negative quantity defined in the interval $\nu \in(0,1)$ the solutions (\ref{eq:balance13}),(\ref{eq:balance23}) exist only if the following conditions hold
$$ \left( w_I E - w_E I\right)\left( w_E - w_I\right)>0\hspace{2cm}\left( E-I\right)\left( w_E - w_I\right)>0$$ and $$ h_{ext}\in\left[0,\text{min}\left(\frac{w_E-w_I}{w_I E-w_E I},\frac{w_E-w_I}{E-I} \right)\right]$$

\chapter{Study of a Random Balanced Network of Inhibitory Neurons}
\label{chap:2}

As a preliminary study to the main body of this work, that will deal with memory storage in balanced neural networks, we introduce the model that is going be used for all the rest of the thesis. The balanced network that is proposed contains the most important features that are needed to obtain biologically plausible results: asymmetric and diluted synaptic links, a continuous transduction function, balanced activity. Moreover, the model respects Dale's Law of neuroscience describing a single population of neurons, chosen to be inhibitory (the main framework that has led to single population description is contained in \cite{ref:mongi}). Our first aim is to study in detail the balanced regime where the network is supposed to work and to show the derivation of the the mean field equations for the system. Hence we are going to deal with a preparatory version of the final model, that is the random network: synaptic efficacies are randomly generated to be all of the same sign with no Hebbian learning embedded in the network. In order to derive an exact mean field description of the state of the network as dependent on the control parameters of the model, it has been decided to work in the DGZ dilution limit. 
\\The stability of the fixed point of the dynamics is later investigated, aiming to recover, even for our model, a well known condition that permits to draw the separation line in the parameters space dividing the chaotic phase of the network from the stable one.\\ 
Numerical simulations will be eventually executed to perform a comparison between the state of a finite sized random network and the theoretical mean field predictions.
\section{Description of the Model}
\begin{figure}[h!!]
    \centering
		\includegraphics[width=13cm]{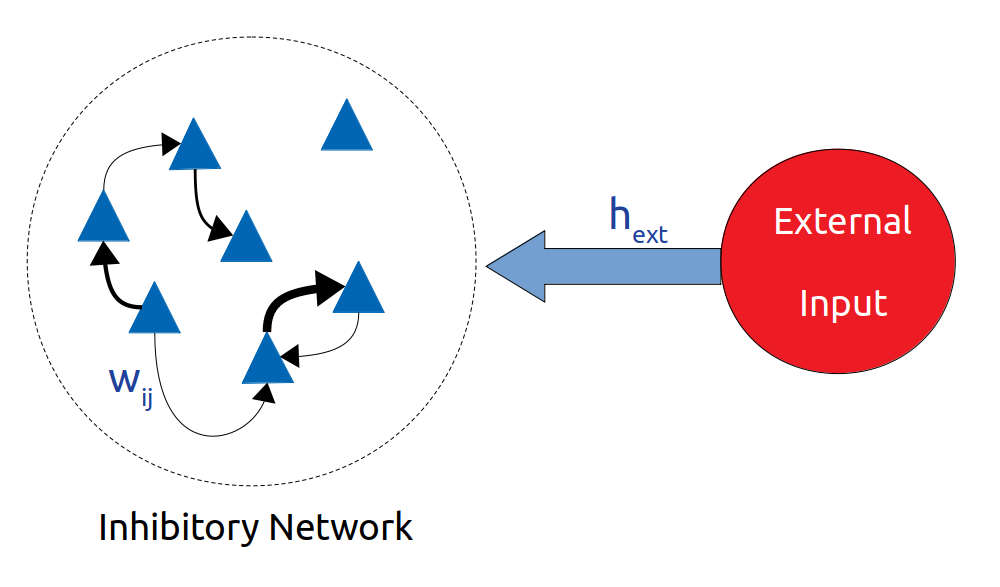}
		\caption{\textbf{Artistic representation of the studied system.} An inhibitory population where neurons (triangles) reciprocally interact with themselves through sparse asymmetric synapses (arrows). All cells also receive an external input from the environment.}
\end{figure}
Let us consider a population of $N$ inhibitory neurons. Each cell is described by a continuous variable $h_i$, with $i=1,..,N$, namely the $\textit{local field}$, that is physically equivalent to the presynaptic input to the neuron. The total incoming input to the each neuron is given by the contribution of two different components: the input from the other cells of the inhibitory network, an external stimulus coming from outside the system.\\
The equation of the dynamics of the model, provided below, includes all the properties of the network. 
\begin{equation}
\label{eq:dynamics}
    \dot h_i(t) = -h_i(t) +\sqrt{C}h_{ext} - \frac{1}{\sqrt{C}}\sum_{j=1}^{N}c_{ij}w_{ij}\nu_j(t) 
\end{equation}
Let us now describe the variables appearing in the equation. 
\begin{itemize}
    \item C \textit{mean connectivity} of the network. It represents the average number of neighbours of any neuron in the network.
    \item $h_{ext}$ $\textit{external input}$. It is a variable of order $O(1)$. In principle it can assume both positive and negative values. 
    \item $c_{ij}$ $\textit{dilution variables}$. They are the anatomical links between pairs of neurons and they are random variables independently generated according to the the Derrida-Gardner-Zippelius regime of extreme dilution (see Chapter \ref{chap:1} Section \ref{sec:bio}). Hence, dilution variables are drawn from the following probability density
    \begin{equation}
    \label{eq:c}
        P(c_{ij}) = \frac{C}{N}\delta(c_{ij}-1) + (1-\frac{C}{N})\delta(c_{ij})
    \end{equation}
    \item $w_{ij}$ $\textit{synaptic efficacies}$. They are strictly positive so that the interaction is inhibitory when a minus sign is placed in front of the sum.   
    \item $\nu_j$ $\textit{activity level}$ of the neuron. It is a function of the local field $h_j$ called $\textit{f-I function}$ or $\textit{trasduction function}$. We choose a sigmoid-shaped function that is limited in the the interval $(0,1)$
    \begin{equation}
    \label{eq:sigmoid}
    \phi(h) = \frac{1}{1 + e^{-\beta[h - \theta]}}
    \end{equation}
    where $\beta$ is called $\textit{gain}$ parameter and $\theta$ is a $\textit{threshold}$ of order $O(1)$. In particular, for $\beta \rightarrow\infty$ the f-I function tends to a step function centered around the threshold.  
    
\end{itemize}

The particular scaling of the synaptic efficacies and the external input in equation (\ref{eq:dynamics}) is consistent with the theory of balanced networks, guaranteeing the existence of fixed points of the dynamics at which the mean value of the local field and its variance are $O(1)$, independently on the choice of the threshold. 
In addition to this, finding a balance condition for the neural state of the population gives the necessary condition for the state of the network to keep the mean activity level $\langle \nu \rangle$ fixed in the thermodynamic limit. 
\\ In the random version of the network quantities $c_{ij}$, $w_{ij}$ are randomly and independently generated $\forall i,j$. Dilution variables are drawn from the distribution expressed by equation (\ref{eq:c}) while synaptic efficacies $w_{ij}$ are positive i.i.d. variables generated according to an arbitrary distribution in such a way to be all positive. The functional shape of their probability density function is only relevant to simulations of the model. 

\section{Mean Field Analysis of the Balanced Network}
\label{sec:2regimes}
We define $\langle \cdot \rangle$ as the statistical average over the sites of the network when $N\rightarrow\infty$. Equation (\ref{eq:dynamics}) at the fixed point can be rewritten as
\begin{equation}
\label{eq:fixed0}
h_i = \sqrt{C}h_{ext} - \frac{1}{\sqrt{C}}\sum_{j=1}^N w_{ij}c_{ij}\nu_j 
\end{equation}
The mean local field in the limit of $N\rightarrow\infty$ and fixed $C$ is computed below 
\begin{equation}
\label{eq:balance_o1}
   \langle h \rangle =  \sqrt{C}\left(h_{ext} - \langle w \rangle\langle \nu \rangle\right)
\end{equation}
Equation (\ref{eq:balance_o1}) can be equivalently rewritten as
\begin{equation}
\label{eq:general_balance}
\langle \nu \rangle = \frac{h_{ext}}{\langle w \rangle} + \varepsilon(C)
\end{equation}
with $$\varepsilon(C) \propto \frac{\langle h \rangle}{\sqrt{C}}$$ where in general $\langle h \rangle$ can scale with $C$. We are now interested in performing the limit $C \rightarrow \infty$ to reach the DGZ extreme dilution regime.
Since $\nu$ is bounded in the interval $(0,1)$ two regimes of activity of the network do emerge: 
\begin{itemize}
   \item $h_{ext} \notin [0, \langle w \rangle]$  $\textit{unbalanced regime}$: $\langle \nu \rangle$ must saturate at: $+1$ for $h_{ext} > \langle w \rangle$, $0$ for $h_{ext} < 0$. This means there is homogeneous activity: all neurons together display the same activity level. Consequently, in order to adjust the mean activity level to the saturation value, $\varepsilon$ must be constant for $C \rightarrow\infty$ and $\langle h \rangle = O(\sqrt{C})$.
    \item $h_{ext} \in [0, \langle w \rangle]$ $\textit{balanced regime}$: in this case $\varepsilon(C)$ must vanish for $C\rightarrow \infty$. Hence $\langle h \rangle = O(1)$ and  $\varepsilon = \langle \nu \rangle - \frac{h_{ext}}{\langle w \rangle} = O\left(\frac{1}{\sqrt{C}}\right)$.\\ 
    In the balanced regime equation (\ref{eq:general_balance}) becomes
    \begin{equation}
    \label{eq:balance}
    h_{ext} - \langle w \rangle \langle \nu \rangle = O\left( \frac{1}{\sqrt{C}}\right)
    \end{equation}
    that we will name $\textit{balance condition}$. This condition implies $$\langle \nu \rangle \xrightarrow[]{C\rightarrow \infty} \frac{h_{ext}}{\langle w \rangle}$$ 
 
\end{itemize}
It has been shown that given the right value of the external input, the network will satisfy the balance condition, working in the balanced regime. This phenomenon is not trivial, since it implies a  collective behaviour of the neurons that balances the network and adjusts the mean local field in such a way it is $O(1)$. This result is equivalent to set the mean local field around the threshold of the neurons, that is also $O(1)$ by construction of the model, whatever the threshold is. 
\\Computing the variance of the local fields at finite $C$ and $N\rightarrow\infty$ one gets
\begin{equation}
\label{eq:var_bef}
\sigma^2 = \langle w^2 \rangle \langle \nu^2 \rangle
\end{equation}
where it has been exploited that $\frac{C}{N}\rightarrow 0$ in the DGZ dilution regime. This expression of the variance is identically valid in the $C\rightarrow\infty$ limit. 
As a consequence, in the unbalanced regime, where $\langle h \rangle$ diverges with $C\rightarrow\infty$, the variance will be subleading with respect to the mean, and relative fluctuations will vanish in the thermodynamic limit. This implies the state of the network to be entirely known.\\ On the other hand, the neural state in the balanced regime, that is for sure particularly interesting for us, is more complicated to be determined. We thus resort to a mean field approach to determine the state of the network at the fixed point.\\The extreme dilution regime permits to consider the single neurons uncorrelated with each other. This is not enough, however, to infer the statistics of the fields. It must be noticed, though, that our connectivity matrix is asymmetric. Consequently, terms in the sum appearing in equation (\ref{eq:fixed0}) are mutually independent and Central Limit Theorem can be invoked. We can conclude that local fields $h_i$ are distributed, in the thermodynamic limit, according to a Gaussian density function having cumulants $\mu,\sigma^2$, such that $$ h_i = \mu + z_i\sigma$$ with $z_i$ being a Gaussian with $0$ mean and unit variance. 
Exploiting the Gaussianity of the local fields and by consistency with equations (\ref{eq:balance}) and (\ref{eq:var_bef}) the mean field equations of the model are derived
\begin{equation}
\label{eq:dgz_mf1}
\frac{h_{ext}}{\langle w \rangle} = \int_{-\infty}^{+\infty} Dz \phi(\mu + \sigma z) 
\end{equation}
\begin{equation}
\label{eq:dgz_mf2}
\sigma^2 = \langle w^2 \rangle \int_{-\infty}^{+\infty} Dz \phi^2(\mu + \sigma z) 
\end{equation}
with $$Dz = \frac{e^{-\frac{z^2}{2}}}{\sqrt{2\pi}}dz$$  \\being the standard Gaussian measure. 
Moreover, one can keep $C$ finite and assume that fields have already reached a high degree of Gaussianity: numerical simulations in the next subsection will give a demonstration of this assumption for $C$ high enough. In this case the corresponding equations can better predict the state at finite $C$ and they are
\begin{equation}
\label{eq:dgz_mf3bis}
\langle \nu \rangle = \int_{-\infty}^{+\infty} Dz \phi(\mu + \sigma z) 
\end{equation}
\begin{equation}
\label{eq:dgz_mf3}
\mu  = \sqrt{C}\left(h_{ext} - \langle w \rangle\langle \nu \rangle \right)
\end{equation}
\begin{equation}
\label{eq:dgz_mf4}
\sigma^2 = \langle w^2 \rangle \int_{-\infty}^{+\infty} Dz \phi^2(\mu + \sigma z) 
\end{equation}
\section{An Analysis of the Stability of the Fixed Points}
\label{sec:stab}
We now want to study the stability of the fixed point of the network dynamics. 
\\Consider a small perturbation $\frac{\eta_i}{\sqrt{C}}$ with $\eta_i = O(1)$ random variable applied to each $i$ neuron. One obtains
\begin{equation}
\label{eq:newfixedpt}
\Tilde{h_i} = \sqrt{C}h_{ext} - \frac{1}{\sqrt{C}}\sum_{j}^{N}w_{ij}c_{ij}\phi(\Tilde{h}_j) + \frac{\eta_i}{\sqrt{C}}
\end{equation}
We then define $\delta h_i$ such that
\begin{equation}
\label{eq:newdelta}
\delta h_i = \Tilde{h_i} - h_i = -\frac{1}{\sqrt{C}}\sum_{j}^{N}w_{ij}c_{ij}\delta \nu_j + \frac{\eta_i}{\sqrt{C}}
\end{equation}
where we can expand $$\delta \nu_{j} = \phi(\Tilde{h}_j) - \phi(h_j) = \phi(h_j)^{'}(\Tilde{h_j}-h_j)$$ for a small perturbation around the fixed point of the dynamics. Using equation (\ref{eq:newdelta}) we obtain
\begin{equation}
\label{eq:deltanu}
\delta \nu_{i} = \phi(h_i)^{'}\left( -\frac{1}{\sqrt{C}}\sum_{k}^{N}w_{ik}c_{ik}\delta \nu_k + \frac{\eta_i}{\sqrt{C}}\right)
\end{equation}
$\delta \nu_i$, which represents the reaction of the network to the perturbation $\frac{\vec{\eta}}{\sqrt{C}}$, can be expanded in powers of $\frac{1}{\sqrt{C}}$ as $$\delta \nu_i = \frac{\delta \nu_{i,1}}{\sqrt{C}} + \frac{\delta \nu_{i,2}}{C} + O\left(\frac{1}{C^{3/2}}\right)$$ 
That, inserted in equation (\ref{eq:deltanu}), yields
\begin{equation}
\label{eq:comparison}
\frac{\delta \nu_{i,1}}{\sqrt{C}} + \frac{\delta \nu_{i,2}}{C} =  \phi(h_j)^{'}\left[ -\frac{1}{C}\sum_{j}^{N} w_{ij}c_{ij}\delta \nu_{j,1} -\frac{1}{C^{3/2}}\sum_{j}^{N}w_{ij}c_{ij}\delta \nu_{j,2} + \frac{\eta_i}{\sqrt{C}} \right]
\end{equation}
\\The first term on right hand side of equation (\ref{eq:comparison}) is $O(1)$. By comparison with the terms on left hand side, we infer that this quantity is null, that implies $\langle \delta \nu_{1} \rangle = 0$. Also by comparison, terms of order $\frac{1}{\sqrt{C}}$ are isolated leading to
\begin{equation}
\label{eq:sqrt}
\delta \nu_{i,1} = \phi(h_i)^{'}\left[ -\frac{1}{C}\sum_j^{N} w_{ij}c_{ij}\delta \nu_{2,j} + \eta_i\right]
\end{equation}
Since $\langle \delta \nu_{1} \rangle = 0$, it is implied that 
\begin{equation}
\label{eq:delta_id}
\langle \delta \nu_2 \rangle \langle w \rangle = \langle \eta \rangle    
\end{equation}

By exploiting these recent results one can rewrite equation (\ref{eq:comparison}) multiplying both terms by $\sqrt{C}$, substituting (\ref{eq:delta_id}) on r.h.s. and adding the quantity $\ \frac{\langle w \rangle\langle c \rangle}{\sqrt{C}}\sum_j\delta \nu_{1,j}$, that is null in the $N\rightarrow \infty$ limit, on r.h.s., gaining the following expression
\begin{equation}
\label{eq:deltanu1}
\delta \nu_{i,1} = \phi(h_i)^{'}\left( -\frac{1}{\sqrt{C}}\sum_j^{N}\delta \Tilde{w}_{ij}\delta \nu_{j,1} + \delta\eta_i\right)
\end{equation}
where $\delta\eta_i = \eta_i - \langle \eta \rangle$ and $\delta \Tilde{w}_{ij} = w_{ij}c_{ij} - \langle w \rangle\langle c \rangle$.\\
Squaring and averaging the expression we obtain $$\langle \delta \nu_{1}^2 \rangle = \langle {\phi^{'}}^2 \rangle\left( \langle w^2 \rangle\langle \delta \nu_{1}^2 \rangle +\sigma_{\eta}^2\right)$$
\begin{equation}
\label{eq:deltanusq}
\langle \delta \nu_1^2 \rangle = \frac{\langle {\phi^{'}}^2\rangle \sigma_{\eta}^2}{1-\langle {\phi^{'}}^2\rangle \langle w^2 \rangle}
\end{equation}
Since the response of the system to the perturbation $\langle \delta \nu_1^2 \rangle$ must be finite for the fixed point to be stable we find the stability condition, that is 
\begin{equation}
\label{eq:stability}
\langle w^2 \rangle \langle {\phi^{'}}^2\rangle < 1
\end{equation}
This result is coherent with previous studies on the stability of fixed points of the dynamics in random neural networks \cite{ref:crisanti}, \cite{ref:juelich}.

\section{Comparing Theory with Numerical Simulations} 
\label{sec:firstnum}
Numerical simulations of the random network are performed implementing the dynamics described by equation (\ref{eq:dynamics}). The equation is integrated by means of the Euler method with time steps $\Delta t = O\left( \frac{1}{N}\right)$. The algorithm automatically stops when all neurons relax at the fixed point, that is, when $ |\dot{h}_i| < \delta$ $\forall i$, with $\delta = 10^{-6}$. \\ The probability density function of the synaptic efficacies is chosen to be lognormal
\begin{equation}
    \label{eq:pdf_w}
    P(w) = \frac{1}{{\sqrt{2\pi\sigma_z^2}w}} \exp\left[{-\frac{\left(\ln(w) - \mu_z\right)^2}{2\sigma_z^2}}\right]
\end{equation} 
With the moments of the distribution expressed by
\begin{equation}
\label{eq:momw}
\langle w^n \rangle = \exp\left[n\left( \mu_z + n\frac{\sigma_z^2}{2}\right) \right]
\end{equation}  
$\mu_z$ and $\sigma_z$ are two parameters that can be tuned to fix the moments to a desired value. In the simulations we have considered $w_{ii} = 0$ $\forall i$, even though no significant modification to the results is observed when autapses are included at $C/N \leq 0.1$. As well as ensuring the the synapses to be positive, the choice of a lognormal distribution for the synaptic efficacies is supported by studies on real cortical data \cite{ref:song}. The choice done for the control parameters is the following: $$\beta = 2\hspace{2cm}\theta = 0$$ and $\langle w \rangle = 1$ is imposed by tuning the parameters of the lognormal distribution as $$\mu_z = - \sigma_z^2/2$$ By fixing $\sigma_z = 1$ the variance of the distribution is set to $\sigma_w^2 = e - 1$.\\  
\subsection{The two Regimes of the Balanced Network}

\begin{figure}[h]
		\includegraphics[width=13cm]{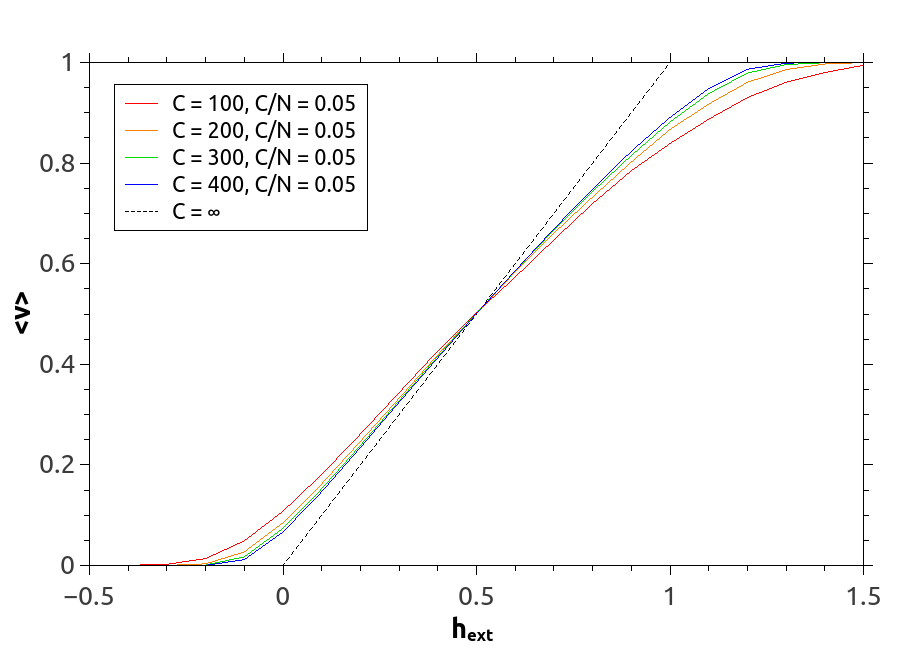}
		\centering
		\label{fig:nuh}
		\caption{\textbf{Measure of $\langle \nu \rangle$ as a function of $h_{ext}$ when increasing $C$ at fixed C/N $= 0.05$.} Notice the two regimes: the balanced regime for $h_{ext}\in[0,1]$ and the unbalanced regime outside the interval, where the mean activity level tends to saturation.}
		 \label{fig:2regimes}  
		
	\end{figure}
The plot of $\langle \nu \rangle$ in figure (\ref{fig:2regimes}) shows how the difference between the two regimes becomes more evident as $C$ increases.
This is due to the $C$ dependent correction $\varepsilon(C) = \langle \nu \rangle - h_{ext}$ that is plotted in the next figures. For $C \rightarrow \infty$ a line with unit angular coefficient is expected, because $\langle \nu \rangle \rightarrow h_{ext}$ from the balance condition (\ref{eq:balance}).\\ Now two values of the external input are chosen to run numerical simulations in both the balanced and unbalanced regimes and compare the results with theory. 
\begin{multicols}{2}
    \centering
    \begin{figure*}[ht!!!]
        \includegraphics[width=.51\textwidth]{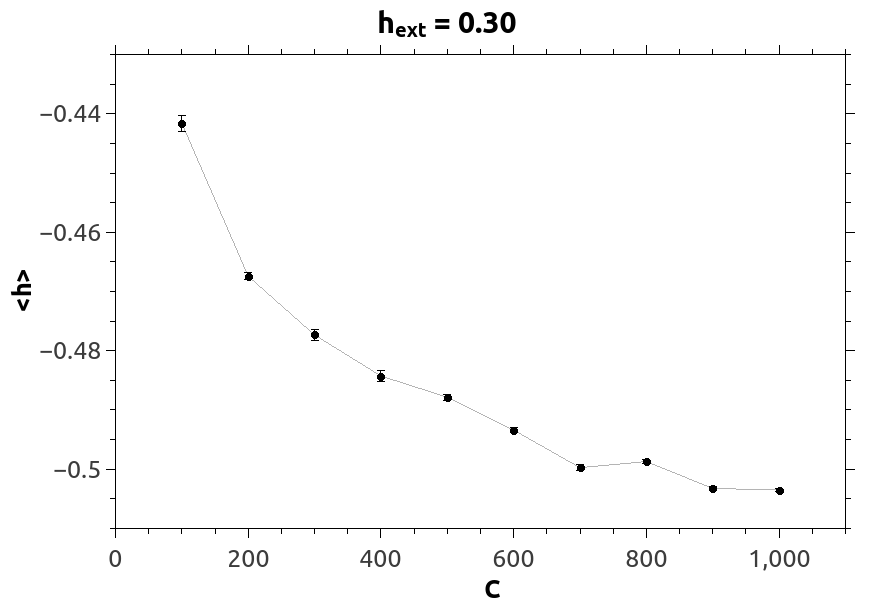}\hfill
        \includegraphics[width=.51\textwidth]{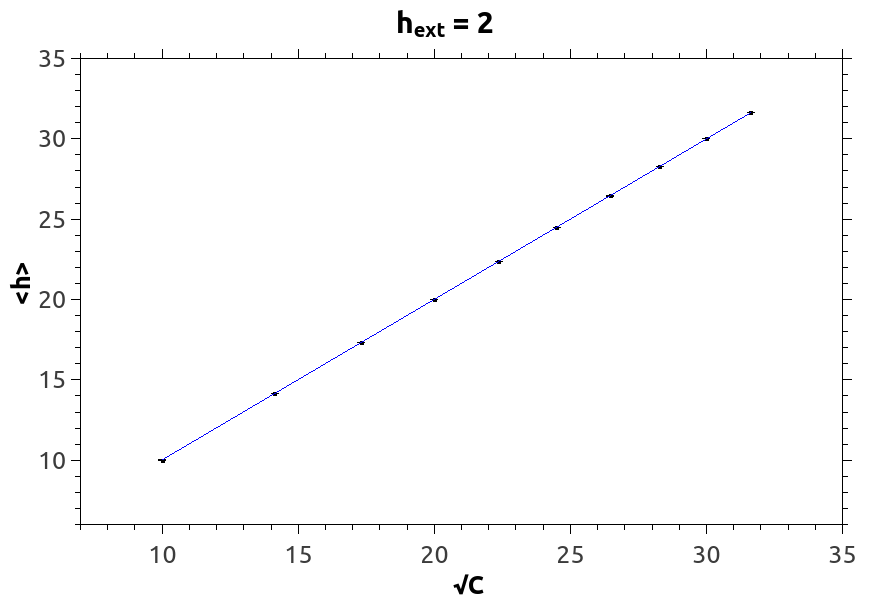}\hfill
       
        \caption{\textbf{Behaviour of $\langle h \rangle$ in the balanced and unbalanced regimes at fixed C/N $= 0.1$.} Left: in the balanced regime the mean voltage input $\langle h \rangle$ decreases until reaching a stable value corresponding to the mean field in the thermodynamic limit.  Right: In the unbalanced regime the behaviour of $\langle h \rangle$ fits a trend scaling like $\sqrt{C}$. \\Points reported in the plot are the average of the measures collected from five replicas of the random network and errorbars are the standard deviations of the means.}
        \label{fig:h_regimes}
           
\end{figure*}
\end{multicols} 
\begin{multicols}{2}
  \centering
    \begin{figure*}[h!!!]
       
        \includegraphics[width=.51\textwidth]{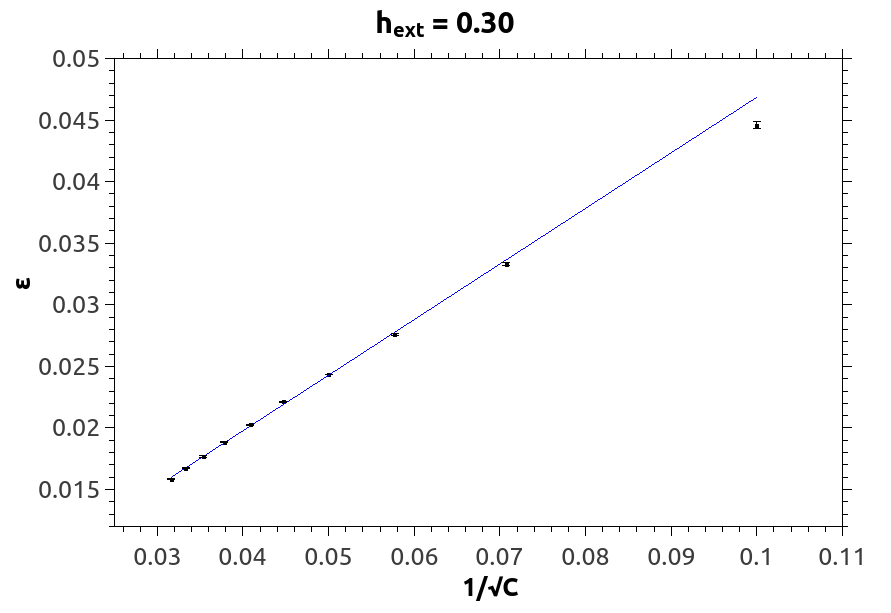}\hfill
        \includegraphics[width=.49\textwidth]{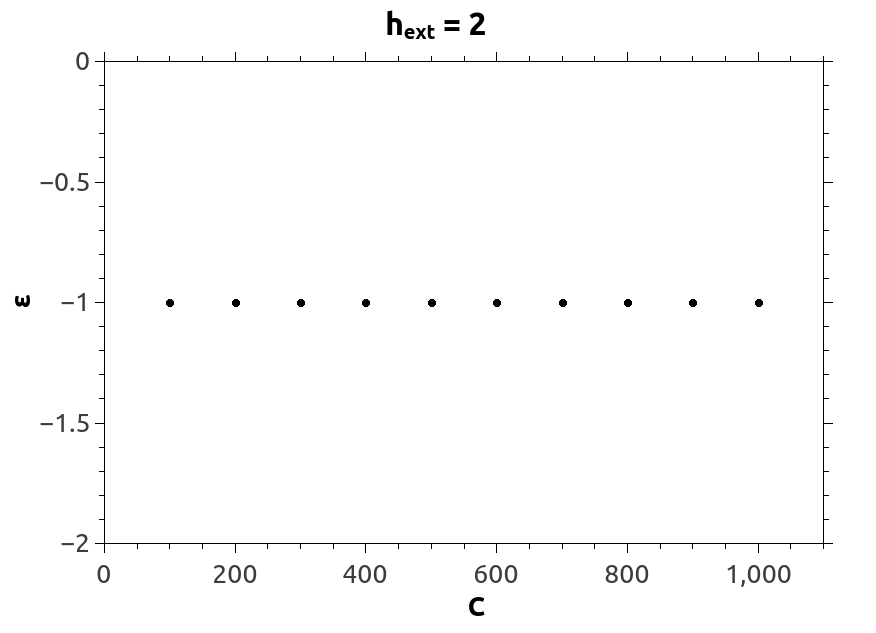}\hfill
       
        \caption{\textbf{Behaviour of $\varepsilon$ in the balanced and unbalanced regimes at fixed C/N $= 0.1$.} Left: In the balanced regime the correction $\varepsilon$ increases as $O\left(\frac{1}{\sqrt{C}}\right)$.  Right: the correction $\varepsilon$ assumes a constant trend in the unbalanced regime, as a consequence of the balance condition: $\varepsilon = -1$ because $\langle \nu \rangle$ has correctly saturated to 1.\\Points reported in the plot are the average of the measures collected from five replicas of the random network and errorbars are the standard deviations of the means.}
        \label{fig:eps_regimes}
           
\end{figure*}
\end{multicols}
Figure (\ref{fig:h_regimes}, left) displays the mean field $\langle h \rangle$ converging to a constant $O(1)$ value which is going to be called $\mu$ in the limit of $C\rightarrow\infty$, corresponding to the mean of the Gaussian distribution of fields in the thermodynamic limit. In addition, figure (\ref{fig:h_regimes}, right) reports the behaviour of $\langle h \rangle$ for $h_{ext} = 2$ that increases as $O(\sqrt{C})$ as expected from the study of the unbalanced regime.\\ Figure (\ref{fig:eps_regimes}, left) depicts the behaviour of the correction $\varepsilon(C)$ at $h_{ext} = 0.30$, that is in the balanced regime. The predicted $O\left(\frac{1}{\sqrt{C}}\right)$ trend emerges as shown from the comparison with the best fit line. Eventually, figure (\ref{fig:eps_regimes}, right) shows that $\varepsilon$ is constant and equal to $-1$, in the unbalanced regime, which is implied by $\nu$ saturating at $+1$ as expected from equation (\ref{eq:balance}).\\
\begin{figure}[h!!]
       \centering
		\includegraphics[width=12cm]{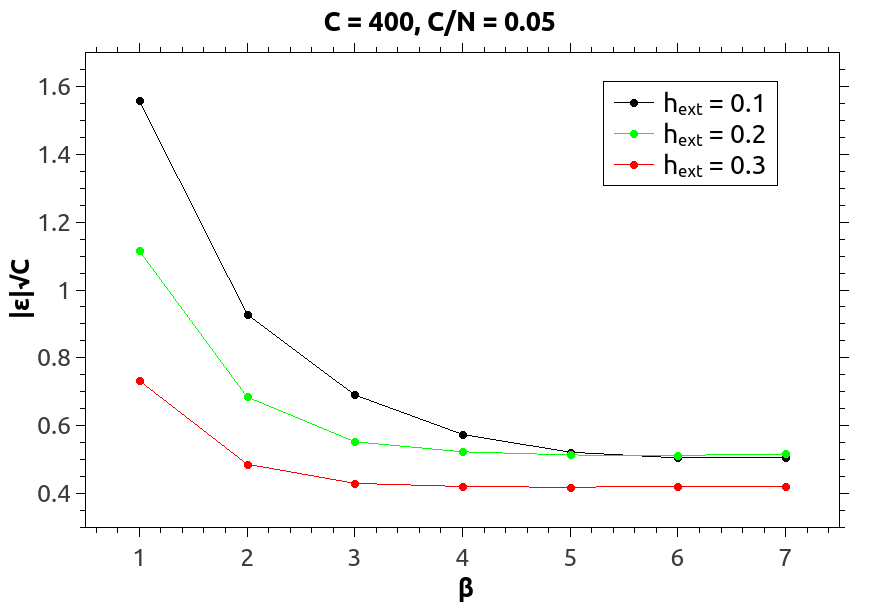}
		\label{fig:beta1}
		\caption{\textbf{The quantity $|\varepsilon|\sqrt{C}$ as a function of $\beta$ at $C = 400$ and a fixed C/N $= 0.05$}. It is clear, by increasing $\beta$, that the response of the model to the external field (depicted in figure (\ref{fig:2regimes})) becomes sharper at higher values of the gain parameter. The points at different values of $h_{ext}$ are reported. Measures are averaged over five simulations of the random network.}
		\label{fig:beta_study}
		\centering
	\end{figure}
It's also interesting to study the behaviour of the correction $\varepsilon(C)$ with respect to the variation of the gain $\beta$. It's known that, increasing $\beta$, the response of the neuron to the voltage input becomes sharper. Hence, it's also expected that the correction to quantity $\langle \nu \rangle$ becomes smaller. The absolute value of the correction multiplied by $\sqrt{C}$ at $C = 400$ is plotted while changing $\beta$ in figure (\ref{fig:beta_study}). That this quantity decreases with $\beta$ implies that finite size effects affecting the network are less important when $\beta$ is large. Nevertheless Section (\ref{sec:stab}) has proved that, given a certain statistics for the synapses, fixed points cease to be stable when $\beta$ is increased over a certain value.

\subsection{Mean Field Equations}
In this subsection we report a comparison between the estimates obtained by solving the mean field equations and numerical simulation at different values of $C,N$.
Equations (\ref{eq:dgz_mf1}),(\ref{eq:dgz_mf2}) are implicit in $\mu$ and $\sigma$. They have been solved by implementing an iterative algorithm based on an initial guess over the values of the variables. A damping convergence method is usually necessary for the success of the algorithm. Gaussian integrals have been computed making use of the Gauss-Hermite quadrature technique \cite{ref:press}. \\
\begin{figure}[h!!]
    \centering
   
		\includegraphics[width=12cm]{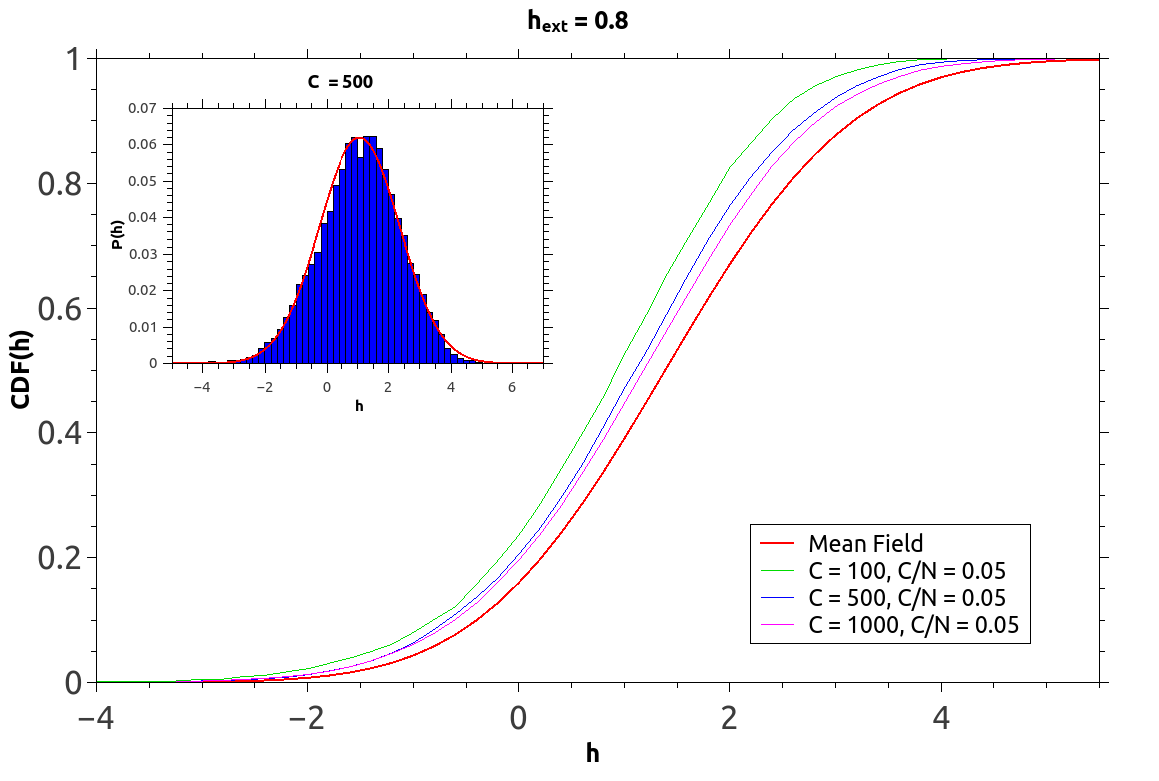}
		\caption{\textbf{Comparison of the experimental CDFs at different values of C with the mean field CDF at $h_{ext} = 0.8$ and fixed C/N $= 0.05$.} The plot shows the experimental lines reaching the theoretical limit as $C$ increases.  The subplot represents the histogram of the fields from the simulation at $C = 500$ with the best fitting Gaussian curve overplotted (namely the rescaled Gaussian featuring the experimental mean as the mean and the experimental variance as the variance). The Gaussianity of the fields is a good proof of the good prediction made by the mean field approach even at small values of the connectivity $C$.}
		\label{fig:comp}
\end{figure}
As a first test of consistency between the experiment and the mean field predictions, cumulative density functions of the fields measured in simulations have been plotted and compared to the theoretical $CDF$ expected from a  Gaussian with $\mu$ and $\sigma$ from equations (\ref{eq:dgz_mf1}), (\ref{eq:dgz_mf2}) when $h_{ext} = 0.8$. Figure (\ref{fig:comp}) reports this analysis. In the smaller subplot the histogram of the fields from the simulation shows a good agreement with a Gaussian distribution even at $C = 500$: this will allow us to implement equations (\ref{eq:dgz_mf3bis}),  (\ref{eq:dgz_mf3}),(\ref{eq:dgz_mf4}) as a tool to predict the state of the network at the fixed point at finite $C$.  Since the lines get closer to the theoretical function and fields become Gaussian as $C$ increases we can conclude that the simulated model correctly tends to the mean field behaviour in the thermodynamic limit.\\ 
\begin{figure}[h!!]
    \centering
		\includegraphics[width=13cm]{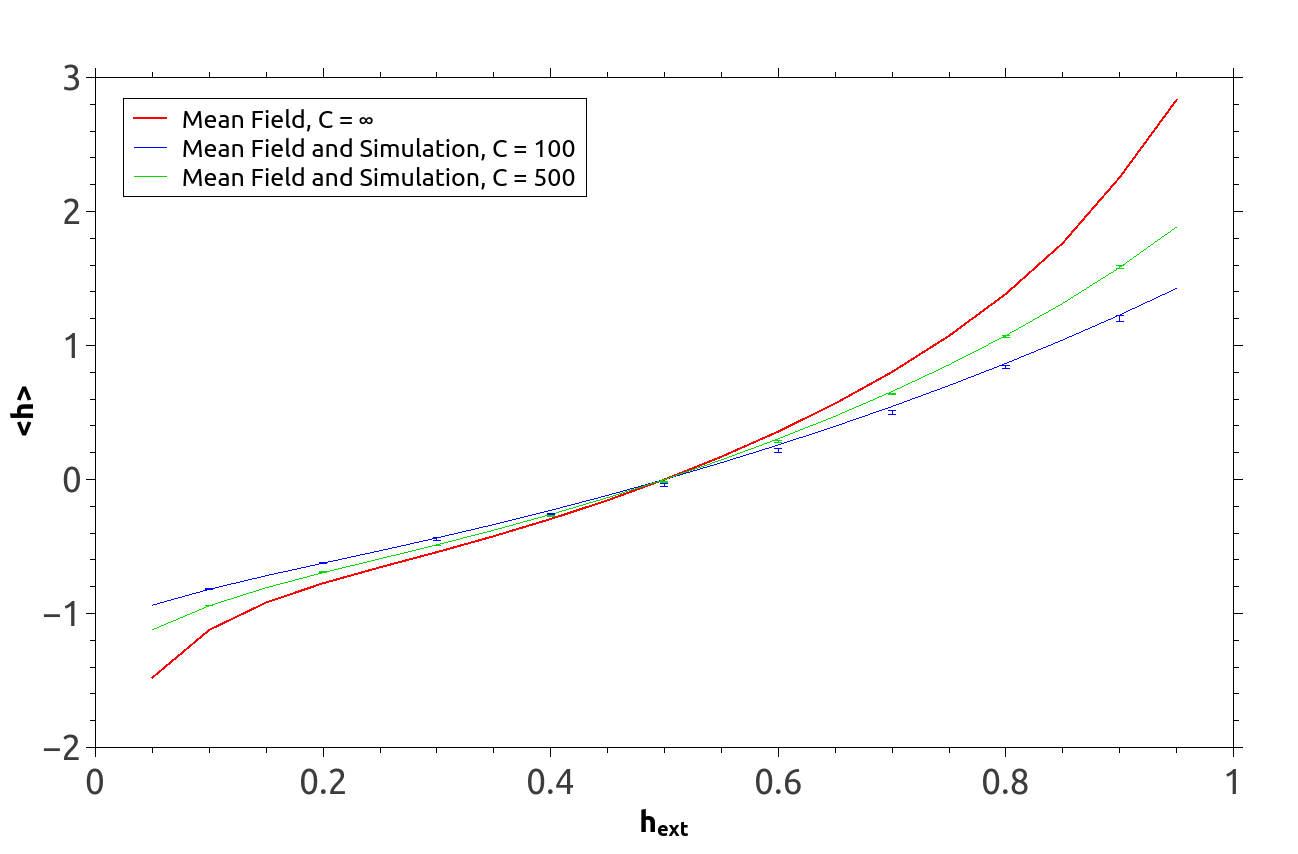}
		\caption{\textbf{The mean value of the local field as a function of $h_ {ext}$ as $C$ increases at fixed $C/N = 0.05$.} The plot reports a comparison between the mean field predictions in the thermodynamic limit, the mean field predictions at finite $C$ with the results obtained from the simulations. Points are means of five measured collected from five distinct simulations and three times the standard deviations of the means are used as errorbars.  When increasing $C$ both experimental points and the mean field predictions at finite $C$ get closer to the theoretical line showing a good consistency between theory and simulations. Experimental measures are consistent with the theoretical line at finite $C$, displaying a high degree of Gaussianity.}
		\label{fig:comp_h}
\end{figure}
\begin{figure}[h!!]
    \centering
		\includegraphics[width=13cm]{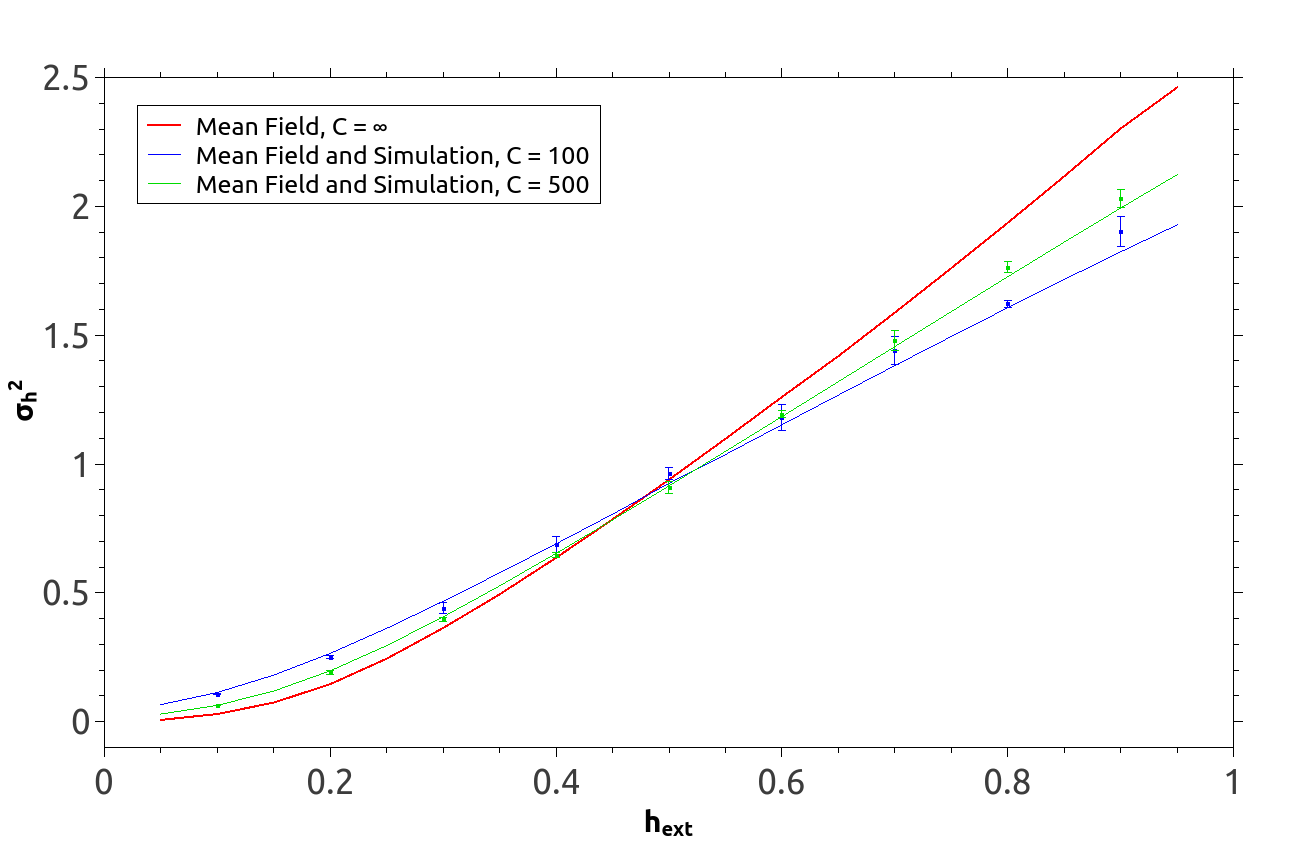}
		\caption{\textbf{Variance of the fields as a function of $h_{ext}$ while as $C$ increases at fixed $C/N = 0.05$.} The plot reports a comparison between the mean field predictions in the thermodynamic limit, the mean field predictions at finite $C$ with the results obtained from the simulations. Points are means of five measured collected from five distinct simulations and three times the standard deviations of the means are used as errorbars.  When increasing $C$ both experimental points and mean field predictions at finite $C$ get closer to the theoretical line showing a good consistency between theory and simulations. Experimental measures are consistent with the theoretical line at finite $C$, displaying a high degree of Gaussianity.}
			\label{fig:comp_sig}
\end{figure}
Figure (\ref{fig:comp_h}) and (\ref{fig:comp_sig}) represent, respectively, $\mu$ and $\sigma^2$ as functions of $h_{ext}$: the thicker line is the mean field prediction while the thinner ones are obtained from mean field at finite $C$. Points are averages computed over five simulations of the random network, while we used three times the standard deviation of the mean as errorbars. Comparing the theoretical lines with the experimental data at different values of $C$ we see that numerical results approach the theoretical predictions in the limit for $C\rightarrow\infty$ and fit well the theoretical line at finite $C$. \\Though the system is still far from satisfying the validity condition of the $DGZ$ regime of extreme dilution we can conclude that the simulated network behaves consistently with the mean field equations.
\subsection{Stability of the Fixed Point}
\begin{figure*}[ht!!!]
    \centering
        \includegraphics[width=13cm]{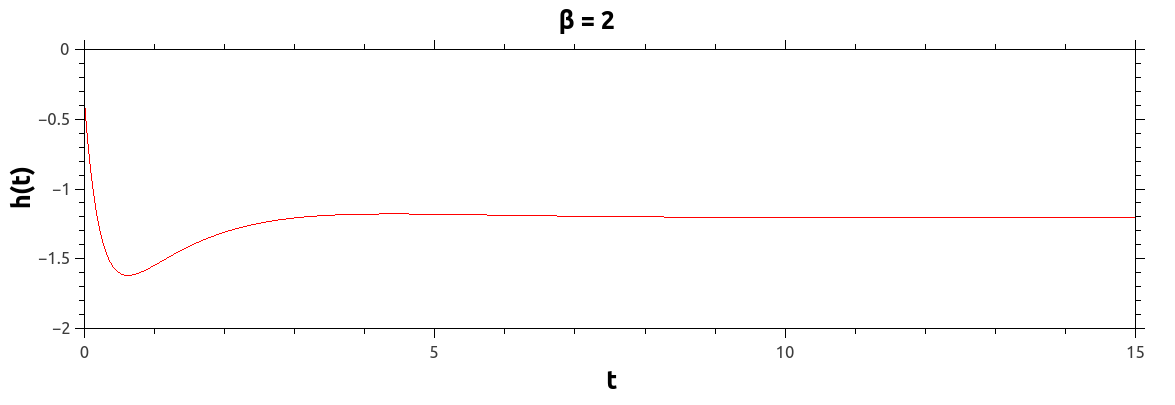}\hfill
        \includegraphics[width=13cm]{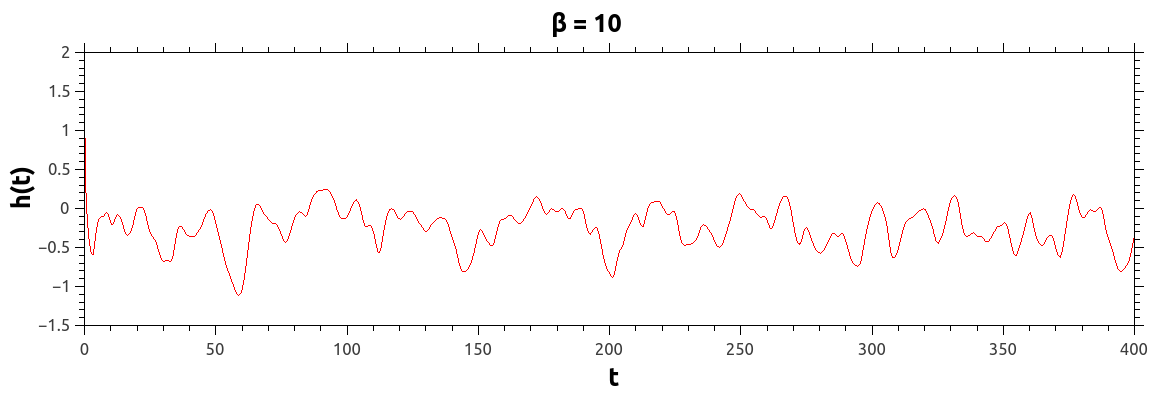}\hfill
       
        \caption{\textbf{Evolution in time of the local field of a random neuron for $h_{ext} = 0.8$, $C = 500$ and $C/N = 0.05$.} Top: Local field of a random neuron in the network for $\beta = 2$ as a function of time. The system effectively reaches the equilibrium as proved by the dynamics of the single cell. Bottom: Local field of a random neuron in the network for $\beta = 10$ as a function of time. The system does not manage to reach the equilibrium because it has collapsed in a chaotic state. Time in the plot is measured in single time units of the numerical simulation. }
        \label{fig:chaos}   
\end{figure*}

\begin{figure}[ht!!]
\centering
		\includegraphics[width=13cm]{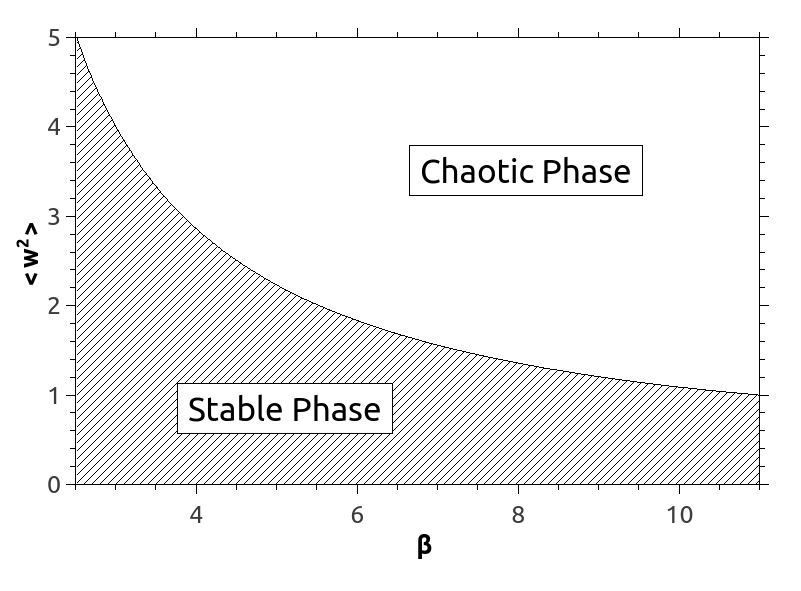}
		\label{fig:beta2}
		\caption{ \textbf{Phase diagram of the random network in the $\beta$, $\langle w^2 \rangle$ space.} The critical line has been obtained solving the mean field equations at different combinations of ($\beta, \langle w^2 \rangle$) and checking the stability condition (\ref{eq:stability}). The border of the two regions is the line satisfying equation (\ref{eq:criticaline}).}
		\label{fig:phase_diag}
		
	\end{figure}

\begin{figure}[ht!!]
\centering
		\includegraphics[width=13cm]{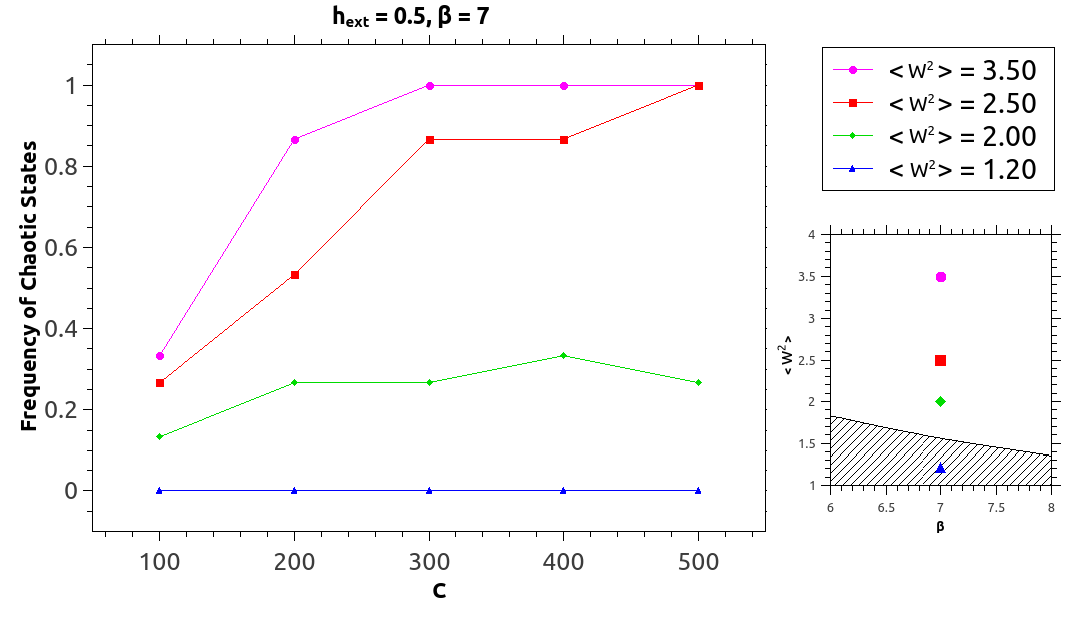}
		\label{fig:finitesize}
		\caption{\textbf{Effect of the finite size of the system on the stable-chaotic transition of the random network.} Frequency of occurrence of chaotic states of the network over $15$ consequent trials run at fixed $C/N = 0.1$ as $C$ increases. Simulations have made run for $4$ different combinations of ($\beta, \langle w^2 \rangle$) as indicated in the subplot at the bottom-right. It can be noticed that fixing $\beta$ and increasing $C$ the frequency of occurrence of chaotic states grows faster depending on the distance of the system from the critical line. It's also expected that all the lines reach the $1$ frequency at $C\rightarrow\infty$ letting finite-size effects vanish in the thermodynamic limit.}
			\label{fig:freq_chaotic}
	
	\end{figure}

Figure (\ref{fig:chaos}) depicts the evolution of the field $h_i(t)$ for a random neuron $i$ for two different values of the gain parameter $\beta$ at fixed $h_{ext}$ and $C,N$. In the upper panel the system starts the dynamics out of the equilibrium and eventually reaches a fixed point where the local field is expressed by equation (\ref{eq:fixed0}). In the lower panel, instead, the fixed point is unstable, the network is chaotic, that means it never reaches such a fixed point. 
\begin{multicols}{2}
\end{multicols}	
Once the stability condition for the fixed point is known from equation (\ref{eq:stability}), the procedure implemented to study the theoretical phase diagram of the random network consists of varying the quantities $\beta$ and $\langle w^2 \rangle$ in order to find the couples $(\beta, \langle w^2 \rangle)$ such that
\begin{equation}
\label{eq:criticaline}
\langle {\phi^{'}}^2\rangle \langle w^2 \rangle = 1
\end{equation}
by solving the mean field equations of the model (\ref{eq:dgz_mf1}),(\ref{eq:dgz_mf2}). This will give us the separation line between the chaotic phase of the network and the stable one. \\
By applying this simple procedure we have recovered the phase diagram depicted in figure (\ref{fig:phase_diag}). The network is now simulated to verify the existence of a critical line. We find very intense finite size effects at the small values of $C$ that we have been able to simulate at $C/N = 0.1$. The coexistence of both equilibrium and chaos in the region right above the critical line is likely to emerge, as already pointed out by previous works on random networks \cite{ref:crisanti}. Figure (\ref{fig:freq_chaotic}) shows an analysis conducted at $h_{ext} = 0.5$ and $\beta = 7$ and different values of $\langle w^2 \rangle$ while keeping $\langle w \rangle = 1$ and increasing $C$ keeping $C/N = 0.1$. We have decided to set such values for $\beta$ and $h_{ext}$ in order to reduce pathological finite size effects: it was previously found out that corrections are proportional to $\langle h \rangle$, which is null at $h_{ext} = 0.5$ (see figure (\ref{fig:comp_h})) and they decrease as $\beta$ increases (as shown in figure (\ref{fig:beta_study})). The plot represents the frequency with which the network collapses into a chaotic state as a function of the connectivity $C$. A chaotic state is registered when the dynamics has not reached the equilibrium before a maximum time limit set at $T_{max} = 500$. Time is measured in time steps of the algorithm we used to integrate equation (\ref{eq:dynamics}).  Notice further points in the bulk of the chaotic phase reaching the unit frequency from $C = 500$. An intermediate point, right above the boundary, reaches slower frequencies, exhibiting stronger finite size effects. On the other hand, simulations located right below the critical line admit no chaotic states, consistently with the theory. 

\chapter{A Theory of Memory in Balanced Networks}
\label{chap:3}
In this chapter we fulfill the main goal of the thesis, that is developing a formal theory of memory storage and retrieval for a balanced neural network. For this purpose, we are going to use the same system introduced in Chapter \ref{chap:3}, describing a population of inhibitory neurons modeled through a balanced neural network in the Derrida-Gardner-Zippelius regime of extreme dilution. As a difference with the previous Chapter, random synaptic efficacies are now replaced with a set of efficacies that enclose the Hebbian principle, so that memories can be stored in the system. This work thus aims to derive the mean field equations of a network which presents both the properties of a balanced neural network and the typical multistability of Hopfield-like models. Moreover, we want these equations to be solvable given a set of control parameters in order to predict the macroscopic state of the system in the thermodynamic limit. 
The full characterization of this $\textit{structured}$ balanced network will be achieved by operating in two steps. \\Firstly, Section \ref{sec:one} will be dedicated to the study of the so called one-memory model, that is a simpler case where only one memory is stored in the network. The simplification we are referring to mainly lies in a higher analytical manageability of the calculations involved: the main quantities of the system, as the mean and variance of the local fields at the fixed point of the dynamics, will be derived without taking into account the effect of other patterns. Another convenient aspect of this version of the network is a reduction of the finite size effect that disturb numerical simulations, as we will see in Section \ref{sec:num}. By consistency with the balance condition mean field equations will be deduced. \\Secondly, Section \ref{sec:multi} upgrades the model b storing a number $P$ of patterns. It will be proved, by an argument based on the dilution property of the interneuronal connections, that mean field equations do not change with respect to the one-memory model. \\Afterwards, Section \ref{sec:beta} will be devoted to the analytical solution of the mean field equations we have previously obtained in the special case of $\beta\rightarrow\infty$. Most importantly, we will also express the critical capacity of the network as a function of the control parameters of the model and evaluate its behaviour in the sparse coding limit.\\ 
At last, Section \ref{sec:num} will compare the mean field theoretical predictions with the numerical simulations of both the one-memory and multimemory models. 
\section{One-memory Model}
\label{sec:one}
The procedure that has been used to assembly the new \textit{structured} synapses, namely, to store memories to be retrieved at the fixed point, consists of applying the Hebbian principle, as it is usually done in the rest of the memory models. 
In particular, in this section only one pattern is going to be stored.
Starting from this elementary architecture of the synaptic efficacies we will seek a solution of the dynamic equation that retrieves the pattern, that is, a configuration of the network activity that describes the state of the system at the fixed point and which explicitly depends on the realisation of the pattern.

We have called this particular version of the model \textit{one-memory model} to differentiate it from the subsequent generalization where an extensive number of patterns will be memorized. 
\subsection{Description of the Model}
The only stored pattern is a vector $\vec{\xi}$ where its components $\xi_i$ are i.i.d. binary variables generated according to the following rule
\[ 
\xi_i = 
\left\{ \begin{array}{l}
1\hspace{1cm}\text{with probability f}\\
0\hspace{1cm}\text{with probability 1 - f}
\end{array} \right. 
\]
with $i = 1,..,N$. We now define $\textit{active}$ sites as all $i$ sites that contribute to the pattern with a coordinate $\xi_i = 1$ and $\textit{inactive}$ sites as the ones being associated to $\xi_i = 0$. The probability $f$ is called $\textit{coding level}$ of the network and it also represents the average number of active sites in the pattern.\\The Hebbian theory expects memories to be embedded in the architecture of the network, namely, into the synaptic efficacies $w_{ij}$. To do this, we introduce the \textit{Hebbian terms} $z_{ij}$ that are defined as it follows
\begin{equation}
\label{eq:hebbian}
z_{ij} = \Tilde{z}_{ij}-\frac{1}{\sqrt{\alpha C}}\frac{(\xi_i-f)(\xi_j-f)}{f(1-f)}
\end{equation}
where $\Tilde{z}_{ij}$ is a random variable drawn from a Gaussian distribution with $0$ mean and unit variance. $\alpha$ is a control parameter of the model that should recall a well-known parameter from associative memory models. $C$ is the mean connectivity of the network, as introduced in the previous Sections.\\The minus sign in front of the second term in equation (\ref{eq:hebbian}), the one depending on the stored pattern, indicates that an anti-Hebbian learning is actually adopted: stronger synapses are the ones that link one active cell to an inactive one, because they correctly inhibit each other. This method of storing memories is in contrast with the Hebbian rule used in Hopfield-like models, where the synaptic efficacy was enhanced when two neurons fired together.\\Since $z_{ij}$ can be both positive and negative, they are not good synaptic efficacies to be used, since we are modeling an inhibitory population of neurons. Therefore, efficacies are defined as $$w_{ij} = F(z_{ij})$$ with $F$ non negative function that respects the Dale's Law of Neuroscience. A possible choice for the $F(x)$ function is
\begin{equation}
\label{eq:F}
F(z_{ij}) = \exp\left( \mu_z + z_{ij}\sigma_z\right)
\end{equation}
This choice is equivalent to consider $w_{ij}$ as a lognormal variable when $C\rightarrow \infty$. 
Yet again we use dilution variables $c_{ij}$ drawn from the probability distribution expressed by equation (\ref{eq:c}), so that the system works in the $DGZ$ regime of extreme dilution.
\subsection{Mean Field Equations}
\label{subsec:mf_one}
Once the pattern has been stored in the network, we are interested in finding that fixed point of the dynamics that is correlated with the memory, performing its retrieval. The local field at the fixed point is
\begin{equation}
\label{eq:fixed}
h_i = \sqrt{C}h_{ext} - \frac{1}{\sqrt{C}}\sum_{j=1}^N w_{ij}c_{ij}\nu_j 
\end{equation}
When the retrieval condition is fulfilled, the activity of each site $\nu_i$ must be dependent on the realization of the pattern on that site $\xi_i$. It follows that the local field $h_i$ in equation (\ref{eq:fixed}) will be also dependent on the realization of the pattern on the same site. Depending on the pattern configuration on sites $i$ and $j$, synaptic efficacies can be rewritten as
\[ 
w_{ij} = e^{\mu_z + \sigma_z \tilde{z_{ij}}}\cdot 
\left\{ \begin{array}{l}
\exp\left[ +\frac{\sigma_z}{\sqrt{\alpha C}}\frac{(\xi_i-f)}{(1-f)}\right]\hspace{1.15cm}\text{when}\hspace{0.5cm} \xi_j = 0\\
\exp\left[ -\frac{\sigma_z}{\sqrt{\alpha C}}\frac{(\xi_i-f)}{f}\right]\hspace{1.15cm}\text{when}\hspace{0.5cm}\xi_j = 1\\
\end{array} \right. 
\]
Hereafter active sites are going to be indicated with a symbol $(+)$ and inactive ones with $(-)$. We can separate the sum in equation (\ref{eq:fixed}) in two pieces, one for $\xi_j = 1$ and the other for $\xi_j = 0$, obtaining
\begin{multline}
\label{eq:newfixed}
    h_i = \sqrt{C}h_{ext} -\frac{1}{\sqrt{C}}\sum_{j:\xi_j = 1}c_{ij}w_{ij}\nu_j -\frac{1}{\sqrt{C}}\sum_{j:\xi_j = 0}c_{ij}w_{ij}\nu_j = \\ = \sqrt{C}h_{ext} -\frac{1}{\sqrt{C}}e^{\mu_z -\frac{\sigma_z}{\sqrt{\alpha C}}\frac{(\xi_i-f)}{f}}\sum_{j:\xi_j = 1}c_{ij}e^{\sigma_z\tilde{z_{ij}}}\nu_j -\\-\frac{1}{\sqrt{C}}e^{\mu_z +\frac{\sigma_z}{\sqrt{\alpha C}}\frac{(\xi_i-f)}{(1-f)}}\sum_{j:\xi_j = 0}c_{ij}e^{\sigma_z\tilde{z_{ij}}}\nu_j
\end{multline}
which for $N\rightarrow\infty$ and finite $C$ becomes
\begin{equation}
\label{eq:av_h}
    \langle h \rangle_{C,\xi_i} = \sqrt{C}\left[ h_{ext} - e^{\mu_z + \frac{\sigma_z^2}{2}}\left(f e^{ -\frac{\sigma_z}{\sqrt{\alpha C}}\frac{(\xi_i-f)}{f}}\langle \nu \rangle_+ + (1-f) e^{+\frac{\sigma_z}{\sqrt{\alpha C}}\frac{(\xi_i-f)}{(1-f)}}\langle \nu \rangle_- \right)\right]
\end{equation}
where equation (\ref{eq:momw}) has been used to compute $\langle e^{\mu_z + \sigma_z \tilde{z_{ij}}} \rangle = e^{\mu_z + \frac{\sigma_z^2}{2}}$. One can compute $\langle h^2 \rangle_{C,\xi_i}$ by applying the same reasoning and derive the variance, 
obtaining  
\begin{equation}
\label{eq:sigma_h}
\sigma^2_{C,\xi_i} = e^{2(\mu_z + \sigma_z^2)}\left[ f e^{ -2\frac{\sigma_z}{\sqrt{\alpha C}}\frac{(\xi_i-f)}{f}}\langle \nu^2 \rangle_+ + (1-f) e^{+2\frac{\sigma_z}{\sqrt{\alpha C}}\frac{(\xi_i-f)}{(1-f)}}\langle \nu^2 \rangle_- \right]
\end{equation}
Equations (\ref{eq:av_h}),(\ref{eq:sigma_h}) are exact at any finite $C$ and they represent the mean and the variance of the field experienced by the site $i$ of the network conditioned to the fact that $\xi_i =1$ or $\xi_i = 0$: depending on the activity/inactivity of the site we have two different statistics with mean $\langle h \rangle_{C,\pm}$ and variance $\sigma_{C,\pm}^2$. 
\\These equations can be rewritten in the $C\rightarrow\infty$ limit by Taylor expanding the exponentials for large values of $C$. We hence find
\begin{equation}
\label{eq:av_h_C}
\langle h \rangle_{\xi_i} = \mu + \sigma_z \frac{\langle w \rangle}{\sqrt{\alpha}}(\xi_i - f)\left[ \langle \nu\rangle_+ - \langle \nu \rangle_-\right]
\end{equation}
\begin{equation}
\label{eq:sigma_h_C}
\sigma^2 = \langle w^2 \rangle \langle \nu^2 \rangle
\end{equation}
where the variance no more depends on the configuration of the pattern $\xi_i$ and both the mean and the variance do not depend on the connectivity $C$. The following expressions of the mean activity and the mean squared activity in the $N\rightarrow\infty$ limit have been exploited to recover equations (\ref{eq:av_h_C}),(\ref{eq:sigma_h_C}). 
\begin{equation}
\label{eq:av_nu}
\langle \nu \rangle = f\langle \nu \rangle_+ + (1-f)\langle \nu \rangle_-    
\end{equation}
\begin{equation}
\label{eq:av_nusq}
\langle \nu^2 \rangle = f\langle \nu^2 \rangle_+ + (1-f)\langle \nu^2 \rangle_-    
\end{equation}
and also the lognormal limit of $w_{ij}$ when $C\rightarrow\infty$.
$\mu$ is a consequence of the balance condition. In fact, at finite $C$, we request that
\begin{equation}
\label{eq:balance_struct}
f\langle h \rangle_{C,+} + (1-f)\langle h \rangle_{C,-} = \sqrt{C}\left( h_{ext} - e^{\mu_z + \frac{\sigma_z^2}{2}}\langle \nu \rangle\right) = O(1)
\end{equation} 
which implies, in the $C\rightarrow\infty$ limit, that
\begin{equation}
\label{eq:balance_struct2}
    \langle \nu \rangle = \frac{h_{ext}}{\langle w \rangle}
\end{equation}
and
\begin{equation}
\label{eq:muu}
f\langle h \rangle_{+} + (1-f)\langle h \rangle_{-} = \mu
\end{equation}
We have found that in the thermodynamic limit the local fields are distributed according to a combination of two Gaussians having $\langle h \rangle_{\pm}$, $\sigma^2$ as cumulants. \\It is now useful to introduce the order parameter of the model, the $\textit{overlap}$, defined as
\begin{equation}
\label{eq:over}
m = \frac{1}{N}\sum_{j=1}^N \frac{(\xi_j-f)}{f(1-f)}\nu_j
\end{equation}
This quantity gives a measure of the 
correlation of the fixed point $\vec{\nu}$ with the stored pattern. 
Notice that, by substituting $\nu_j = \pm\xi_j$ one finds $m = \pm1$. In particular, when $m \neq 0$ the system is in the retrieval phase, otherwise the network has not recalled the memory (or its complementary state) because the activity configuration is orthogonal to the pattern. Notice that the overlap, when $N\rightarrow\infty$, can be expressed as
\begin{equation}
\label{eq:new_over}
m = \frac{1}{Nf}\sum_{j:\xi_j = 1}\nu_j -\frac{1}{N(1-f)}\sum_{j:\xi_j = 0}\nu_j = \langle \nu\rangle_+ - \langle \nu \rangle_-     
\end{equation}
From here it is evident that the overlap signals the retrieval of the memory when $\langle \nu \rangle_+ \neq \langle \nu \rangle_-$.\\
Substituting the expression found in (\ref{eq:new_over}) in equation (\ref{eq:av_h_C}) and from the Gaussianity of the fields implied by the both the dilution limit and the asymmetry of the synaptic efficacies, one can express the fixed point of the dynamics as
\begin{equation}
\label{eq:lf3}
h_i = \mu + \eta_i\sigma + \frac{(\xi_i-f)}{\sqrt{\alpha}}A m 
\end{equation}
which depends on the realisation of the pattern as long as $m \neq 0$. We have defined 
\begin{equation}
\label{eq:A}  
 A = \langle \frac{dF}{dz}\rangle = \sigma_z\langle w \rangle
\end{equation}
By consistency with equations (\ref{eq:av_nu}), (\ref{eq:av_nusq}), (\ref{eq:new_over}), and exploiting the balance of the network (\ref{eq:balance_struct2}) one can finally write the mean field equations of the one-memory model:
\begin{multline}
\label{eq:mf1}
\frac{h_{ext}}{\langle w \rangle} = f\int_{-\infty}^{+\infty} Dz\phi\left(\mu + z\sigma + \frac{(1-f)}{\sqrt{\alpha}}Am\right) + \\ +(1-f)\int_{-\infty}^{+\infty} Dz\phi\left(\mu + z\sigma - \frac{f}{\sqrt{\alpha}}Am\right)
\end{multline}
\begin{multline}
\label{eq:mf3}
\sigma^2  = \langle w^2 \rangle[f\int_{-\infty}^{+\infty} Dz\phi^2\left(\mu + z\sigma + \frac{(1-f)}{\sqrt{\alpha}}Am\right) + \\ + (1-f)\int_{-\infty}^{+\infty} Dz\phi^2\left(\mu + z\sigma - \frac{f}{\sqrt{\alpha}}Am\right)]
\end{multline}
\begin{equation}
\label{eq:mf2}
m = \int_{-\infty}^{+\infty} Dz\phi\left(\mu + z\sigma + \frac{(1-f)}{\sqrt{\alpha}}Am\right) - \int_{-\infty}^{+\infty} Dz\phi\left(\mu + z\sigma - \frac{f}{\sqrt{\alpha}}Am\right)
\end{equation}
\\These equations can be solved numerically by computing the Gaussian integrals of the $\phi$ function. Shared solutions among the three equations solve the system, representing the state of the network at the fixed point of the dynamics. The network state will depend on the particular combination of the control parameters, namely $h_{ext}$, $f$, $\alpha$ and the chosen statistics for the synaptic efficacies. The threshold $\theta$ does not appear in the mean field equations, as a consequence of the balance condition, so it is not a relevant parameter of the model.  
\\ \\
\subsection{Deriving the exact Statistics of the Synaptic Efficacies for the One-Memory model}
\begin{figure*}[h!!!]
         \includegraphics[width=.51\textwidth]{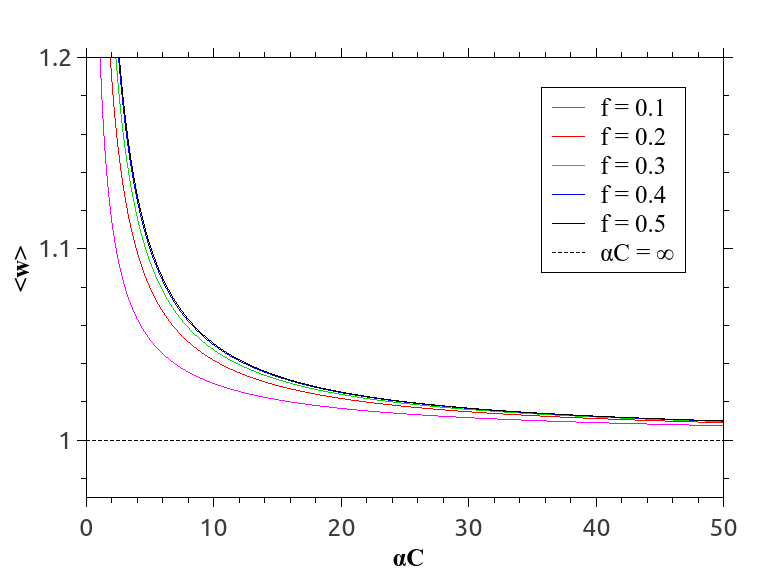}\hfill
            \includegraphics[width=.51\textwidth]{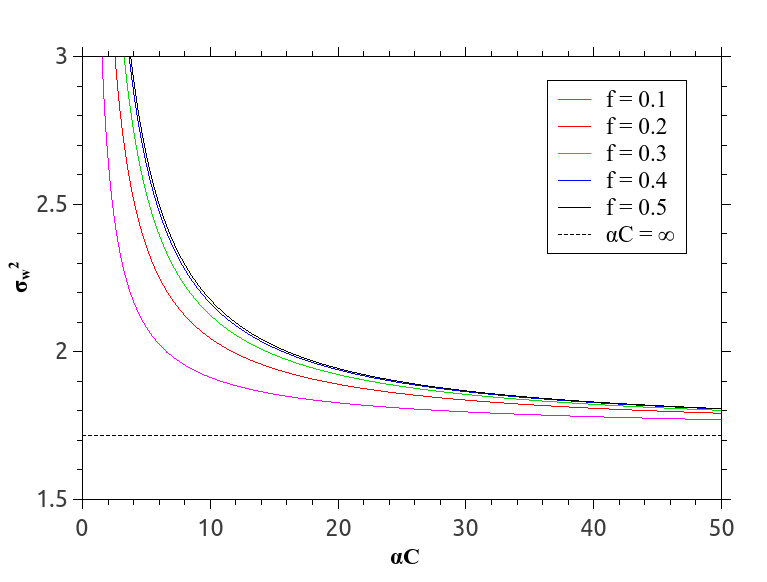}\hfill
             \caption{\textbf{Behaviour of $\langle w \rangle$ and $\sigma_w^2$ as functions of $\alpha C$ in the one-memory model.} The parameters of the distribution are $\sigma_z = 1$ and $\mu_z = -\frac{\sigma_z^2}{2}$. Left: the mean synaptic efficacy as a function of $\alpha C$ at different values of the coding level $f$.  Notice that for all $f$ the system approaches the asymptotic value of $\langle w \rangle$ very fast: at $\alpha C = 20$ the relative error is $< 10\%$. Right: variance of the synaptic efficacies as a function of $\alpha C$ at different values of the coding level $f$.  Also in this case the variance reaches the asymptotic limit very fast, showing a relative error that is $< 20\%$ at $\alpha C = 20$.}
            \label{fig:w_one}           
           \end{figure*}

Let us rewritte the Hebbian term as 
\begin{equation}
\label{eq:newhebb}
z_{ij} = \tilde{z}_{ij} + \epsilon_{ij}
\end{equation}
where 
\begin{equation}
\label{eq:eps_hebb}
\epsilon_{ij} = -\frac{1}{\sqrt{\alpha C}}\frac{(\xi_i-f)(\xi_j-f)}{f(1-f)}
\end{equation}
From the statistics of the $\xi_i$ one can deduce that
\[ 
\epsilon_{ij} = 
\left\{ \begin{array}{l}
-\frac{1}{\sqrt{\alpha C}}\frac{(1-f)}{f}\hspace{1.5cm}p_1 = f^2\\
-\frac{1}{\sqrt{\alpha C}}\frac{f}{(1-f)}\hspace{1.5cm}p_2 = (1-f)^2\\
\frac{1}{\sqrt{\alpha C}}\hspace{2.7cm}p_3 = 2f(1-f)
\end{array} \right. 
\]
Synaptic efficacyies are rewritten as
\begin{equation}
\label{eq:w_stat}
w_{ij} = e^{\left(\mu_z + \sigma_z \tilde{z}_{ij}\right)}\exp\left(\sigma_z \epsilon_{ij}\right)
\end{equation}
where $\tilde{z}_{ij}$ is by definition a Gaussian variable with $0$ mean ad unit variance. Since $\tilde{z}_{ij}$ is generated independently from $\epsilon_{ij}$, the  moments of order $n$ of the distribution of $w_{ij}$ at any finite value of $\alpha C$ are
\label{eq:w_n}
\begin{equation}
\langle w^n \rangle = e^{n\left(\mu_z + n\frac{\sigma_z^2}{2}\right)} \left[ f^2 e^{ -n\frac{\sigma_z}{\sqrt{\alpha C}}\frac{(1-f)}{f}} + (1-f)^2 e^{ -n\frac{\sigma_z}{\sqrt{\alpha C}}\frac{f}{(1-f)}} + 2f(1-f) e^{n\frac{\sigma_z}{\sqrt{\alpha C}}}\right]
\end{equation}
from which we can compute the mean and the variance of the synaptic efficacies as
\begin{equation}
\label{eq:w_av}
\langle w \rangle = e^{\left(\mu_z + \frac{\sigma_z^2}{2}\right)} \left[ f^2 e^{ -\frac{\sigma_z}{\sqrt{\alpha C}}\frac{(1-f)}{f}} + (1-f)^2 e^{ -\frac{\sigma_z}{\sqrt{\alpha C}}\frac{f}{(1-f)}} + 2f(1-f) e^{\frac{\sigma_z}{\sqrt{\alpha C}}}\right]
\end{equation}
\begin{multline}
\label{eq:var_w}
\sigma_w^2 = e^{2\left(\mu_z + \sigma_z^2\right)} \left[ f^2 e^{ -\frac{2\sigma_z}{\sqrt{\alpha C}}\frac{(1-f)}{f}} + (1-f)^2 e^{ -2\frac{\sigma_z}{\sqrt{\alpha C}}\frac{f}{(1-f)}} + 2f(1-f) e^{\frac{2\sigma_z}{\sqrt{\alpha C}}}\right] - \\ - e^{2\left(\mu_z + \frac{\sigma_z^2}{2}\right)} \left[ f^2 e^{ -\frac{\sigma_z}{\sqrt{\alpha C}}\frac{(1-f)}{f}} + (1-f)^2 e^{ -\frac{\sigma_z}{\sqrt{\alpha C}}\frac{f}{(1-f)}} + 2f(1-f) e^{\frac{\sigma_z}{\sqrt{\alpha C}}}\right]^2
\end{multline}
\\By definition, $$A = \sigma_z \langle w \rangle$$ Consequently, the behaviour of $\langle w \rangle$ in the thermodynamic limit will determine the behaviour of $A$. \\
Taylor expanding  the exponential in equations ($\ref{eq:w_av}$), (\ref{eq:var_w}) when $C\rightarrow\infty$ one recovers the cumulants of a lognormal distribution, in accordance with the Gaussian limit achieved by $z_{ij}$.
\begin{equation}
\label{ref:cum_w_inf}
\langle w \rangle = \text{exp}\left(\mu_z + \frac{\sigma_z^2}{2}\right) \hspace{2cm}\sigma_w^2 = \text{exp}\left[2\left(\mu_z + \sigma_z^2\right)\right]
\end{equation}
Figure (\ref{fig:w_one}) illustrates the behaviour of $\langle w \rangle$ and $\sigma_w^2$ as functions of $\alpha C$ showing the lines reaching the asymptotic limit very fast for any chosen $f$. Reduced finite size effects are thus expected in simulations of the system at finite values of $\alpha C$.   
\begin{multicols}{2}
        
    \end{multicols} 
\newpage
\section{Multimemory Model}
\label{sec:multi}
Passing to the multimemory model permits to generalize our theory to the more realistic case where an extensive number of memories are stored, as it happens in real neural networks and in most studied associative memory models. We are going to prove that the theoretical framework developed in the previous Section is identically valid in the current case of $P$ stored patterns, as a consequence of the extreme dilution hypothesis. 

\subsection{Description of the Model}

By contrast with the one-memory model, we now aim to store $P$ randomly generated patterns. Therefore, patterns $\vec{\xi^{\mu}}$ are assembled with the following rule
\[ 
\xi_i^{\mu} = 
\left\{ \begin{array}{l}
1\hspace{1cm}\text{with probability f}\\
0\hspace{1cm}\text{with probability 1 - f}
\end{array} \right. 
\]
with $i = 1,..,N$ and $\mu = 1,..,P = \alpha C$. The control parameter $\alpha$ achieves now the meaning of $\textit{load parameter}$, indicating the storage capacity of the neural network. In the multimemory model Hebbian terms are defined as
\begin{equation}
\label{eq:hebbian2}
z_{ij} = -\frac{1}{\sqrt{P}}\sum_{\mu=1}^P\frac{(\xi_i^{\mu}-f)(\xi_j^{\mu}-f)}{f(1-f)}
\end{equation}
that is a random variable which tends to a Gaussian with $0$ mean and unit variance in the $C\rightarrow\infty$ limit. Synaptic efficacies are generated as $w_{ij} = F(z_{ij})$ where $F(x)$ is the same defined in equation (\ref{eq:F}).
\subsection{Mean Field Equations}
Also in this case, memory retrieval occurs when the fixed point of the dynamics is strictly dependent on the realisation of the retrieved pattern, namely, there is a strong correlation between the neuronal state and the particular memory recalled by the network.\\  
We firstly generalize the $\textit{overlap}$ defined for the one-memory model to the current case. Since we can singularly retrieve $P$ different patterns, $m$ becomes a vector $m_{\mu}$ expressed by
\begin{equation}
\label{eq:over2}
m_{\mu} = \frac{1}{N}\sum_{j=1}^N \frac{(\xi_j^{\mu}-f)}{f(1-f)}\nu_j
\end{equation}
We will rename $m_1 = m$ considering the $\mu = 1$ pattern as the retrieved (or $\textit{condensed}$) one. It appears evident from equation (\ref{eq:over2}) that $m \neq 0$ and $m_{\mu >1} = O\left(\frac{1}{\sqrt{N}}\right)$.\\In principle, when the system stores an extensive number of patterns, even if memory $\mu = 1$ is recalled, the fixed point might be consistently correlated with the so called $\textit{uncondensed}$ patterns, the non-retrieved ones. However, since we put ourselves in the $DGZ$ extreme dilution limit one can make the following consideration: an estimate of the correlation of the fixed point $\vec{\nu}$ with the uncondensed patterns is given by $$ \sum_{\mu > 1}m_{\mu} = O\left(\frac{C}{\sqrt{N}}\right)\xrightarrow[]{N\rightarrow \infty} 0$$ a vanishing quantity in the $\textit{first step}$ of the thermodynamic limit.  
\\Hence we can rewrite the Hebbian term as
\begin{equation}
\label{eq:hebbian3}
z_{ij} = \Tilde{z_{ij}} - \frac{1}{\sqrt{P}}\frac{(\xi_i^{1} - f)(\xi_j^{1}-f)}{f(1-f)}
\end{equation}
where $\Tilde{z_{ij}} = O(1)$ and it contains the contribution given by the uncondensed patterns, while the second term is $O\left(\frac{1}{\sqrt{P}}\right)$ and it depends on the retrieved memory, that is a known vector. 
Hereafter the procedure that can be used to recover the mean field equations is the same implemented in Section \ref{sec:one} Subsection \ref{subsec:mf_one} replacing $\xi_i$ with $\xi^1_i$. It has just to be born in mind in the computations that, even if $\tilde{z}_{ij}$ is uncorrelated with the rest of $z_{ij}$ thanks to the DGZ limit, it is not Gaussian already, because at finite $C$, $P$ is also finite. Consequently $\langle e^{\mu_z + \sigma_z \tilde{z_{ij}}} \rangle = e^{\mu_z + \frac{\sigma_z^2}{2}}$ only in the $C\rightarrow\infty$ limit and not before. 
Once again it is found that
$$ h_i = \mu + \eta_i\sigma +  \frac{(\xi^1_i-f)}{\sqrt{\alpha}}A m$$
with $A$ defined by equation (\ref{eq:A}) and $\sigma$ is the same reported in equation (\ref{eq:sigma_h_C}).\\
Equations (\ref{eq:mf1}),(\ref{eq:mf3}),(\ref{eq:mf2}) are thus identically recovered for the multimemory model. 
\\ \\
\subsection{Deriving the exact Statistics of the Synaptic Efficacies for the Multimemory Model}
The exact statistics of synapses in the multimemory model is now derived.\\
One can thus rewrite equation (\ref{eq:hebbian2}) as 
\begin{equation}
\label{eq:z2}
    z_{ij} = \sum_{\mu}^P z_{\mu}
\end{equation}
with 
\begin{equation}
\label{eq:zmu}
    z_{\mu} =  -\frac{(\xi_i^{\mu} -f)(\xi_j^{\mu}-f)}{\sqrt{P}f(1-f)}
\end{equation}
which is a random variable that is distributed accordingly to the following discrete distribution
\[ 
z_{\mu} = 
\left\{ \begin{array}{l}
-\frac{(1-f)}{\sqrt{P}f}\hspace{1.5cm}p_1 = f^2\\
-\frac{f}{\sqrt{P}(1-f)}\hspace{1cm}p_2 = (1-f)^2\\
\frac{1}{\sqrt{P}}\hspace{2.1cm}p_3 = 2f(1-f)
\end{array} \right. 
\]
We can then derive the distribution of the sum \begin{multline}
\label{eq:pdf1}
P(z_{ij} = x) = \frac{P!}{n_1!n_2!n_3!}f^{2n_1}(1-f)^{2n_2}[2f(1-f)]^{n_3}\times \\ \times\delta\left( n_1 + n_2 + n_3 - P\right)\delta\left( -n_1\frac{(1-f)}{\sqrt{P}f} - n_2\frac{f}{\sqrt{P}(1-f)} + \frac{n_3}{\sqrt{P}}-x \right) = 
\end{multline}
\begin{multline}
\label{eq:pdf2}
 = \frac{P!}{n_1!n_2!(P-n_1-n_2)!}f^{2n_1}(1-f)^{2n_2}[2f(1-f)]^{P-n_1-n_2}\x \\ \x\delta\left( \frac{n_1}{\sqrt{P}f} + \frac{n_2}{\sqrt{P}(1-f)} - \sqrt{P} + x \right)
\end{multline}
Synaptic efficacies are defined by equation (\ref{eq:F}). Hence
\begin{multline}
\label{eq:w_av1}
\langle w \rangle = \sum_{n_1,n_2}^P\int\frac{P!}{n_1!n_2!(P-n_1-n_2)!}f^{2n_1}(1-f)^{2n_2}\left[2f(1-f)\right]^{P-n_1-n_2} e^{\mu_z + \sigma_z x}\times \\ \times\delta\left( \frac{n_1}{\sqrt{P}f} + \frac{n_2}{\sqrt{P}(1-f)} - \sqrt{P} + x \right)dx = 
\end{multline}
\begin{multline}
\label{eq:w_av2}
 = e^{\mu_z + \sigma_z\sqrt{P}}\sum_{n_1,n_2}^P\frac{P!}{n_1!n_2!(P-n_1-n_2)!}\left(f^2e^{-\frac{\sigma_z}{\sqrt{P}f}}\right)^{n_1}\x \\ \x\left((1-f)^2e^{-\frac{\sigma_z}{\sqrt{P}(1-f)}}\right)^{n_2}\left[2f(1-f)\right]^{P-n_1-n_2} 
\end{multline}
Invoking the multinomial theorem we obtain
\begin{equation}
\label{eq:w_av3}
\langle w \rangle = e^{\mu_z + \sigma_z\sqrt{P}}\left[f^2e^{-\frac{\sigma_z}{\sqrt{P}f}} + (1-f)^2e^{-\frac{\sigma_z}{\sqrt{P}(1-f)}} + 2f(1-f)\right]^P 
\end{equation}
In the same manner one can compute $\langle w^2 \rangle$ so that the variance is expressed by
\begin{multline}
\label{eq:sigma_w1}
\sigma_w^2 = e^{2(\mu_z + \sigma_z\sqrt{P})}\{\left[f^2 e^{-\frac{2\sigma_z}{\sqrt{P}f}} + (1-f)^2 e^{-\frac{2\sigma_z}{\sqrt{P}(1-f)}} + 2f(1-f)\right]^P - \\ -\left[f^2e^{-\frac{\sigma_z}{\sqrt{P}f}} + (1-f)^2e^{-\frac{\sigma_z}{\sqrt{P}(1-f)}} + 2f(1-f)\right]^{2P} \}  
\end{multline}
Analogously, the quantity $A$ appearing in the mean field equations measures $$A = \sigma_z \langle w \rangle$$ hence it shows the same behaviour of
$ \langle w \rangle$ deformed by a factor $\sigma_z$.\\ This result is exact at finite values of $P$. We can test that the statistics of the $w_{ij}$ correctly reaches the lognormal limit for $C\rightarrow\infty$ by Taylor expanding the exponentials in equations (\ref{eq:w_av3}),(\ref{eq:sigma_w1}) for small values of the exponent. Let us show it for equation (\ref{eq:w_av3}).
$$ \langle w \rangle = e^{\mu_z + \sigma_z\sqrt{P}}\left( 1 - \frac{\sigma_z}{\sqrt{P}} + \frac{\sigma_z^2}{P} + O\left(P^{-3/2}\right)\right)^P = $$
\begin{multline}
\label{eq:expand_w}
 = \exp\left( \mu_z + \sigma_z\sqrt{P} + P\ln\left(1 - \frac{\sigma_z}{\sqrt{P}} + \frac{\sigma_z^2}{P} +  O\left(P^{-3/2}\right)\right)\right) \rightarrow \\ \xrightarrow[]{C\rightarrow\infty} \exp\left( \mu_z + \frac{\sigma_z^2}{2}\right)
\end{multline}
To sum up, we have obtained
\begin{equation}
\langle w \rangle = \text{exp}\left( \mu_z + \frac{\sigma_z^2}{2}\right)      
\end{equation}
Proceeding in the same identical way the variance of the synaptic efficacy in the thermodynamic limit results from
\begin{equation}
\label{eq:cum_w_2}
\sigma_w^2 = \text{exp}\left[2\left( \mu_z + \sigma_z^2\right)\right]
\end{equation}
that are the cumulants of a lognormal distribution with parameters $\mu_z $ and $\sigma_z$ as expected. So the $\langle w \rangle$, $\sigma_w^2$ and $A$ change also while varying $C$. This will make the comparison between the simulated system and the mean field predictions more difficult, as a consequence of the finite size effects that are going to affect the network at finite $C$. 
\begin{multicols}{2}
  \begin{figure*}[ht!]
         \includegraphics[width=.51\textwidth]{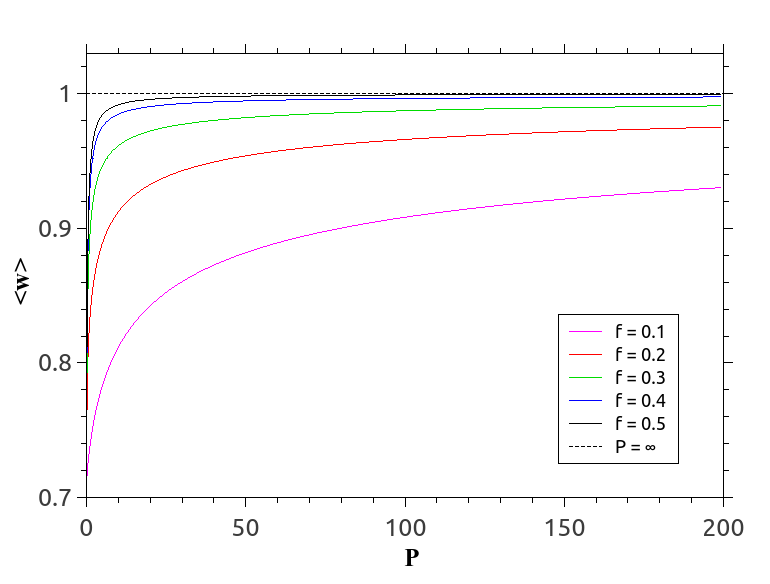}\hfill
            \includegraphics[width=.51\textwidth]{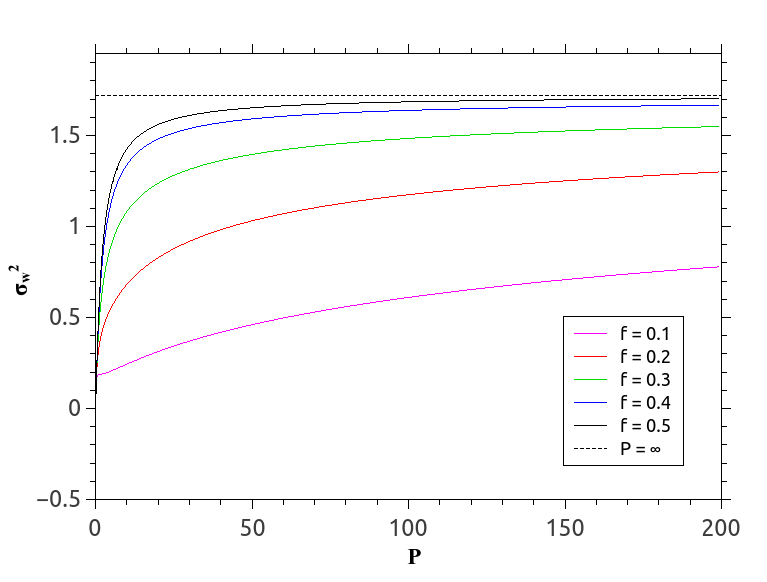}\hfill
            
            \caption{\textbf{Behaviour of $\langle w \rangle$ and $\sigma_w^2$ as functions of $P$ in the multimemory model.} The parameters of the distributions are $\sigma_z = 1$ and $\mu_z = -\frac{\sigma_z^2}{2}$. Left: $\langle w \rangle$ approaches the asymptotic limit quite fast for almost every value of $f$. The highest relative error at $P = 50$ is $<15\%$ signaling a good agreement of the synaptic statistics at finite $C$ with the mean field one even at little numbers of stored patterns.  Right: $\sigma_w^2$ seems to tend much more slower to the asymptotic limit, especially at low values of $f$. This might be a source of stronger finite size effects in numerical simulations, with respect to the one-memory case. }
           \label{fig:syn_stat}
        \end{figure*}
    \end{multicols}

It can be observed from figure (\ref{fig:syn_stat}) that, by contrast with the one-memory model, where the cumulants of the synaptic efficacies tended to the asymptotic value quite fast (reminding that $P = \alpha C$), in the multimemory model they seem to approach the same limit slower, especially at low values of $f$. This is because the multinomial distribution slowly converges to the Gaussian limit, except for the case of $f = 0.5$, where it becomes a binomial probability distribution and the Gaussian regime is reached faster. This point will corroborate the choice of $f = \frac{h_{ext}}{\langle w \rangle} = 0.5$ in the course of the future numerical simulations in Section \ref{sec:num}. This particular choice will reduce the finite size effects to better study the consistency of the simulations with the theory. 
  
\section{$\beta \rightarrow\infty$ limit of the Mean Field Equations}
\label{sec:beta}
In the limit $\beta \rightarrow \infty$ calculations simplify enormously. In fact $\phi(x), \phi^2(x) \longrightarrow \theta(x)$ with $\theta(x)$ being the Heaviside function that permits to compute Gaussian integrals analytically. This implies $\langle \nu \rangle = \langle \nu^2 \rangle$ and mean field equations can be rewritten in a very compact way. This particular limit permits to obtain one of the main results of this work, namely an explicit expression for the critical capacity of the network in terms of the control parameters of the model. In the course of this Section we are going to make use of the definition of $\textit{error-function}$
\begin{equation}
\label{eq:erf}
\text{erf}\left( x\right) = 2\int_{0}^{x}\frac{e^{-z^2}}{\sqrt{\pi}}dz
\end{equation}
\\Mean field equations (\ref{eq:mf1}), (\ref{eq:mf2}), (\ref{eq:mf3}) assume the following expressions   
\begin{equation}
 \label{eq:1_1}
   1 - 2\frac{h_{ext}}{\langle w \rangle} = f\text{erf}\left(\frac{x + y}{\sqrt{2}}\right) + (1-f)\text{erf}\left(\frac{x}{\sqrt{2}}\right) 
 \end{equation}
 \begin{equation}
 \label{eq:3}
     \sigma^2 = h_{ext}\frac{\langle w^2 \rangle}{\langle w \rangle}
 \end{equation}
\begin{equation}
\label{eq:2_2}
\frac{\sqrt{\alpha}}{B}y = \text{erf}\left(\frac{x + y}{\sqrt{2}}\right) - \text{erf}\left(\frac{x}{\sqrt{2}}\right) 
\end{equation}	
with $$B = \frac{A}{2\sigma}\hspace{1cm}x = -\frac{1}{\sigma}\left(\mu -f\frac{A}{\sqrt{\alpha}}m\right)\hspace{1cm}y = -\frac{2B}{\sqrt{\alpha}}m$$ 
Notice that equation (\ref{eq:1_1}) defines a monotonic, and thus invertible, $x$-dependent function for $y$ because the $\textit{error-function}$ is monotonic. This function for $y$ is written below
\begin{equation}
\label{eq:y_invert}
y = -x +\sqrt{2}\text{erf}^{-1}\left[\frac{1}{f}\left(1-2\frac{h_{ext}}{\langle w \rangle} - (1-f)\text{erf}\left(\frac{x}{\sqrt{2}}\right)\right)\right]    
\end{equation}
Notice that it becomes $$ y = -2x$$ when $\frac{h_{ext}}{\langle w  \rangle} = f = 0.5$. Therefore, there is only one $x^*$ such that $y = 0$. From equation (\ref{eq:y_invert}) it can be found that $$ x^* = \sqrt{2}\text{erf}^{-1}\left(1 - 2\frac{h_{ext}}{\langle w \rangle}\right)$$
On the other hand the implicit function in equation (\ref{eq:2_2}) is not bijective. Whereas $y=0$ always verifies the equations, other solutions can be obtained intersecting the line on left hand side with the sigmoid-shaped function on right hand side. A non-zero solution exists if and only if the following condition is satisfied
\begin{equation}
\label{eq:cond}
\frac{1}{\sqrt{2\pi}}e^{-\frac{x^2}{2}} > \frac{\sqrt{\alpha}}{2B}
\end{equation}
We can therefore define $$ \alpha_{max} = \frac{2B^2}{\pi}$$ that is the maximum possible $\alpha$ such that equation (\ref{eq:2_2}) shows other solutions apart from the null one. Since $y$ is linearly dependent on the order parameter $m$ we are interested in the value of $\alpha$ for which $m = 0$, meaning that the system does not manage to retrieve memories any more. From equation (\ref{eq:cond}) it can be inferred that
\begin{equation}
 \label{eq:critical1}  
\alpha_c = \alpha_{max}e^{-{x^*}^2}
\end{equation} 
that represents the value of $\alpha$ for which $y$ reaches $0$ in a continuous fashion. The fact that $y$ becomes null in such a way is proved by the fact that equation (\ref{eq:2_2}) can be Taylor expanded the second order around small values of $y$ displaying the same result for $\alpha_c$. Specifically, one gets
\begin{equation}
\frac{\sqrt{\alpha}}{B}y = \sqrt{\frac{2}{\pi}}e^{-\frac{x^2}{2}}y - \frac{x}{\sqrt{2\pi}}e^{-\frac{x^2}{2}}y^2 + O(y^3)
\end{equation}
which leads to a solution that is identically $0$ $\forall x$, and another one deriving from
\begin{equation}
\left( \frac{\sqrt{\alpha}}{B} - \sqrt{\frac{2}{\pi}}e^{-\frac{x^2}{2}}\right) = O(y)  \end{equation}
where $y$ continuously reaches the $0$ for $\alpha \rightarrow \alpha_c$. As a result
\begin{equation}
\label{eq:ac}
\alpha_c = \alpha_{max}\exp\left\{-2\left[\text{erf}^{-1}\left(1 - 2\frac{h_{ext}}{\langle w \rangle}\right)\right]^2\right\}
\end{equation}
Notice that one gets $\alpha_c = \alpha_{max}$ at $\frac{h_{ext}}{\langle w \rangle} = 0.5$.
\begin{figure}[h!!]
    
		\includegraphics[width=11cm]{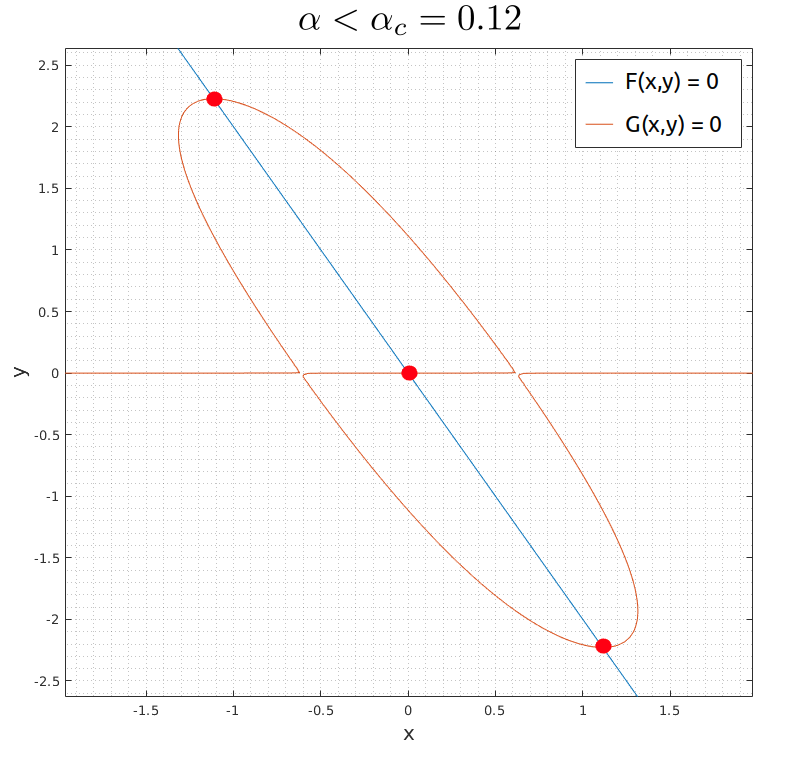}
	    \centering
		\caption{\textbf{Plot of the implicit functions derived from the mean field equations at $\beta\rightarrow\infty$ with $f = 0.5$, $\frac{h_{ext}}{\langle w \rangle} = 0.5$, $\sigma_z = 1$, $\mu_z = -\frac{\sigma_z^2}{2}$ and $\theta = 0$.} Red circles indicate the intersection of the implicit functions relative to equations (\ref{eq:1_1}) and (\ref{eq:2_2}). In \textit{blue} the $\alpha$-independent implicit function given by equation (\ref{eq:1_1}) that can be exactly expressed by $y = -2x$; In \textit{orange} the solutions to the equation (\ref{eq:2_2}) for $\alpha < \alpha_c$. Solutions of the bifurcation problem are symmetric and get closer to the $y=0$ line as $\alpha$ increases. In this case we have $\alpha_c = \alpha_{max} = 0.12$.}
		\label{fig:bif_1}
		
	\end{figure}
\\In figure (\ref{fig:bif_1}) implicit functions relative to equations (\ref{eq:1_1}), (\ref{eq:2_2}) are plotted while changing $\alpha$ at $f = 0.5$ and $\frac{h_{ext}}{\langle w \rangle} = 0.5$. Fixed points of the dynamics are represented by intersections of the implicit functions as it happens in typical bifurcation problems. We have to bear in mind that, even though figures report all the range of $x$, the solutions of the system of mean fields equation cannot exceed the range identified by the domain of the $\alpha$-dependent implicit function, namely the $F(x,y) = 0$ in the plots. In this particular case three solutions initially appear for $\alpha < \alpha_c$, two symmetric ones for $y>0$ and $y < 0$ and the other one at $y = 0$. As $\alpha$ increases the closed curve shrinks around the point $(0,0)$. The intersection between the curves reaches $y = 0$ at $\alpha_c = 0.12$ as predicted by equation (\ref{eq:ac}), undergoing a second order phase transition from the retrieval regime to the non-retrieval one. In fact, the order parameter vanishes continuously. 

\begin{multicols}{4}

    \begin{figure*}[ht!]
     \includegraphics[width=.49\textwidth]{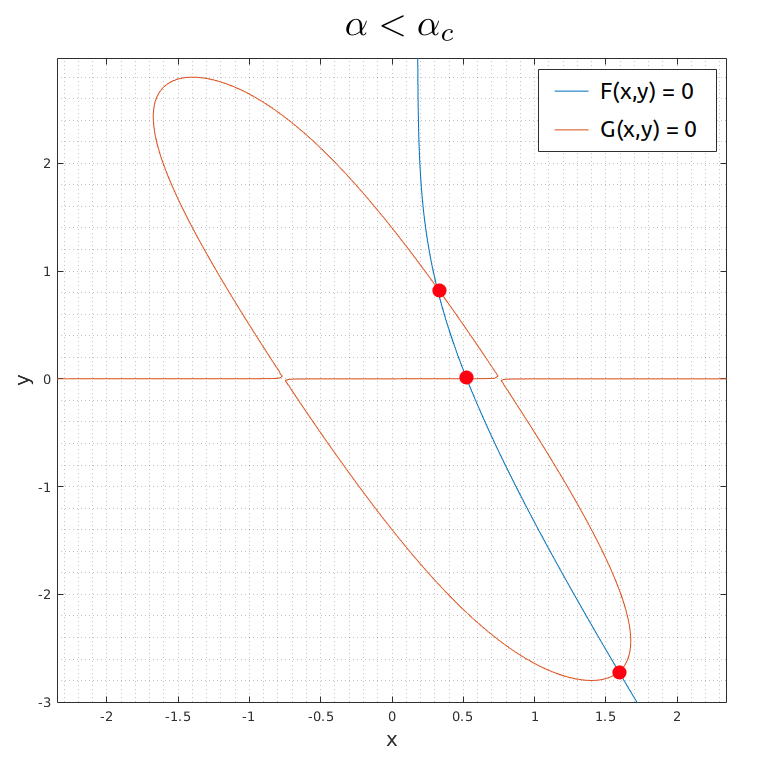}\hfill
        \includegraphics[width=.49\textwidth]{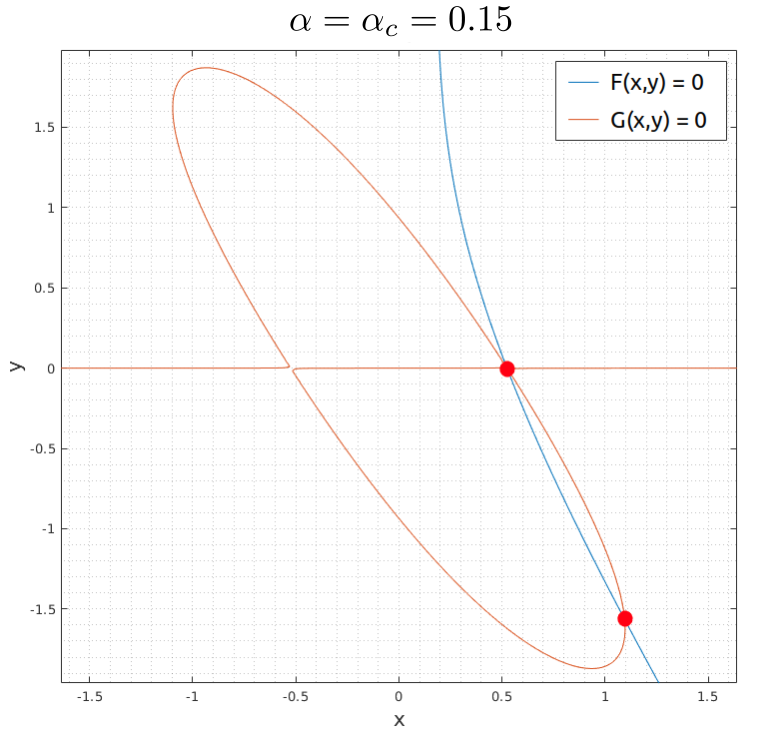}\hfill
        \includegraphics[width=.48\textwidth]{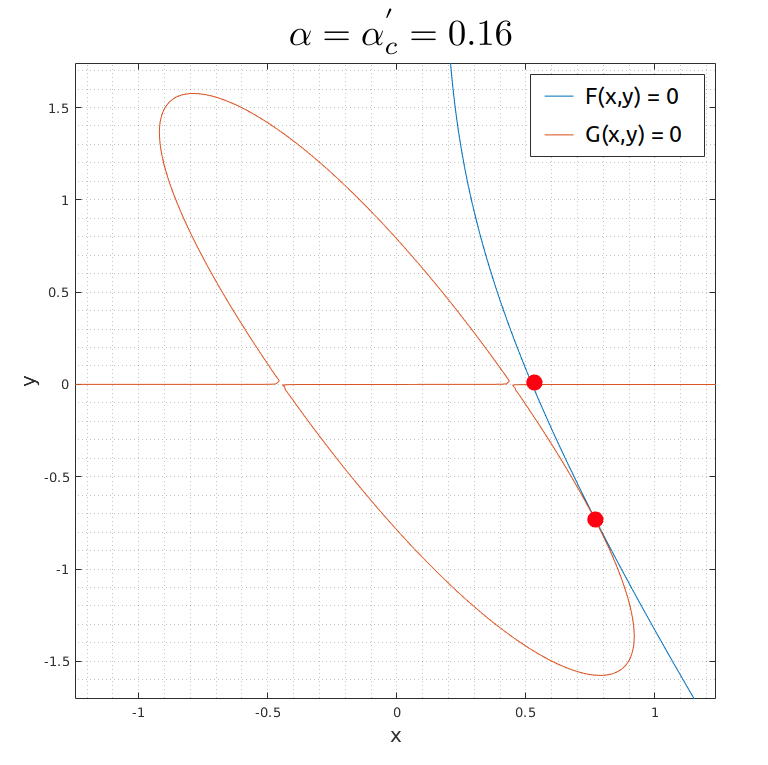}\hfill
        \includegraphics[width=.49\textwidth]{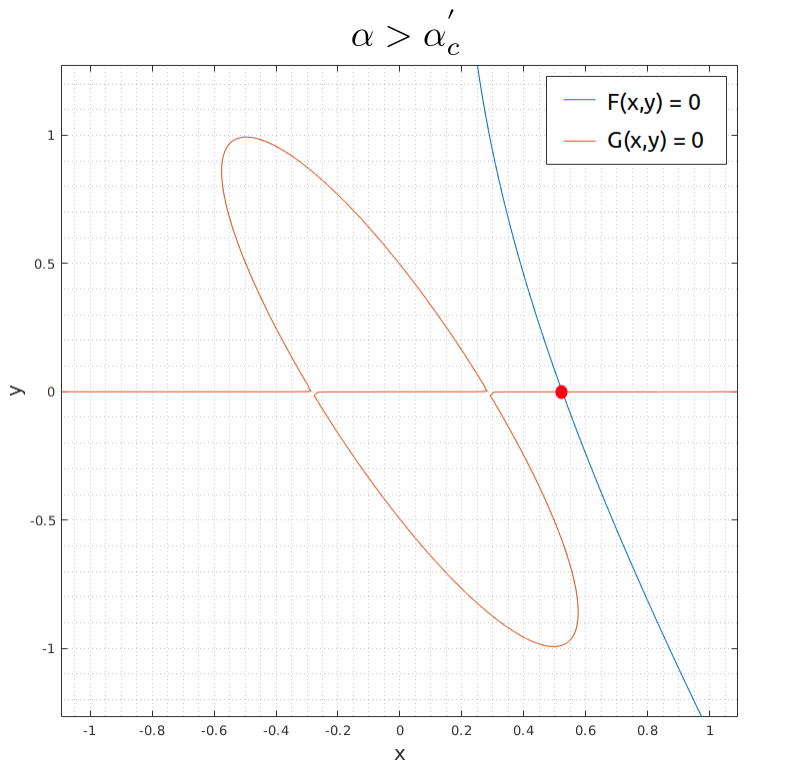}\hfill
        
    \caption{\textbf{Plot of the implicit functions derived from the mean field equations at $\beta\rightarrow\infty$ with $f = 0.3$, $\frac{h_{ext}}{\langle w \rangle} = 0.3$, $\sigma_z = 1$, $\mu_z = -\frac{{\sigma_z}^2}{2}$ and $\theta = 0$.} Red circles indicate the solutions of the system of mean field equations (\ref{eq:1_1}) and (\ref{eq:2_2}) while changing the load parameter $\alpha$. Top Left: $\alpha = \alpha_c < {\alpha_c}^{'}$, opposite solutions are asymmetric. Top Right: $\alpha = \alpha_c =  0.15$, the network undergoes a second order phase transition where the overlap vanishes continuously. Bottom Left: $\alpha = {\alpha_c}^{'} = 0.16$, the system already has still retrieval of patterns having a positive overlap with the network configuration. Bottom Right: $\alpha > {\alpha_c}^{'}$, the system has reached the non-retrieval phase where the only solution of the mean field equations is $m = 0$.}
    \label{fig:bif_2}
    \end{figure*}
\end{multicols}
On the other hand, figure (\ref{fig:bif_2}) illustrates the solutions of the system of the mean field equations for $f = 0.3$,  $\frac{h_{ext}}{\langle w \rangle} = 0.3$ when varying $\alpha$. For $\alpha < \alpha_c$ there are three asymmetric solutions. In $\alpha_c = 0.15$ the closed curve relative to (\ref{eq:2_2}) and the line of solutions of equation (\ref{eq:1_1}) intersect in $y=0$ and $x = x^*$, showing a second order phase transition. Moreover, at $\alpha_c^{'} = 0.16$ the two lines are tangent in one point: it consists of another bifurcation where the system undergoes an abrupt transition to $y = 0$. Hence, there is another critical $\alpha_c^{'}$ at which the system undergoes a first order transition to the non-retrieval phase which, however, will not be object of our study. Eventually, $y = 0$ is left as the only solution of the system and the closed curve shrinks until disappearing at $\alpha = \alpha_{max}$.

	\begin{multicols}{2}
        \begin{figure*}[h!!!]
         \includegraphics[width=.51\textwidth]{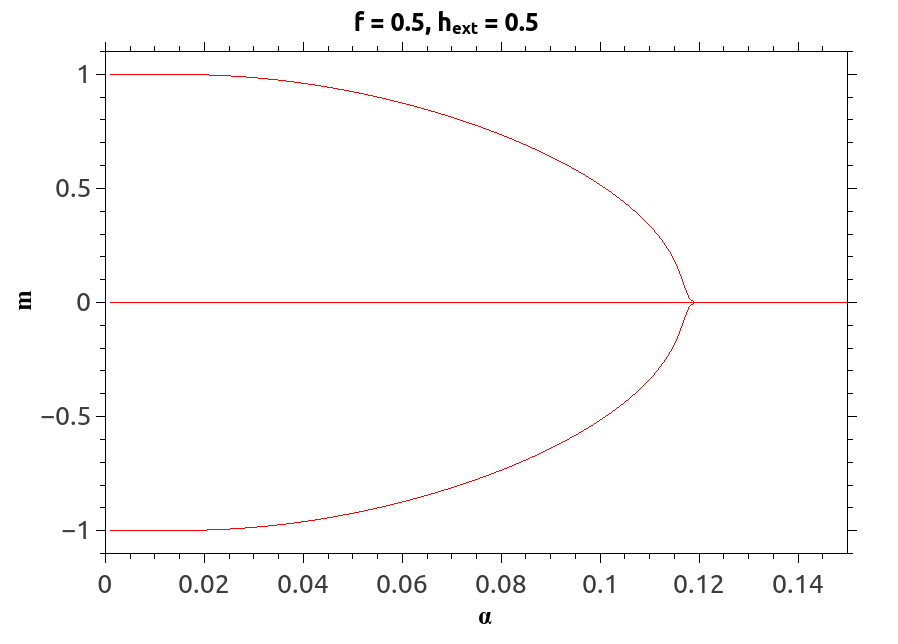}\hfill
            \includegraphics[width=.51\textwidth]{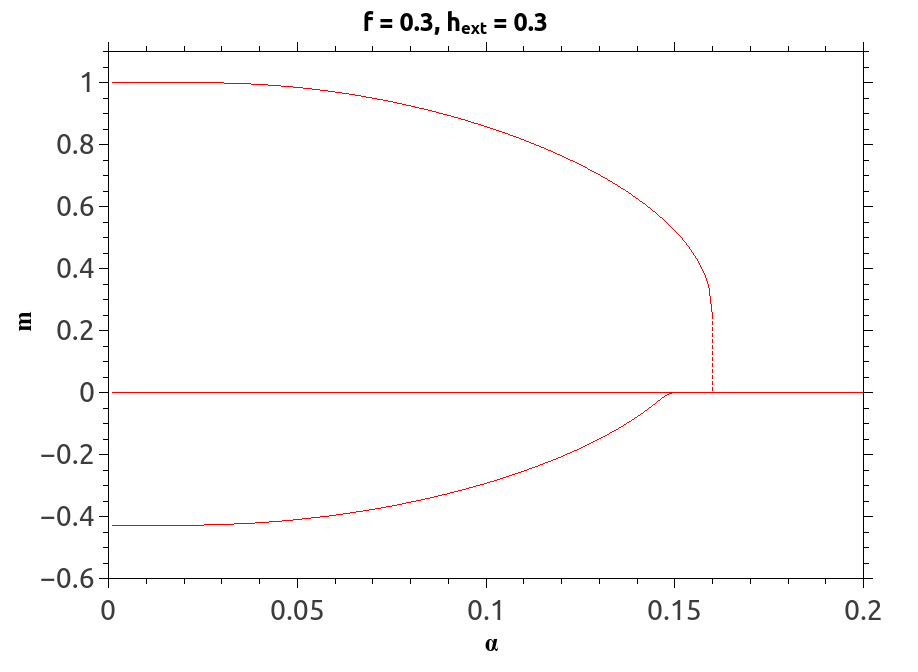}\hfill
             \caption{\textbf{Bifurcation plot representing the overlap $m$ as a function of the control parameter $\alpha$ for the two analysed cases with $\sigma_z = 1$, $\mu_z = -\frac{\sigma_z^2}{2}$ and $\theta = 0$.} Left: bifurcation plot of the network for $f = 0.5$, $\frac{h_{ext}}{\langle w \rangle} = 0.5$. A symmetric second order phase transition occurs at $\alpha_c = 0.12$. Right: bifurcation plot of the network for $f = 0.3$, $\frac{h_{ext}}{\langle w \rangle} = 0.3$. Along with a second order phase transition at $\alpha_c = 0.15$ the system undergoes a first order phase transition at $\alpha_c^{'} = 0.16$.}
            \label{fig:bif_3}           
           \end{figure*}
            
    \end{multicols} 
Phase transitions are clearly represented in figure (\ref{fig:bif_3}) where the value of the order parameter, namely the overlap relative to the intersections of the implicit functions, is plotted as a function of the control parameter $\alpha$. 

\subsection{The Sparse Coding Limit}
We are now interested in what happens to the system, and so to its memory performance, when the sparse coding limit is performed at $\beta\rightarrow\infty$. This limit consists of decreasing the average number of active sites in the pattern to $0$. 
This can be achieved setting $f = \frac{h_{ext}}{\langle w \rangle}$ and performing the limit $f \rightarrow 0$. In particular, the constraint over $h_{ext}$ implies the maximum similarity between the network configuration and the retrieved pattern at a given $\alpha$.\\ In this case the variance of the fields vanishes, since $$ \sigma^2  = \langle w^2 \rangle f \xrightarrow[]{f\rightarrow 0} 0$$ 
As a consequence the expression for $x^*$ can be approximated in the $x\rightarrow\infty$ limit, making use of the following asymptotic expansion for the error-function
\begin{equation}
\label{eq:erf_trick}
\text{erf}(x) \xrightarrow[]{x\rightarrow \infty}1 - \frac{e^{-x^2}}{\sqrt{\pi}x}
\end{equation}
and obtaining
\begin{equation}
f = \frac{1}{x^*\sqrt{2\pi}}e^{-\frac{{x^*}^2}{2}}    
\end{equation}
At the leading order one can express $x^*$ as
\begin{equation}
x^* \simeq \sqrt{-2\ln(f)} 
\end{equation}
From equation (\ref{eq:critical1}) the critical capacity can be rewritten in the sparse coding limit as
\begin{equation}
\label{eq:flnf}
    \alpha_c = \frac{2B^2}{\pi}e^{{-x^*}^2} = \left(2 B f x^*\right)^2 \simeq \frac{2A^2}{\langle w^2 \rangle}f|\ln(f)|
\end{equation}
This conclusion implies the existence of an optimal coding level, that is an optimal number of sites of the network that must be active, on average in each retrieved pattern, to maximize the memory capacity of the system. 
\begin{figure}[h!!]
\centering
		\includegraphics[width=12cm]{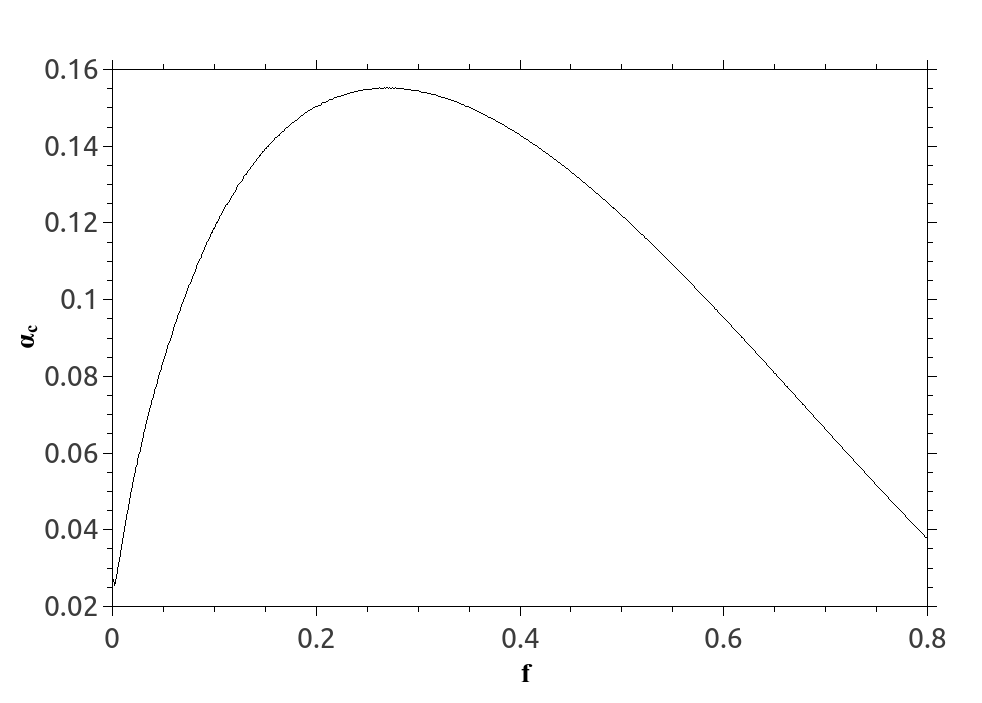}
		\label{fig:1}
		\caption{\textbf{Critical capacity $\alpha_c$ as a function of the coding level $f$.} The plot has been obtained solving mean field equations (\ref{eq:1_1}), (\ref{eq:3}), (\ref{eq:2_2}) fixing $f = \frac{h_{ext}}{\langle w \rangle}$ and choosing $\sigma_z = 1$, $\mu_z = -\frac{\sigma_z^2}{2}$ and $\theta = 0$. The line fits well the behaviour described by equation (\ref{eq:flnf}) presenting a maximum in $f \simeq 0.27$.}
		\label{fig:optimal}
		\centering
	\end{figure}
\\Figure (\ref{fig:optimal}) represents $\alpha_c$ as a function of the coding level $f$, in the conditions defined by $\sigma_z = 1$, $\mu_z = -\frac{\sigma_z^2}{2}$, obtained solving the mean field equations with $f = \frac{h_{ext}}{\langle w \rangle}$. Notice the critical capacity having a peak at $f\simeq0.27$. Hence the maximum storage is related to an average number of active sites roughly equal to $\sim 0.27C$. The critical capacity doesn't vanish completely at $f = 0$ due to neglected corrections in the computation.\\ This result is different from what was obtained in previous memory models, such as the one  developed by Tsodyks and Feigelman \cite{ref:tsodyks} where the critical capacity diverges when $f\rightarrow 0$. The difference with this specific model mainly lies in the absence of the threshold $\theta$ in our mean field equations, as a consequence of the balance condition which adjusts the fields at the fixed point near any chosen threshold of the network.  
\section{Comparing Theory with Numerical Simulations}
\label{sec:num}
Mean Field equations obtained in Section \ref{sec:one} are now compared with numerical simulations. Simulations are performed on a network of $N$ neurons with mean connectivity $C$.
The equation of dynamics (\ref{eq:dynamics}) is integrated by means of Euler method with time steps $\Delta t = O\left(\frac{1}{N}\right)$. The algorithm automatically stops when all neurons relax at the fixed point, that is, when $ |\dot{h}_i| < \delta$ $\forall i$, with $\delta = 10^{-6}$. 
\\Control parameters have been set to the following values: \vspace{0.1cm}
\begin{equation}
\label{eq:choice}
   \sigma_z = 1 \hspace{1cm} \mu_z = -\frac{\sigma_z^2}{2}  \hspace{1cm} \beta = 2 \hspace{1cm} \theta = 0 \hspace{1cm} f = \frac{h_{ext}}{\langle w \rangle} = 1/2 
\end{equation}
The choice of $\mu_z,\sigma_z$ has been made in order to obtain $\langle w \rangle = 1$ while the choice of $f,h_{ext}$ should help to reduce the finite size effects and get closer to the theoretical regime, especially for what concerns the multimemory model.\\Patterns are randomly generated according to a Bernoulli process with probability $f$, as required by the theory. Since both the one-memory model and the multimemory model have been simulated, synaptic efficacies have been assembled in a different way depending on the case, making use of the patterns previously produced. In both models we have used $w_{ij} = \exp\left(\mu_z  + z_{ij}\sigma_z\right)$ while Hebbian terms $z_{ij}$ are computed according to equation (\ref{eq:hebbian}) for the one-memory model and equation (\ref{eq:hebbian2}) for the multimemory one.  In the simulations we have considered $w_{ii} = 0$ $\forall i$, even though no significant modification to the results is observed when autapses are included at $C/N \leq 0.1$.
\subsection{Local Fields}
The first numerical study is devoted to the evaluation of the statistics of local fields. We know from the theory that fields are meant to be Gaussian in the thermodynamic limit. Moreover, in this limit the distributions associated to the active and inactive sites must have symmetric means with the same variance when $f=1/2$. 
    \begin{figure*}[h!!!]

       \centering
		\includegraphics[width=12cm]{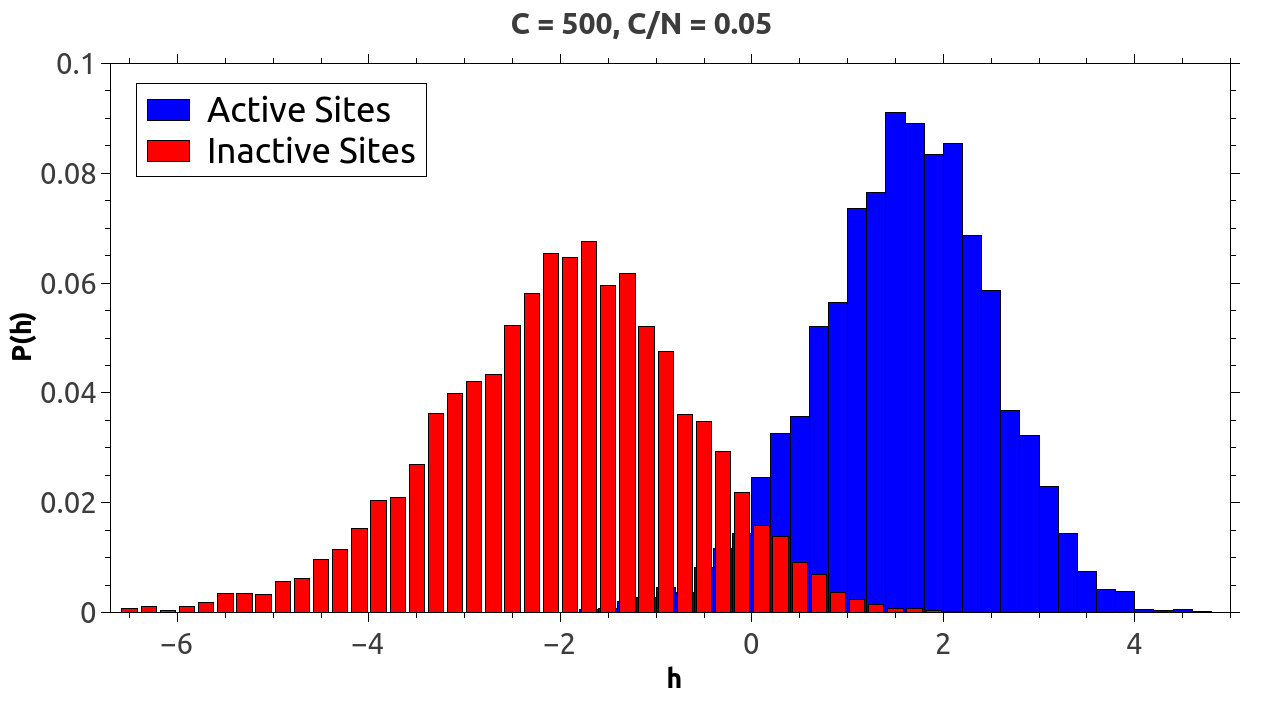}
       
        \caption{\textbf{Histograms of the fields over the active and inactive sites in the one-memory model from a simulation at $C = 500$, $C/N = 0.05$ and $\alpha = 0.05$.} Notice the asymmetry of the distributions in both mean and variance. Particularly evident is the way inactive sites show a larger dispersion of the fields, hence an overestimation of the theoretical variance, while the active ones present the opposite behaviour.}
        \label{fig:histograms1}
           
\end{figure*}
\begin{multicols}{1}
    \begin{figure*}[h!!!]
         \centering
		\includegraphics[width=12cm]{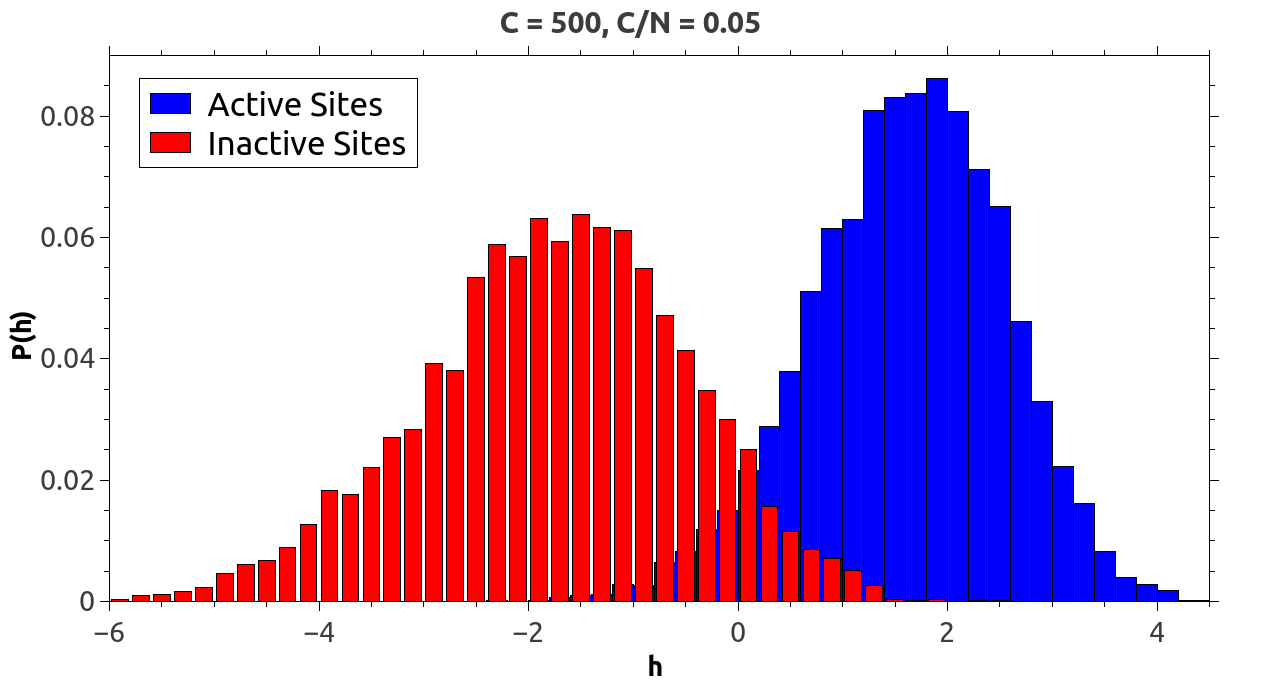}
    
        \caption{\textbf{Histograms of the fields over the active and inactive sites in the multimemory model from a simulation at $C = 500$, $C/N = 0.05$ and $\alpha = 0.05$.} As well as in the one-memory case, the asymmetry of the distributions in both mean and variance is pretty evident. Inactive sites show a larger dispersion of the fields, hence an overestimation of the theoretical variance, while the active ones present the opposite behaviour.}
        \label{fig:histograms2}
           
\end{figure*}
\end{multicols} 
Figures (\ref{fig:histograms1}$) and (\ref{fig:histograms2}$) report the histograms of the local fields associated to the active and inactive fields from, respectively, one simulation of the one-memory model and one of the multimemory model. Both simulation have been run at $\alpha = 0.05$, $C = 500$ and $C/N = 0.05$. An evident asymmetry in the distributions can be noticed from the figures, in contrast with the mean field predictions.  
Let us consider, for instance, the one-memory model. 
From five repetitions of the simulation at $C=500$, $C/N = 0.05$, $\alpha = 0.05$ we have performed the following measures

$$\overline{h_+} = 1.620\pm0.012\hspace{2cm}\overline{\sigma_+^2} = 0.834\pm0.008 $$

\begin{equation}
\label{eq:measure1}
\overline{h_-} = -2.000\pm0.020\hspace{1.95cm}\overline{\sigma_-^2} = 1.690\pm0.008
\end{equation}
Where $\overline{h} $ represents the empirical mean over the sites. They are compared with the mean field values
\begin{equation}
\label{eq:measure_th}
\langle h \rangle_+ = 1.77\hspace{2cm}\langle h \rangle_- = -1.77\hspace{2cm} \sigma^2 = 1.19
\end{equation}
Since theoretical values are all distant more than three times the standard deviations of the mean from the experimental measures, we conclude the estimates are not consistent with the mean field predictions. This effect was accurately explained by the statistics of the fields at finite $C$ that resulted from equations (\ref{eq:av_h}$), (\ref{eq:sigma_h}$). 
The mean input and its variance to the active and inactive sites at the current experimental conditions are reported below 
\begin{equation}
\label{eq:h_act}
    \langle h \rangle_{C,+} = \frac{\sqrt{C}}{2}\left[1 - \left( e^{ -\frac{1}{\sqrt{\alpha C}}}\langle \nu \rangle_+ +  e^{+\frac{1}{\sqrt{\alpha C}}}\langle \nu \rangle_- \right)\right]
\end{equation}
\begin{equation}
\label{eq:sigma_h_act}
\sigma^2_{C,+} = \frac{e}{2}\left[ e^{ -\frac{2}{\sqrt{\alpha C}}}\langle \nu^2 \rangle_+ + e^{+\frac{2}{\sqrt{\alpha C}}}\langle \nu^2 \rangle_- \right]
\end{equation}
\begin{equation}
\label{eq:h_inact}
    \langle h \rangle_{C,-} = \frac{\sqrt{C}}{2}\left[ 1 - \left( e^{ +\frac{1}{\sqrt{\alpha C}}}\langle \nu \rangle_+ + e^{-\frac{1}{\sqrt{\alpha C}}}\langle \nu \rangle_- \right)\right]
\end{equation}
\begin{equation}
\label{eq:sigma_h_inact}
\sigma^2_{C,-} = \frac{e}{2}\left[ e^{ +\frac{2}{\sqrt{\alpha C}}}\langle \nu^2 \rangle_+ + e^{-\frac{2}{\sqrt{\alpha C}}}\langle \nu^2 \rangle_- \right]
\end{equation}
 \begin{multicols}{2}
    \begin{figure*}[h!!!]
         \includegraphics[width=.51\textwidth]{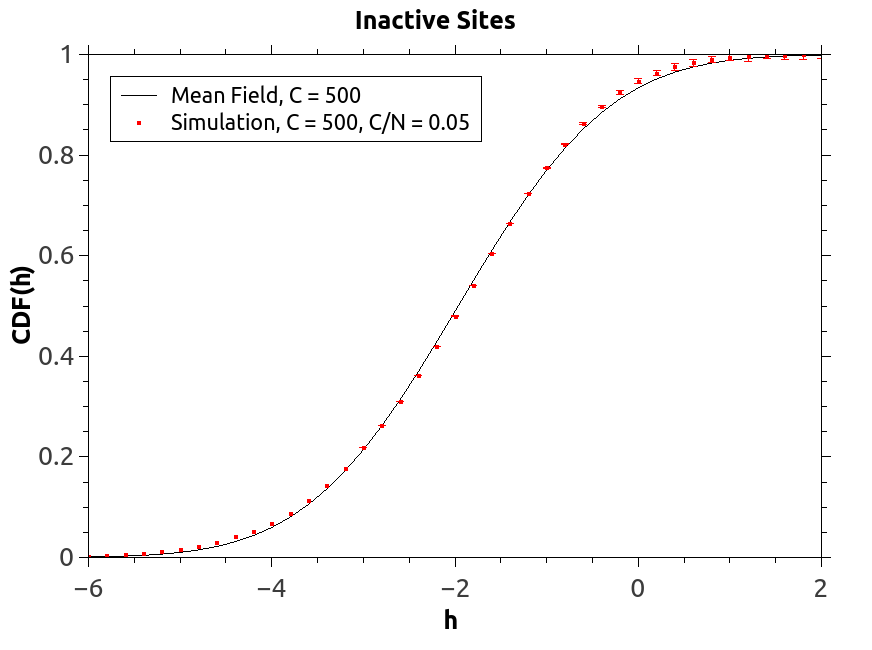}\hfill
        \includegraphics[width=.51\textwidth]{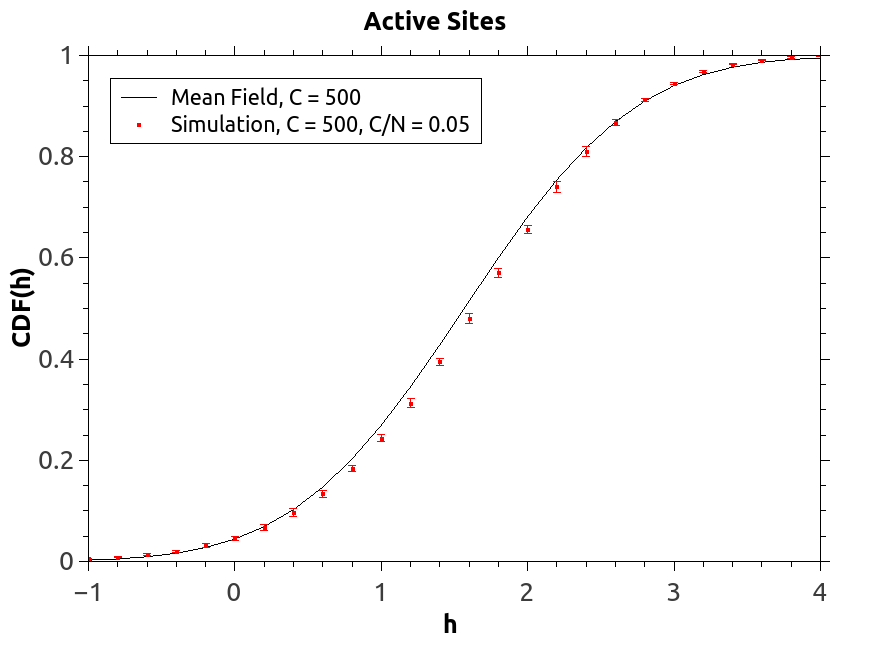}\hfill
        
        \caption{\textbf{CDF of the fields from the simulations compared to the expected one from theory at $C = 500$, $C/N = 0.05$ and $\alpha = 0.05$ in the one-memory model.} Left: Experimental points report measures of the CDF of the fields associated to the inactive sites averaged over five replicas of the network, while the errorbars are three times the standard deviation of the mean. Right: Experimental points report measures of the CDF of the fields associated to the active sites averaged over five replicas of the network, while the errorbars are three times the standard deviation of the mean. \\In both cases the test shows a good agreement between theory and the experiment at finite $C$.}
        \label{fig:finite_C}
           
\end{figure*}
\end{multicols}
Points in figure (\ref{fig:finite_C}) report the experimental cumulative density function of the local fields for the active and inactive sites of a one-memory model evaluated over five replicas of the system reproduced at $C=500$, $C/N = 0.05$, $\alpha = 0.05$. The errorbars have been set as three times the standard deviations of the mean. The continuous line, instead, represents the theoretical $CDF$ where we have assumed the theoretical behaviour to be a Gaussian with the moments expressed by equations (\ref{eq:h_act}), (\ref{eq:sigma_h_act}), (\ref{eq:h_inact}), (\ref{eq:sigma_h_inact}).
Small deviations from the line are visible, especially in figure (\ref{fig:finite_C}, right) for middle values of $h$: they are due to the fact that fields, at finite $C$, can be approximated as Gaussians but they are still not completely Gaussian (as it can be observed from figure (\ref{fig:histograms1})). Nevertheless, it is evident that equations found at finite $C$ for the statistics of the fields predict well the experimental behaviour of the system.\\
 \begin{multicols}{2}
    \begin{figure*}[h!!!]
         \includegraphics[width=.51\textwidth]{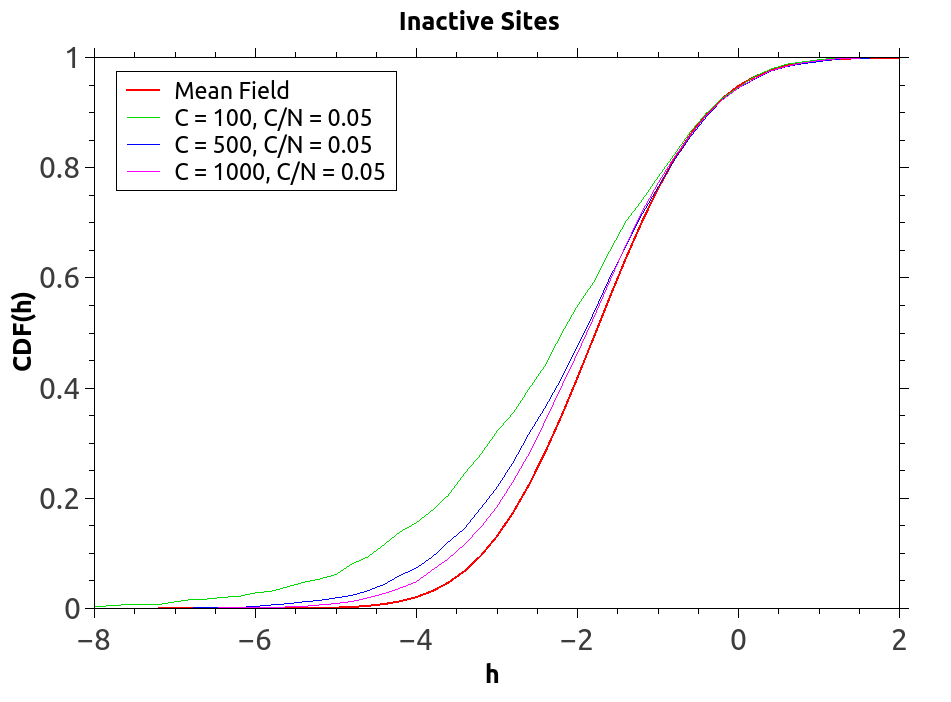}\hfill
        \includegraphics[width=.51\textwidth]{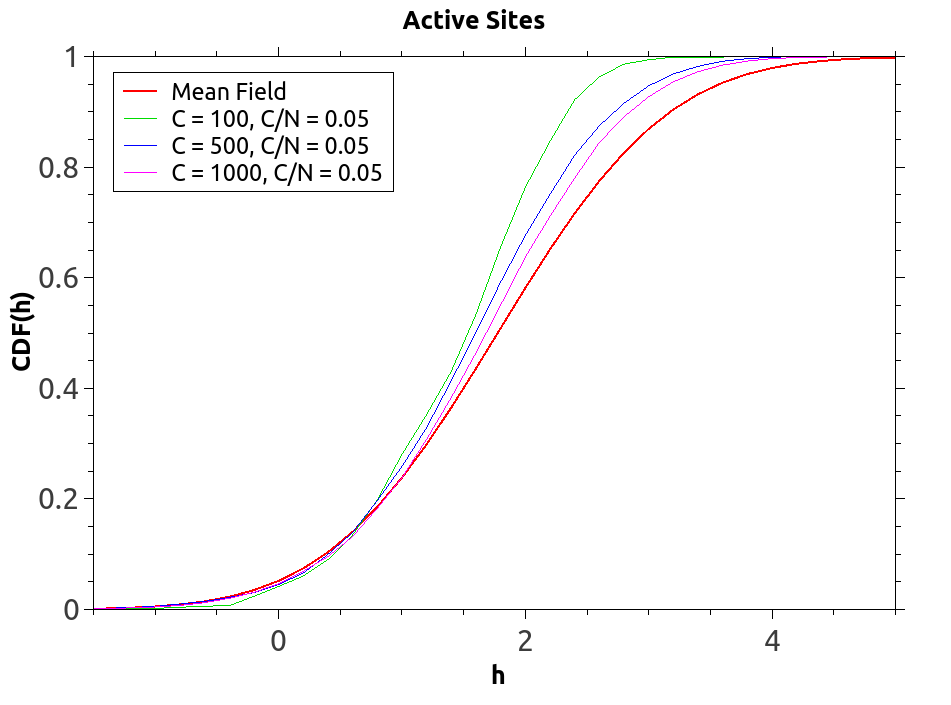}\hfill
        
        \caption{\textbf{The CDFs of fields relative to the active and inactive sites in the one-memory model are compared to the theoretical one at $\alpha = 0.05$ as $C$ increases at fixed $C/N = 0.05$.} Left: Experimental CDFs of the inactive sites at different values of $C$ are compared to the mean field CDF. Right: Experimental CDFs of the active sites at different values of $C$ are compared to the mean field CDF. The lines approach the expected trend proving a good consistency between the mean field approach and simulations. }
\label{fig:cdf_onememo}           
\end{figure*}
\end{multicols}

 \begin{multicols}{2}
    \begin{figure*}[h!!!]
        \includegraphics[width=.51\textwidth]{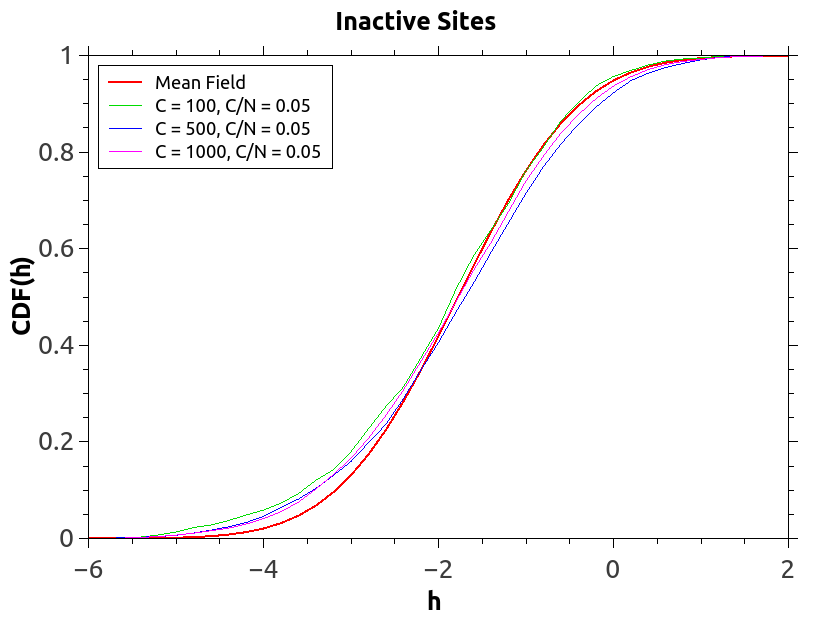}\hfill
        \includegraphics[width=.51\textwidth]{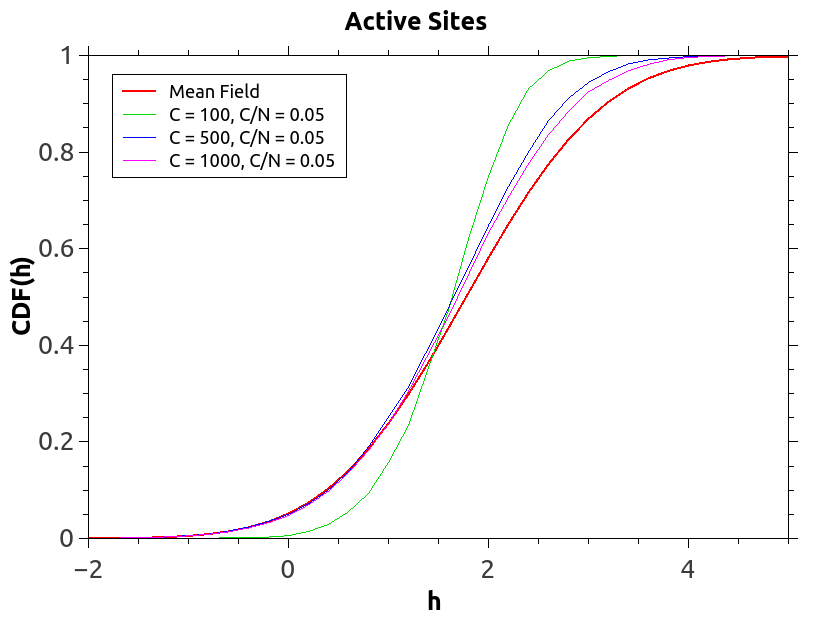}\hfill
        
        \caption{\textbf{The experimental CDFs of fields relative to the active and inactive sites in the multimemory model are compared to the theoretical CDF at $\alpha = 0.05$ as $C$ increases at fixed $C/N = 0.05$.} Left: Experimental CDFs of the inactive sites at different values of $C$ are compared to the mean field CDF. Right: Experimental CDFs of the active sites at different values of $C$ are compared to the mean field CDF. In both cases the lines approach the expected trend but the action of finite size effects on the system looks more intense than the one-memory case.}
         \label{fig:cdf_multi}  
\end{figure*}
            
\end{multicols}
Eventually, the statistics of the local field is studied for increasing values of $C$. Figures (\ref{fig:cdf_onememo}$), (\ref{fig:cdf_multi}$) report the experimental $CDF$s of the fields relative to inactive and active sites from one simulation of, respectively, the one-memory model and the multimemory one at $\alpha = 0.05$, $C = 500$ and fixed $C/N = 0.05$. The lines have been plotted and overlayed to the mean field $CDF$ relative to the asymptotic Gaussian of the local fields. In both the figures the experimental lines tend to the theoretical one in the thermodynamic limit, exhibiting a good consistency between theory and experiment. However, the accordance looks better in the one-memory model with respect to the multimemory one, due to stronger finite size effects affecting the latter. 
\begin{figure}[h!!]
\centering
		\includegraphics[width=12cm]{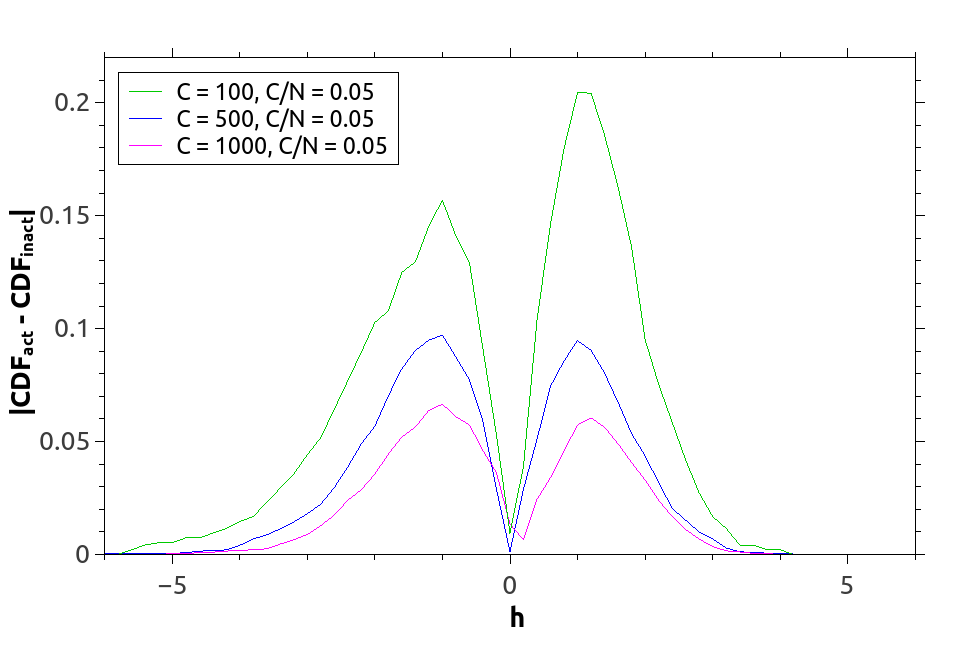}
	
		\caption{\textbf{CDFs of fields relative to active and inactive sites in the one-memory model are shifted with respect to the experimental mean and mutually compared through a difference in absolute value, at $\alpha = 0.05$ and $C/N = 0.05$.} As $C$ increases the lines flatten, suggesting that the asymmetry measured in the simulations vanishes as the mean field equations exactly predict the state of the network.}
		\label{fig:sym_onememo}
		\centering
	\end{figure}

\begin{figure}[h!!]
\centering
		\includegraphics[width=12cm]{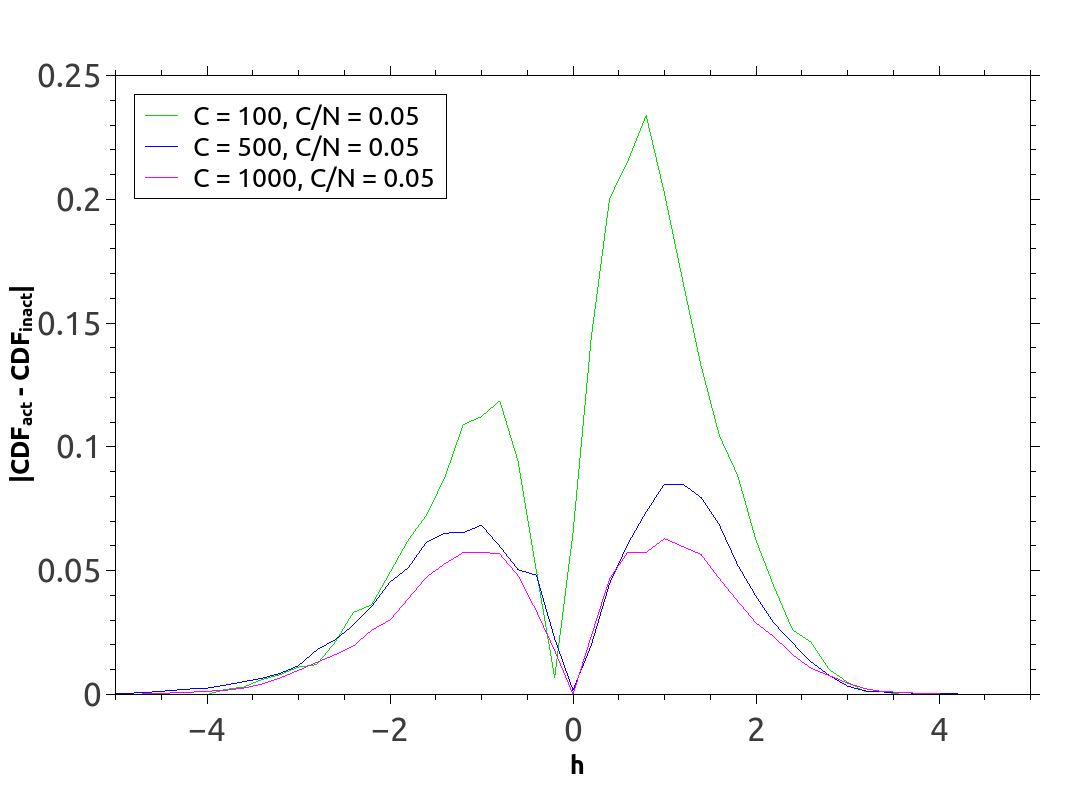}
	
		\caption{\textbf{CDFs of fields relative to active and inactive sites in the multimemory model are shifted with respect to the experimental mean and mutually compared through a difference in absolute value, at $\alpha = 0.05$ and $C/N = 0.05$.} Just like in the one-memory model, the lines flatten as $C$ increases, suggesting that the asymmetry measured in the simulations vanishes as the mean field equations exactly predict the state of the network.}
		\label{fig:comp_cdf_multi}
		\centering
	\end{figure}
It is also pointed out that the asymmetry affecting the distribution of the local fields at finite $C$ vanishes when approaching the thermodynamic limit, reaching the symmetry requested by the mean field equations. This particular trend is exhibited in figures (\ref{fig:sym_onememo}$), (\ref{fig:comp_cdf_multi}$) for the two models. In these plots the cumulative density functions of active and inactive sites from one simulation of the network have been shifted with respect to the experimental mean and reciprocally substracted. The absolute value of the substraction is reported at different values of $C$, with $C/N = 0.05$ and $\alpha = 0.05$. In both the models the curves tend to flatten as $C$ increases, meaning that distributions become symmetric in the thermodynamic limit. 
	
\subsection{Overlap}
A second numerical study consists of comparing the experimental overlap, with the mean field predictions. Five simulations of the one-memory and multimemory models ran at $C = 100$, $C/N = 0.05$. Every run is performed starting from an initial configuration that coincides with: the only stored pattern if we are simulating the one-memory model; one of the $P$ stored patterns, picked at random, if we are working with the multimemory model. The overlap is measured at the fixed point of the dynamics making use of the formulas (\ref{eq:over}), (\ref{eq:over2}) depending on the case. When the multimemory model is analysed, we consider the overlap relative to the initial pattern. Since the bifurcation plot at $f = \frac{h_{ext}}{\langle w \rangle} = 1/2$ is symmetric, as represented in figure (\ref{fig:phase_diag}), we are taking the absolute value of the overlap and focus on the positive branch of the solutions.
\begin{figure}[h!!]
\centering
		\includegraphics[width=13cm]{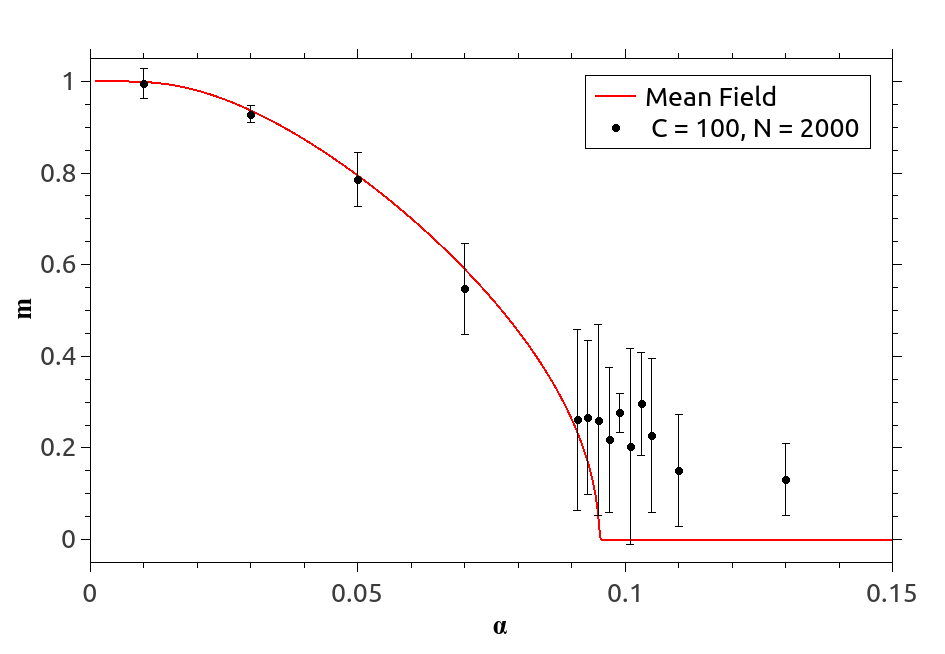}
	
		\caption{\textbf{Plot of the experimental overlap overlayed with the theoretical trend predicted by mean field equations at $C = 100$, $C/N = 0.05$.} Experimental points are the mean of measures from five replicas of the network and errorbars are three times the standard deviations of the mean. Even at low values of $C$ there is a good agreement between theory and simulations. As expected from the statistical mechanics the fluctuations of the order parameter grow near the second order phase transition.}
		\label{fig:over_onememo}
		\centering
	\end{figure}

\begin{figure}[h!!]
\centering
		\includegraphics[width=13cm]{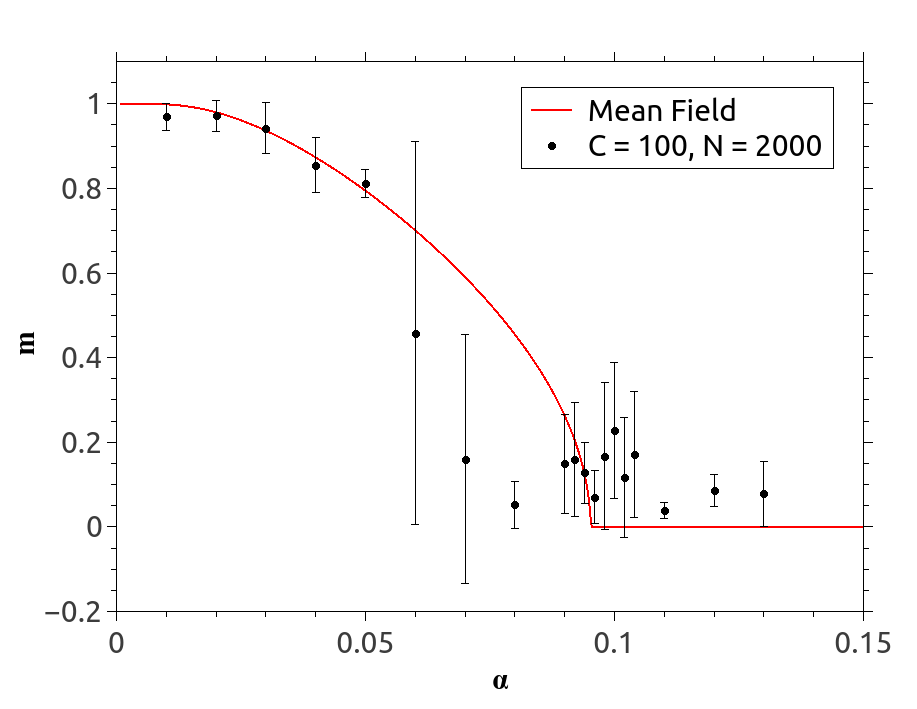}
	
		\caption{\textbf{Plot of the experimental overlap overlayed with the theoretical trend predicted by mean field equations at $C = 100$, $C/N = 0.05$.} Experimental points are the mean of measures from five replicas of the network and errorbars are three times the standard deviations of the mean. The accordance seems much worse than the one-memory case. This is probably mainly due to the more evident non-gaussianity of the Hebbian terms at finite values of $C$. }
		\label{fig:over_multi}
		\centering
	\end{figure} 
The experimental overlaps are reported in figures (\ref{fig:over_onememo}) and (\ref{fig:over_multi}) as functions of $\alpha$. Points are given by the mean over the five simulations of the network at each value of $\alpha$ and errorbars are three times the standard deviation of the mean.\\
As for the one-memory model (figure \ref{fig:over_onememo}), measures follow the theoretical line up to the critical capacity $\alpha = \alpha_c$ where the fluctuations of the points grow as dictated from the statistical mechanics of critical phenomena. On the other hand, the multimemory model (figure \ref{fig:over_multi}) appears to fit the theoretical trend until $\alpha = 0.05$. After that value of the storage capacity points deviate from the mean field predictions. This behaviour is attributable to finite size effects being stronger in the multimemory case, since the slowness with which Hebbian terms reach the Gaussian limit implies the statistics of the efficacies not to be lognormal. As a consequence of these effects, we might suppose that fixed points become unstable and the system retrieves some kind of spurious state. 
	\begin{figure}[h!!]
\centering
		\includegraphics[width=13cm]{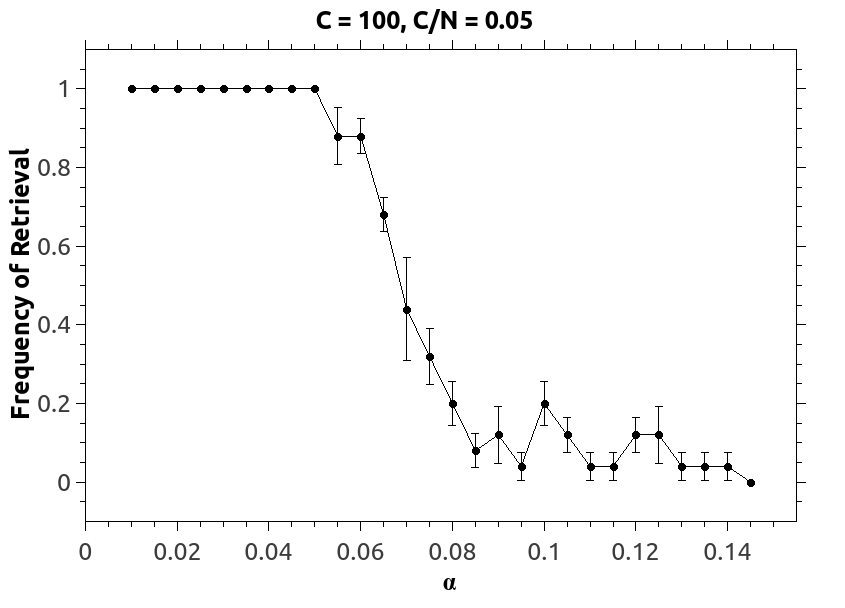}
	
		\caption{\textbf{Plot of the frequency of retrieval of the initial pattern as a function of $\alpha$ at $C = 100$ and $C/N = 0.05$.} Each point is the mean of five measures of the frequency on different networks with the same parameters and initial condition, whereas the errorbars are the standard deviation of the mean. Notice the frequency to be identically one up to $\alpha = 0.05$. For larger values of $\alpha$ the the fixed points become unstable due to the finite size of the network.}
		\label{fig:over_freq}
		\centering
	\end{figure}	
Concerning this particular point, figure (\ref{fig:over_freq}) reports a study performed on the multimemory model at the same experimental conditions used to plot figure (\ref{fig:over_multi}). The frequency of retrieval of the initial pattern is reported as a function of $\alpha$. Points are the mean of the measures over five repetitions of the simulations and errors consist of the standard deviation of the mean. It should be noticed that the frequency is identically one up to $\alpha = 0.05$ and then decreases dramatically, signaling an instability of the fixed points. We know that the plotted curve should fit a sigmoid function (see supplementary notes from \cite{ref:brunel}) having the critical capacity as its inflection point. It is expected the line to become steeper shifting its inflection point towards $\alpha = \alpha_c$ as the thermodynamic limit is reached. In the analysed case, where $\alpha_c \simeq 0.095$, we can estimate the inflection point by looking at the $\alpha$ at which $\text{Freq} = 1/2$: we obtain $\overline{\alpha_c}\simeq 0.07$. This yields a relative error of $\sim 30\%$ to an estimation of the critical capacity. We thus expect this error to get lower when $C$ increases and $C/N$ decreases. 

\chapter*{Conclusions}
\addcontentsline{toc}{chapter}{Conclusions}
In this work we have proposed a model of balanced neural network describing a population of inhibitory neurons which is able to retrieve memories. \\Chapter \ref{chap:2} presents a preliminary study of the model where no patterns are still stored in the network. Synaptic efficacies are thus generated at random in such a way to ensure strictly inhibitory interactions among neurons. It has been proved both analytically and numerically that, when the external input does belong to a precise interval, the system correctly operates in the balanced regime. In this case the statistics of the local fields present both the mean and the variance being $O(1)$ quantities. Furthermore, from the balance condition, we have a linear expression of $\langle \nu \rangle$ in terms of the external input that is
$$ \langle \nu \rangle = \frac{h_{ext}}{\langle w \rangle}$$
The extreme dilution of connectivity matrix, along with its being asymmetric, imply the Gaussian distribution of the local fields. This last property of the network, consistently with the balance, has permitted to derive a set of mean field equations for the model. These equations consist of a powerful tool to predict the macroscopic state of the network in the thermodynamic limit given a set of control parameters of the model. \\At last we have recovered, from the study of the stability of the fixed points of the dynamics, the known stability condition $$ \langle w^2 \rangle \langle {\phi^{'}}^2 \rangle < 1$$ When this condition is respected the network remains balanced and predictions done through the mean field equations are correct, otherwise the network does not manage to reach an equilibrium state described by the theory.\\ 
In Chapter \ref{chap:3} the balanced network has been integrated by embedding random patterns in the synaptic efficacies, consistently with the Hebbian rule that is implemented in Hopfield-like models. The peculiarity of our model, though, is that the network preserves its balance property in the thermodynamic limit.\\ As a first original result of our research, mean field equations for the new structured network were derived by treating the correlation between the network configuration at the fixed point and the retrieved pattern. As for the random network, mean field equations can be solved numerically and permit to recover the neural state of the network in the thermodynamic limit at the fixed point of the dynamics. The retrieval state of the network is indicated by the non zero value of an order parameter we called overlap. \\In the particular case of $\beta \rightarrow \infty$ Gaussian integrals of the f-I function have been computed analytically and the expression of the critical capacity $\alpha_c$ of the model has been obtained from the mean field equations (\ref{eq:ac}). We got
$$ \alpha_c = \frac{\langle w \rangle A^2}{2\pi \langle w^2 \rangle h_{ext}}\exp\left\{-2\left[\text{erf}^{-1}\left(1 - 2\frac{h_{ext}}{\langle w \rangle}\right)\right]^2\right\}$$
This quantity represents the capacity at which the network undergoes a continuous phase transition to the phase of non-retrieval of the memories. \\Another important theoretical result has emerged from the sparse coding limit of the mean field equations in the $\beta\rightarrow\infty$ limit. According to our theory the critical capacity vanishes when $f\rightarrow0$ and $f = \frac{h_{ext}}{\langle w \rangle}$ (see figure (\ref{fig:optimal})) implying the existence of an optimal number of active sites in the pattern to maximize the capacity of the balanced network. This outcome is different from many known memory models where the critical capacity diverges at $f = 0$ \cite{ref:tsodyks} \cite{ref:will}. 
\\These results have been proved to be valid for two versions of the structured balanced network. The first version has been called $\textit{one-memory}$ model. This network concretely stores one single pattern and it has been introduced as a first manageable version of the network where the statistics of the fields at finite values of $C$
can be exactly predicted without worrying about spurious correlations with eventual uncondensed patterns. Another improvement brought by this model is the fact that synaptic efficacies are closer to be described as lognormal variables at low values of $C$, reducing the finite size effects that might disturb the comparison theory-experiment.\\The second version, that is also the most biologically plausible one, is the $\textit{multimemory}$ model, where $P$ patterns are randomly generated and stored in the network. From the extreme dilution of the connectivity matrix the system loses its correlation with the uncondensed patterns in the thermodynamic limit, implying the multimemory model to correctly respect the same mean field equations found for the single memory case. 
\\Eventually, the comparison between theory and the numerical simulations of both the one-memory model and the multimemory has sanctioned the consistency between the measures over the system and the the theoretical results in the limit $C,N \rightarrow\infty$. Even though the system is far from respecting the DGZ dilution limit that originally justifies the mean field equations to be exact in the thermodynamic limit, simulations show a behaviour that is fully coherent with the one predicted by the theory. Nevertheless, by construction of the model, the multimemory case shows stronger finite size effects with respect to the one-memory version of the network, as it was envisaged from the analysis of the statistics of the synaptic efficacies at finite $C$.\\
\\
This work makes a prediction over the optimal number of active sites in patterns encoded from real neural networks. Our findings give room to interesting future comparisons between these results and real data available from the experiments. It is reasonable to expect the number of active sites in the memories retrieved by an inhibitory population of neurons to be consistent with the value obtained from the mean field equations, that is $\simeq 30\%$ of the mean connectivity $C$ of the network. \\ \\
Another point that has not been developed by the present work, but that certainly completes the analysis of our model, is the study of the stability of the solutions of the system of mean field equations found in Chapter \ref{chap:3} Section \ref{sec:one}. The same procedure implemented in Chapter \ref{chap:2} Section \ref{sec:stab} can be applied. When deriving the solutions of the mean field equations, represented for instance in figure (\ref{fig:bif_3}), stable solutions have not been discriminated from unstable ones. From such a study we should expect to find a fully unstable branch at $m = 0$ and two stables ones relative to the non-zero overlap solutions, at least up to the critical capacity.\\In the case of the random network we have seen the stability as dependent on the gain parameter. It would be then interesting to search for an optimal $\beta$ such that the balanced network is both stable and maximizes its storage capacity. This estimate would give us the maximum biological capacity of the balanced network. We might thus set ourselves on the critical capacity and increase $\beta$ until this optimal value is measured. 
\backmatter
\phantomsection

\end{document}